\documentclass[a4paper,12pt]{article}
\usepackage{jheppub2} 
\usepackage{hyperref}
\usepackage{adjustbox}
\hypersetup{
	bookmarks=true,         
	unicode=false,          
	pdftoolbar=true,        
	pdfmenubar=true,        
	pdffitwindow=false,     
	pdfstartview={FitH},    
	pdftitle={My title},    
	pdfauthor={Author},     
	pdfsubject={Subject},   
	pdfcreator={Creator},   
	pdfproducer={Producer}, 
	pdfkeywords={keyword1} {key2} {key3}, 
	pdfnewwindow=true,      
	colorlinks=true,       
	linkcolor=red,          
	citecolor=purple,        
	filecolor=magenta,      
	urlcolor=blue,           
	linktocpage=true
}

\usepackage{graphics}
\usepackage{rotating}
\usepackage{subfigure}
\usepackage{physics}
\usepackage{amssymb}
\usepackage{xcolor}
\usepackage{pstricks}
\usepackage{amsfonts}
\usepackage{mathrsfs,amsmath}
\usepackage{epsfig}
\usepackage{amsmath,amssymb,amsthm,graphicx,latexsym}
\usepackage{float}
\usepackage{textcomp}
\usepackage{array}
\usepackage{booktabs}
\usepackage{dsfont}

\usepackage{ulem}

\newcommand{\be}{\begin{equation}}
\newcommand{\ee}{\end{equation}}
\newcommand{\beno}{\begin{equation*}}
\newcommand{\eeno}{\end{equation*}}
\newcommand{\bc}{\begin{center}}
\newcommand{\ec}{\end{center}}
\newcommand{\bea}{\begin{eqnarray}}
\newcommand{\eea}{\end{eqnarray}}
\newcommand{\beano}{\begin{eqnarray*}}
\newcommand{\eeano}{\end{eqnarray*}}
\newcommand{\bml}{\begin{subequations}}
\newcommand{\eml}{\end{subequations}}
\newcommand{\bfig}{\begin{figure}}
\newcommand{\efig}{\end{figure}}
\newcommand{\ag}{\alpha}
\newcommand{\bg}{\beta}
\newcommand{\gam}{\gamma}

\newcommand{\slashed}{\hspace{-1.1ex}/}

\newcommand{\bmat}{\begin{pmatrix}}
	\newcommand{\emat}{\end{pmatrix}}
\begin{document}
	
	\title{ \textcolor{red}{
	Relating the curvature of De Sitter Universe to Open Quantum Lamb Shift Spectroscopy 	
	}}
	
	
 \author[a]{Hardik Bohra,}	
\affiliation[a]{Department of Physics and Astronomy,
National Institute of Technology Rourkela,
Odisha - 769008, India.}		
\author[b,c,d]{Sayantan Choudhury,
	\footnote{\textcolor{red}{\bf Corresponding author, Alternative
			E-mail: sayanphysicsisi@gmail.com}. ${}^{}$}
			\footnote{\textcolor{blue}{\bf NOTE: This project is the part of the non-profit virtual international research consortium ``Quantum Aspects of the Space-Time \& Matter~(QASTM)" }. ${}^{}$}	}
\affiliation[b]{Quantum Gravity and Unified Theory and Theoretical Cosmology Group, Max Planck Institute for Gravitational Physics (Albert Einstein Institute),
	Am M$\ddot{u}$hlenberg 1,
	14476 Potsdam-Golm, Germany.}
	\affiliation[c]{National Institute of Science Education and Research, Jatni, Bhubaneswar, Odisha - 752050,India.}
\affiliation[d]{Homi Bhabha National Institute, Training School Complex, Anushakti Nagar, Mumbai-400085, India.}	
 \author[e]{
Prashali
Chauhan,}	
\affiliation[e]{Ashoka University, Sonepat - 131029,
India.}
 \author[f,g]{
Purnima Narayan,}	
\affiliation[f]{Theoretical Cosmology Group, Max Planck Institute for Gravitational Physics (Albert Einstein Institute),
	Am M$\ddot{u}$hlenberg 1,
	14476 Potsdam-Golm, Germany.}
\affiliation[g]{Institute of Physics and Astronomy, University of Potsdam, Campus Golm, Haus 28, Karl-Liebknecht-Straße 24/25, 14476 Potsdam-Golm, Germany.}
\author[c,d]{Sudhakar Panda,}	   
			\author[i]{Abinash Swain}	
\affiliation[i]{Department of Physics, Indian Institute of Technology Gandhinagar,
Palaj, Gandhinagar - 382355, India.}

\emailAdd{hardikbohra.nitr@gmail.com, sayantan.choudhury@niser.ac.in, chauhan.prashali@gmail.com,  narayan.purnima0@gmail.com,panda@niser.ac.in,abinashswain2010@gmail.com     }

\abstract{In this paper, we explore the connection between the curvature of the background De Sitter space-time with the spectroscopic study of entanglement of two atoms. Our set up is in the context of an Open Quantum System (OQS), where the two atoms, each having two energy levels and represented by Pauli spin tensor operators projected along any arbitrary direction. The system mimic the role of a pair of freely falling {\it Unruh De-Witt detectors}, which are allowed to non-adiabatically interact with a conformally coupled massless probe scalar field which has the role of  background thermal bath. The effective dynamics of this combined system takes into account of the  non-adiabatic interaction, which is commonly known as the {\it Resonant Casimir Polder Interaction} ({\bf RCPI}) with the thermal bath. Our analysis revels that the {\bf RCPI} of two stable entangled atoms in the quantum vacuum states in OQS depends on the de Sitter space-time curvature relevant to the temperature of the thermal bath felt by the static observer. We also find that, in OQS, {\bf RCPI} produces a new significant contribution appearing in the effective Hamiltonian of the total system and thermal bath under consideration. We find that the {\it Lamb Shift} is characterised by a decreasing inverse square power law behaviour, $L^{-2}$,  when inter atomic Euclidean distance, $L$, is much larger than a characteristic length scale, $k$, which is the inverse surface gravity of the background De Sitter space. If the  background space time would have been Minkowskian this shift decreases as, $L^{-1}$, and is independent of temperature. Thus, we establish a connection between the curvature of the De Sitter space-time with the {\it Lamb Shift} spectroscopy.}


\maketitle
\flushbottom

\section{\textcolor{blue}{\bf \large Introduction}}

Quantum theory of open system has been popular in recent times mainly because it provides a platform to study time evolution of system interacting weakly with its environment. See for examples ref.~\cite{OQS1} for the development of this subject. This development in Open Quantum SYstems (OQS) plays a crucial role for studying physical quantum phenomena in nature because they are mostly not isolated from its environment. It is important to develop a theoretical framework for treating these non-adiabatic interactions in order to obtain an accurate understanding of quantum mechanical systems. Theoretical tools and techniques developed in the context of OQS have been very successful in the context of quantum optics, quantum measurement theory, quantum statistical mechanics, quantum information theory, quantum thermodynamics, quantum cosmology, quantum biological systems. In a most generalised prescription, the time evolution of OQS is described by the non-adiabatic interactions between the physical system and its thermal environment. Consequently, the dynamical behaviour of the OQS cannot be accurately described using unitary time evolution operators alone after integrating out the bath degrees of freedom from the environment. The time evolution of OQS can be explicitly determined by solving the effective master equations of motion, also known as {\it Gorini–Kossakowski–Sudarshan–Lindblad} (GSKL) master equations, from which one can understand the non-unitary time evolution of the reduced density matrix of the system. In this context, the time evolution actually describes the time dependent behaviour of the system at different stages over time and also the dynamical behaviour of the observables that are associated with the OQS. The theory of OQS treats the system with dependent degrees of freedom as a subsystem in a much larger thermal bath \cite{Breuer:2002pc}. Due to the complicated structure of the environmental degrees of freedom finding exact analytical solution of  {\it Gorini–Kossakowski–Sudarshan–Lindblad} (GSKL) master equations are extremely difficult for practical purposes. 

Due to this difficulty, a variety of approaches have been developed in the present context. In this connection, a common objective is to derive the reduced time dependent description of the OQS wherein the dynamical behaviour of the quantum system are considered explicitly and the corresponding bath dynamics are described implicitly to know about the underlying physics of OQS under consideration. 
The seemingly obvious approach to deal with such systems, which is simulating evolution of both system and environment, would be naive as its complexity evolves exponentially. The OQS can no longer be defined by a pure state and to study the modified time evolutionary dynamics in presence of weak interaction between the system and thermal bath, density matrix formalism is required. We can study smaller subsystems to get across this problem by incorporating probabilistic description, where the quantum state of the subsystem is described by density operator. In an OQS, the effects of dissipation and decoherence is introduced by the environment degrees of freedom. The induced effects of decoherence and dissipation owing to the system is introduced by an operator more famously known as the Lindbladian \cite{Gorini:1975nb}. While dealing with OQS, following crucial assumptions play significant role to describe the underlying physics:
\begin{enumerate}
\item \underline{\textcolor{red}{\bf Assumption~I:}} \\
The combination of the system and the thermal environment is treated as a {\it large closed system}. Therefore we can assume that {\it the time evolution is given by a unitary transformation generated by a global Hamiltonian}.
\item \underline{\textcolor{red}{\bf Assumption~II:}} \\
The interaction between system and environment is considered as {\it Markovian}, which describes {\it the state of the quantum mechanical system in the next instant and dependent only on the current moment, not in the past}. In short, the interaction between the system and thermal bath describes a phenomena without memory. This approximation is justified when the OQS under consideration has enough time for the system to achieve equilibrium state before being quantum mechanically perturbed again by non-adiabatic interactions with its thermal environment. 
\item \underline{\textcolor{red}{\bf Assumption~III:}} \\
The interaction between the system and thermal bath is considered to be {\it weak} in nature, which implies that {\it the only change over time we see originates in the open system}. This helps us to treat the time evolution of the system in a perturbative manner. 
\item \underline{\textcolor{red}{\bf Assumption~IV:}} \\
In the present context, additionally we have assumed that {\it system and thermal bath are completely uncorrelated at the initial time}.
\end{enumerate}

In this paper, we have studied the connection of the curvature of de Sitter spacetime with the two atomic spectroscopy using the framework of OQS. To construct the theoretical set up we use a two body entangled system, which is described by two identical atoms. This two entangled OQS is considered as our theoretical probe which mimic the role of a pair of freely falling {\it Unruh de Witt detectors}. In this connection, it is important to note that, the concept of quantum entanglement describes a physical phenomenon which deals with strong correlations between the two atomic quantum states of the entangled particles \cite{Schrodinger:1935ur,Schrodinger:1936re} in OQS. Since this non-local property of quantum mechanics seemed very puzzling, Einstein, Podolsky and Rosen (EPR) argued for the existence of hidden variables in the context of quantum theory \cite{Einstein:1935et}. Later Bell proposed a set of inequalities to test their existence which when untrue, support the non-locality of quantum theory \cite{Bell:1964js}. Bell's inequalities have to be violated in inflationary era to get persistent long range quantum correlation in the early universe cosmology and for this, the concept of quantum entanglement is commonly used. The violation of the Bell-inequality in de Sitter space has been addressed with axionic Bell pair to a great extent in several works \cite{Choudhury:2017qyl,Choudhury:2017bou}. Bell-inequality violation in various cosmological scenario with quenched time dependent mass profiles in de Sitter space has also been extensively studied in \cite{Choudhury:2016pfr,Choudhury:2016cso}.
  
   In the present context, we use mainly the concept of two body quantum entanglement in OQS which currently serves as one of the leading candidates to study the long range quantum correlation and many other unknown physics of the de Sitter space. The idea being that two or more entangled atoms which mimic the role of particle detectors in OQS can be used to measure the spacetime curvature of de Sitter spacetime and the consequent thermal effects, some of which we will discuss in detail in the following sections. In refs.~\cite{Choudhury:2018fpj,Hu:2012ed,Hu:2013ypa,Zhang:2014wua}, the authors have explored many more aspects of quantum entanglement in the background of a fluctuating scalar field in de Sitter space.
   In OQS the interaction of single particle with conformally coupled massless scalar field in de Sitter space is identical to the interaction of the same with thermal bath environment in Minkowski spacetime, making it difficult to distinguish between the two frameworks. This problem has led to further investigations with two body entangled OQS in order to shed light on the quantum structure and nature of the spacetime of our universe.  
 
In a more general prescription, it is important to note that, when an accelerated particle detector moves through an external field in its vacuum state, spontaneous excitation may occur. The particle behaves as in the presence of a thermal bath, giving rise to an {\it Unruh Temperature}, which is proportional to its acceleration \cite{Unruh:1983ms,Unruh:1976db}. The excitation rate, of such a particle detector vanishes when it is at rest, but is found to be non-zero when uniformly accelerated. Same result is achieved if we analyse the vacuum state of the external conformally coupled scalar field in the frame of the accelerating particle detector. This is known as {\it thermaliszation} and shows that these accelerating detectors act as an OQS in the context of two body physics. Very recently, it has been explicitly shown that scalar fields coupled conformally to a de Sitter spacetime can be treated as an out-of-equilibrium system with a fluctuating background that introduced the thermal excitations in de Sitter space \cite{Choudhury:2018rjl}. Such a detector-system combined set up sees the above thermalisation phenomena as a manifestation of the decoherence and dissipation due to the interactions with the surrounding. Furthermore, the approach followed in this work is also used to investigate the {\it Gibbons-Hawking effect} \cite{Gibbons:1977mu}, which can be treated as a consequence of the thermalisation effect of the vacuum fluctuation of the external field in the de Sitter spacetime. These two phenomena encode the thermal behaviour of de Sitter spacetime. 

Another very important outcome of vacuum fluctuations of quantised fields is {\it Casimir effect} which is heavily influenced by the curvature of de Sitter spacetime \cite{Casimir:1948dh,Casimir:1986na,Bordag:2009zzd}. Casimir effect and the associated {\it Casimir-Polder interaction} (CPI) has been verified experimentally in multitudes of systems at microscopic and macroscopic levels. It has been used to study properties of quantum entanglement, long range effect of field correlations, Unruh effect etc. The spacetime curvature can alter the {\it Casimir-Polder interaction} \cite{Emig:2006uh,Bordag:2018log,Bordag:2017qnh} between the atoms \cite{Buhmann:2006ji} in two body OQS. The CPI in the de Sitter spacetime \cite{Tian:2014hwa} has been investigated in detail in refs~\cite{Zhang:2014wua,Zhang:2012nz,Tian:2013lna,Huang:2017fuy}. In this paper, we aim to distinguish the curvature of de Sitter spacetime and Minkowski flat spacetime using \textit{Casimir-Polder interaction} as a theoretical probe. In order to distinguish a conformally coupled massless scalar field with a de Sitter background to that of a Minkowski spacetime interacting with a thermal bath in which the excited \textit{Bell-states} of the pair of atoms interacts with the background field placed at thermal bath via the exchange of real photons \cite{Dalvit:2011tua,Maclay:2001gv,Kong:1998ic,Milonni:1994xx,GellMann:1954fq}. Additionally, in ref.~\cite{Tian:2016uwp} the authors have explicitly shown the use of {\it Resonance Casimir-Polder interaction} (RCPI) to detect curvature of de Sitter spacetime. In this paper, we use this methodology to know about the curvature of de Sitter by analysing the {\it Lamb shift spectra} computed from the two entangled atomic OQS. In \cite{VerSteeg:2007xs}, the authors have provided a methodology to distinguish between thermal Minkowski spacetime and de Sitter spacetime using the entanglement power. However, here we will use RCPI for the same purpose. Entanglement between quantum field modes, have been discussed in \cite{Ball:2005xa} (for scalar fields) and \cite{Fuentes:2010dt} (for fermionic fields). Also Unruh-DeWitt detectors and entanglement harvesting have been studied in case of different spacetimes in these beautiful papers \cite{Kerr, BHvacuum, AdS1, AdS2}. Some other useful references regarding quantum entanglement are \cite{Ent1, Ent2, Ent3, Ent4, Ent5}. An excellent review of entanglment in expanding universe is given in \cite{MartinMartinez:2012sg}, where they also briefly discuss the information encoded in such entanglement and other prospects. Dynamics of such a two qubit system in Unruh spacetime has been studied in \cite{Tian:2014jha}. In \cite{OqeSC}, the authors have calculated som information theoretic measures for such a system.

 The various discussions in this paper can be summarised as follows:
 
In \underline{section \ref{dS}}, we present a brief overview of the dS spacetime. The background metric has a hyperbolic geometry with a cosmological constant $\Lambda$ as a solution of Einstein's field equations. First, we present the static dS metric.  Then compute the two body two point Wightman functions for the external conformally coupled massless scalar field. Once we have these Wightman functions, we calculate the Fourier transform of these which we use later to calculate Lamb Shift Hamiltonian and Lindbladian.

In \underline{section \ref{OpenQFT2atoms}}, we describe our model two atomic~\footnote{By atom we mean simple a qubit. Please see section \ref{OpenQFT2atoms} for more clarity.} system conformally coupled to a scalar field in a thermal bath in de Sitter spacetime. Here, we allow our spins to have any orientation in the spacetime. Then we investigate the reduced dynamics of a pair of {\it Unruh-De-Witt detectors} or atoms to obtain th Master Equation. Then using the Fourier and Hilbert transforms of the Wightman functions (calculated in section \ref{dS}) we write down the matrix elemnts of Lamb shift Hamiltonian and Lindbladian. The details of calculation is given in appendix \ref{coeff} and \ref{int}.

Further, in \underline{section \ref{lss}}, we have derived the expression for the energy shift to explicitly study the role of {\it Lamb Shift spectroscopy} \cite{Bethe:1947id,Welton:1948zz} in the context of the two atomic entangled OQS. To serve this purpose, first of all we actually construct all possible two atomic entangled states out of the individual ground and excited states of the two atoms. By using the tensor product we have constructed four possible combination of the two body entangled states, which are ground, excited, symmetric and antisymmetric states. Further, using these two body entangled quantum states we compute the expectation values of the most relevant part of the effective Hamiltonian, known as the {\it Lamb Shift Hamiltonian}. Using these energy shifts we distinguish between the geometry of de Sitter spacetime from that of Minkowski spacetime. Here we have actually expressed the spectroscopic energy shifts in terms of the Euclidean distance between the two atoms separated in the space time and the inverse of the surface gravity of De Sitter space. Further, we have expressed the temperature of the thermal bath in terms of the surface gravity of de Sitter space. This allows us to connect macroscopic description of the environment in terms of the microscopic quantum mechanical observables of the two entangled OQS under consideration as well the geometry of our universe. Here particularly the microscopic observables can be characterised through few sets of spectroscopic integrals (see appendix \ref{int}) which is in principle divergent in nature in QFT. To collect only the finite contributions out of these integrals we apply the {\it Bethe regularisation technique}. At the end one can express the surface gravity in terms of the curvature of de Sitter space, where curvature is related to the positive cosmological constant. This helps us to measure the curvature of de Sitter space from microscopic quantum spectroscopy.

FInally, we end the article with discussion given in conclusion and presenting all the necessary relevant calculations in the Apendices that follow.

\section{\textcolor{blue}{\bf \large Geometry of de Sitter space}}
\label{dS}
The de Sitter space describes a universe with constant positive curvature and having the same degree of symmetry as a Minkowski spacetime, which also fits in with the picture of our current universe quite well. Since we want to work in this spaceyime, we will have a brief overview of the de Sitter spacetime in this section. Here our prime objective is to compute the two body two point (Wightman) correlation function between two atom conformally coupled with a massless probe external scalar field in de Sitter space. We will use the results obtained in this section in further sections to calculate our desired results. To serve this purpose we start with the background metric which is represented by the surface of the following hyperboloid:
\be \label{kop1}z^2_0-z^2_1-z^2_2-z^2_3-z^2_4=-\alpha^2=-\frac{3}{\Lambda} \ee 
which describes a solution of the {\it Einstein's field equations} with the following radius of the hyperboloid:
\be 
\alpha=\sqrt{\frac{3}{\Lambda}}>0~~~.\ee
Here, 
$\Lambda$ is the cosmological constant with positive signature in De Sitter space and the corresponding embedded metric in the five dimensional Minkowski space is given by \cite{Birrell:1982ix,Polarski:1989bv}:
\be
ds^{2}=\left(dz^{2}_{0}-\sum^{4}_{p=1}dz^{2}_{p}\right)=\left(dz^{2}_{0}-dz^{2}_{1}-dz^{2}_{2}-dz^{2}_{3}-dz^{2}_{4}\right)
\ee
By applying the following appropriate parametrisation one can express the five dimensional Minkowski flat metric in terms of a static four dimensional De Sitter metric, as given by:
\bea
\begin{aligned}
z_{0} &= \sqrt{\alpha^{2}-r^{2}}~\sinh(\frac{t}{\alpha}) \\
z_{1}  &= \sqrt{\alpha^{2}-r^{2}}~\cosh(\frac{t}{\alpha}) \\
z_{2} &= r\cos \theta \\
z_{3}  &= r\sin \theta \cos \phi \\
z_{4}  &= r\sin \theta \sin \phi
\end{aligned}
\eea
which are consistent with the equation of the surface of the hyperboloid in five dimension as expressed in Eq~(\ref{kop1}).

With the above parametrisation the static de Sitter metric in four dimensions can be expressed as:
\be
ds^{2}=\left(1-\frac{r^{2}}{\alpha^{2}}\right)dt^{2}-\left(1-\frac{r^{2}}{\alpha^{2}}\right)^{-1}dr^{2}-r^{2}(d\theta^{2}+\sin^{2}\theta d\phi^{2})
\ee
which is actually characterised by $(t,r,\theta,\phi)$ in spherical polar coordinate in four dimensions. In the present context, to compute the explicit contributions of the two body Wight function of the probe scalar field present in the external thermal bath the geometry of the re-parametrised four dimensional De Sitter space play the most crucial role. In the next subsection, we actually compute these two body Wightman function in detail.

\subsection{\textcolor{blue}{\bf \large Two atomic Wightman functions for probe external field in de Sitter space}}
To compute the expression for the each of the entries of the two body Wight function of the probe scalar field present in the external thermal bath we use the four dimensional static de Sitter geometry of our space-time. In this set of coordinate system in four dimension, the Klein-Gordon field equation for the massless conformally coupled external probe scalar field for the non-adiabatic environment can be expressed as:
\bea 
\left[\frac{1}{\cosh^3\left(\frac{t}{\alpha}\right)}\frac{\partial}{\partial t}\left(\cosh^3\left(\frac{t}{\alpha}\right)\frac{\partial}{\partial t}\right)-\frac{1}{\alpha^2\cosh^2\left(\frac{t}{\alpha}\right)}{\bf L}^2\right]\Phi(t,\chi,\theta,\phi)=0 \eea
where ${\bf L}^2$ is the {\it Laplacian differential operator} in the three dimensions characterised by the coordinate $(\chi,\theta,\phi)$ , which is explicitly defined as:
\be
{\bf L}^2=\frac{1}{\sin^2\chi}\left[\frac{\partial}{\partial\chi}\left(\sin^2\chi\frac{\partial}{\partial\chi}\right)+\frac{1}{\sin\theta}\frac{\partial}{\partial\theta}\left(\sin\theta\frac{\partial}{\partial\theta}\right)+\frac{1}{\sin^2\theta}\frac{\partial^2}{\partial\phi^2}\right]
\ee
where we introduce a new coordinate $\chi$ which is related to the radial coordinate $r$ as:
$ r=\sin\chi.$
Now, in the present context of discussion we are not exactly interested 
to solve the wave function for the probe scalar field. Since our prime objective is to compute the two body two point correlation function for the probe external scalar field we construct the two body Green's function. 

From the geometrical structure of the four dimensional static De Sitter metric it is obvious that the coordinate singularity, $\left(1-r^2/\alpha^2\right)^{-2}\rightarrow \infty$, which appears at the point $r=\alpha$, and this is identified to be the {\it cosmological horizon} in the present context. Here it is important to note that, in flat space there is no problem to define the vacuum state ($i.e.$ Minkowski vacuum) of the quantum field in open quantum mechanical system. But for curved space the definition of the vacuum state is more complicated in OQS than the result obtained in the context of flat space. For the computation of Wightman function from the present two entangled atomic open quantum set up, in the curved space we choose specifically the de Sitter invariant isometric ${\rm SO}(1,4)$ group to connect the idea of spectroscopic energy shift with the geometry of De Sitter space-time. In this computation the isometric vacuum state ($i.e.$ Bunch Davies and $\alpha$-vacua) is actually identified with the quantum mechanical state of open system described by the conformally coupled massless probe scalar field.
Now, the corresponding two point correlation function, often known as the Wightman function for massless probe scalar field takes the following form \cite{Birrell:1982ix}:
\be
G^{+}(x,x')=-\frac{1}{4\pi^{2}}\frac{1}{(z_{0}-z'_{0})^{2}-\Delta z^{2}-i\epsilon}
\ee
where, $\epsilon$ represents an infinitesimal constant, which is appearing in the representation of Wightman function in the $i\epsilon$ prescription. Also, we define the distance between two static atoms localized at the coordinates $(r,\theta,\phi)$ and $(r,\theta^{'},\phi)$, appearing in this computation as:
\bea 
\Delta z^{2}&=&(z_{1}-z'_{1})^{2}+(z_{2}-z'_{2})^{2}+(z_{3}-z'_{3})^{2}+(z_{4}-z'_{4})^{2}\nonumber\\
&=&\left(\alpha^2-r^2\right)\left[\cosh\left(\frac{t}{\alpha}\right)-\cosh\left(\frac{t^{'}}{\alpha}\right)\right]^2+r^2\left[\left(\cos\theta-\cos\theta^{'}\right)^2
+\left(\sin\theta-\sin\theta^{'}\right)^2\right]\nonumber\\
&=&\left(\alpha^2-r^2\right)\left[\cosh\left(\frac{t}{\alpha}\right)-\cosh\left(\frac{t^{'}}{\alpha}\right)\right]^2+4r^2\sin^2\left(\frac{\Delta\theta}{2}\right)\nonumber\\
&=&\left(\alpha^2-r^2\right)\left[\cosh\left(\frac{t}{\alpha}\right)-\cosh\left(\frac{t^{'}}{\alpha}\right)\right]^2+L^2
\eea 
Here  $L$ represents the Euclidean distance between the coordinates $(r,\theta,\phi)$ and $(r,\theta',\phi)$, which is defined as:
\be 
L=2r\sin\left(\frac{\Delta\theta}{2}\right)
\ee
where, the angular difference $\Delta\theta$ is defined as,
$\Delta\theta=\theta-\theta^{'}.$ Here
both the signatures of the $\Delta\theta$ is allowed in the present context depending on the relative ordering of the angular coordinates $\theta$ and $\theta^{'}$. Additionally, the parameter $\epsilon$in the present context  represents an infinitesimal constant.The corresponding two body Wightman function between two space-time points for massless probe scalar field can be expressed as~\footnote{From the two atomic two body system we get four Wightman functions in the present context due to the interaction between the two atoms with the external probe conformally coupled massless scalar field. The diagonal entries of the two body Wightman function represents the auto correlation function of the atom 1 and atom 2 with the external probe scalar field present in the thermal bath respectively. Also it is import to note that,  these diagonal entries are same as we have considered two identical in our computation. On the other hand, the off diagonal entries of the Wightman function represent the cross correlation between the atom 1 and atom 2 with the external probe scalar field. Since in the present context the Wightman function is symmetric  and constructed due to the interaction of two identical atoms with the external probe scalar field, the contributions appearing from the off-diagonal entries are exactly same. }:
\be 
\footnotesize{G(x,x')=\begin{pmatrix} ~
	G^{11}(x,x')~~~ &~~~ G^{12}(x,x')~ \\
	~G^{21}(x,x')~~~ &~~~ G^{22}(x,x') ~
\end{pmatrix}
=\begin{pmatrix} ~
	\langle \Phi({\bf x_{1}},\tau)\Phi({\bf x_{1}},\tau')\rangle~~~ &~~~ \langle \Phi({\bf x_{1}},\tau)\Phi({\bf x_{2}},\tau')\rangle~ \\
	~\langle \Phi({\bf x_{2}},\tau)\Phi({\bf x_{1}},\tau')\rangle~~~ &~~~ \langle \Phi({\bf x_{2}},\tau)\Phi({\bf x_{2}},\tau')\rangle ~
\end{pmatrix}}
\ee
where the components of the two body Wightman function can be expressed as:
\be
\begin{aligned}
&\underline{\textcolor{red}{\bf Two ~atomic~Wightman ~function~representing~auto correlation:}}\nonumber\\
G^{11}(x,x')  &= G^{22}(x,x') \\ 
& =\langle \Phi({\bf x_{1}},\tau)\Phi({\bf x_{1}},\tau')\rangle \\
&=\langle \Phi({\bf x_{2}},\tau)\Phi({\bf x_{2}},\tau')\rangle \\
&=- \frac{1}{4\pi^{2}} \frac{1}{\left\{(z_{0}-z'_{0})^{2}-(z_{1}-z'_{1})^{2}-i\epsilon\right\}} \\
& =- \frac{1}{4\pi^{2}} \frac{1}{\left\{(\alpha^{2}-r^{2})\left[\left\{\sinh\left(\frac{t}{\alpha}\right)-\sinh\left(\frac{t'}{\alpha}\right)\right\}^{2}-\left\{\cosh\left(\frac{t}{\alpha}\right)-\cosh\left(\frac{t'}{\alpha}\right)\right\}^{2}\right]-i\epsilon\right\}}\\
& =- \frac{1}{4\pi^{2}} \frac{1}{\left\{2(\alpha^{2}-r^{2})\left[\cosh\left(\frac{t-t'}{\alpha}\right)-1\right]-i\epsilon\right\}} \\
& =- \frac{1}{4\pi^{2}}\frac{1}{\left\{4(\alpha^{2}-r^{2})\sinh^2\left(\frac{t-t'}{2\alpha}\right)-i\epsilon\right\}}\\  
& =- \frac{1}{4\pi^{2}}\frac{1}{\left\{4(\alpha^{2}-r^{2})\sinh[2](\frac{\Delta \tau}{2\sqrt{g_{00}}\alpha})-i\epsilon\right\}}\\
& = -\frac{1}{4\pi^{2}}\frac{1}{\left\{4k^{2}\sinh[2](\frac{\Delta \tau}{2k})-i\epsilon\right\}}\\         
& =- \frac{1}{16\pi^{2}k^{2}}\frac{1}{\sinh^2\left(\frac{\Delta \tau}{2k}-i\epsilon\right)}
\end{aligned}
\ee
\be
\begin{aligned}
&\underline{\textcolor{red}{\bf Two ~atomic~Wightman ~function~representing~cross correlation:}}\nonumber\\
G^{12}(x,x')   &= G^{21}(x,x') \\
& = \langle \Phi({\bf x_{1}},\tau)\Phi({\bf x_{2}},\tau')\rangle\\
&= \langle \Phi({\bf x_{2}},\tau)\Phi({\bf x_{1}},\tau')\rangle \\
&= -\frac{1}{4\pi^{2}}\frac{1}{(z_{0}-z'_{0})^{2}-\Delta z^{2}-i\epsilon}\\
&=-\frac{1}{4\pi^{2}} \left\{(\alpha^{2}-r^{2})\left[\left(\sinh\left(\frac{t}{\alpha}\right)-\sinh\left(\frac{t'}{\alpha}\right)\right)^{2}-\left(\cosh\left(\frac{t}{\alpha}\right)-\cosh\left(\frac{t'}{\alpha}\right)\right)^{2}\right] - i\epsilon \right. \\
&~~~~~~~~~~~~~~~~~~~~~\left. -r^{2}\left[(\cos \theta-\cos \theta ')^{2} -(\sin \theta -\sin \theta')^{2}\right]\right\}^{-1} \\
&=-\frac{1}{4\pi^{2}} \frac{1}{\left\{2(\alpha^{2}-r^{2})\left[\cosh\left(\frac{t-t'}{\alpha}\right)-1\right]+2r^{2}[\cos(\theta-\theta')-1]-i\epsilon\right\}} \\
&=-\frac{1}{4\pi^{2}}\frac{1}{\left\{2(\alpha^{2}-r^{2})[\cosh(\frac{t-t'}{\alpha})-1]-4r^{2}\sin[2](\frac{\theta-\theta'}{2})-i\epsilon\right\}} \\
&=-\frac{1}{4\pi^{2}}\frac{1}{\left\{4(\alpha^{2}-r^{2})[\sinh[2](\frac{t-t'}{2\alpha})-i\epsilon]-4r^{2}\sin[2](\frac{\theta-\theta'}{2})\right\}}\\
& =-\frac{1}{16\pi^{2}k^{2}}\frac{1}{\left\{\sinh[2](\frac{\Delta \tau}{2k}-i\epsilon)-\frac{r^{2}}{k^{2}}\sin[2](\frac{\Delta \theta}{2})\right\}}
\end{aligned}
\ee
where we use the following identity:
\bea 
\sinh(\frac{\Delta \tau}{2k}-i\epsilon)=\sinh(\frac{\Delta \tau}{2k})\cosh(i\epsilon)-\cosh(\frac{\Delta \tau}{2k})\sinh(i\epsilon)
\eea
where $\epsilon$ being an infinitesimal constant which we have already mentioned earlier. Furthermore, we have used the fact that since $\epsilon$ is an infinitesimal constant then one can approximate:
\bea 
&&\sinh(i\epsilon)=i\sin(\epsilon){\sim i\epsilon} \\
&&\cosh(i\epsilon)=\cos(\epsilon){\sim 1} 
\eea
As a result, we get the following simplified result:
\be
\sinh(\frac{\Delta \tau}{2k}-i\epsilon)\sim\sinh(\frac{\Delta \tau}{2k})-i\epsilon\cosh(\frac{\Delta \tau}{2k})
\ee
Additionally, we have used the following definitions:
\bea 
&&k=\sqrt{g_{00}}\alpha = \sqrt{1-\left(\frac{r}{\alpha}\right)^{2}}\alpha = \sqrt{\alpha^{2}-r^{2}} \\ 
&&\Delta \tau=\tau-\tau'=\sqrt{g_{00}}(t-t')= \sqrt{1-\left(\frac{r}{\alpha}\right)^{2}}(t-t')= \sqrt{\alpha^{2}-r^{2}}\left(\frac{t-t'}{\alpha}\right)=k\left(\frac{t-t'}{\alpha}\right)~~~~~~~~~~~~ \eea with $\tau$ being the proper-time in the co-moving frame of the pair of atoms and $k$ represents the surface gravity in the present context.

Now that we have our Wightman functions, for later use, here we calculate the Fourier transform of the two point field correlation functions in frequency ($\omega$) space for external massless probe scalar field as follows:

\begin{eqnarray} \label{fourier}
\small{\begin{aligned} 
\mathcal{G}^{11}(\omega) = \mathcal{G}^{22}(\omega) &= -\int_{-\infty}^{\infty}d\Delta \tau~\frac{e^{i\omega \Delta \tau}}{16\pi^{2}k^{2}\sinh[2](\frac{\Delta \tau}{2k}-i\epsilon)}
=\frac{1}{2\pi}\frac{\omega}{1-e^{-2\pi k \omega}} \\
\mathcal{G}^{12}(\omega) =\mathcal{G}^{21}(\omega) & =  -\int_{-\infty}^{\infty}d \Delta \tau~\frac{1}{16\pi^{2}k^{2}}\frac{e^{i\omega \Delta \tau}}{\sinh^2(\frac{\Delta \tau}{2k}-i\epsilon)-\frac{r^{2}}{k^{2}}\sin^2(\frac{\Delta \theta}{2})} 
 = \frac{1}{2\pi}\frac{\omega}{1-e^{-2\pi k \omega}}f(\omega,L/2)
\end{aligned}}
\end{eqnarray}

where, we define the spectral function $f(\omega,L/2)$ as:
\bea 
 f(\omega,L/2)=\frac{1}{L\omega\sqrt{1+\left(\frac{L}{2k}\right)^2}}\sin(2k\omega \sinh^{-1}\left(\frac{L}{2k}\right)) 
\eea
In this expression the Euclidean distance $L$ between the coordinates $(r,\theta,\phi)$ and $(r,\theta',\phi)$ is already defined earlier.

\section{\textcolor{blue}{\bf \large Open quantum system (OQS) of two entangled atoms} }\label{OpenQFT2atoms}
\subsection{\textcolor{blue}{\bf \large Two atomic OQS model}}
In this section, we investigate a model of OQS with entanglement. For simplicity, we consider a pair of identical static entangled atoms at an Euclidean distance $L$ apart in the de Sitter spacetime. The two sets of internal energy levels of each atom are represented by: 
\be \left\{|g_{\alpha}\rangle,|e_{\alpha} \rangle \right\}~~ ~\forall \ \ \ \alpha=(1,2)\ee These are the ground-states and the excited-states for two atoms respectively. To avoid any further confusion here we note that, by the word "atom" we actually represented a very simplest spin bath open quantum model where two spin is immersed in the thermal bath.  Here such spin pairs are characterised by the Pauli spin matrices. However, for a more general situation where we are interested in very complicated two atomic models of Hamiltonian the present analysis holds good, but solving that two body OQS problem itself extremely complicated. So for our better understanding we restrict ourself to a pair of spins which we are treating as a pair of atom, using which we will study various underlying physics of OQS. In this discussion, such pair of atoms are conformally coupled to a massless scalar field in the De Sitter background. The scalar field acts as a thermal bath for the pair of entangled atoms. In our discussion, the above two atomic entangled system are represented by a pair of {\it Unruh-De-Witt detectors}. These two identical atoms interact weakly with a quantised conformally coupled massless probe scalar field in its quantum mechanical vacuum state. Consequently, the corresponding two energy levels of the two atoms are identified as:
 \be E^{(\pm)}_{\alpha}=\pm \frac{1}{2}\omega~~\forall \alpha=(1,2).\ee 
 Here, $\omega$ represents the renormalized energy level for two atoms, given by:
\be\begin{array}{lll}
	 	 \displaystyle \omega=\left\{\begin{array}{ll}
		      		   		      		       \displaystyle   \omega_{0}+i[\mathcal{K}^{(11)}(-\omega_{0})-\mathcal{K}^{(11)}(\omega_{0})] ~~~~~~ &
		      		   		      		    \mbox{\small \textcolor{red}{\bf Atom~1}}  \\ 
		      		   		      		   	\displaystyle  \omega_{0}+i[\mathcal{K}^{(22)}(-\omega_{0})-\mathcal{K}^{(22)}(\omega_{0})]~~~~~~ & \mbox{\small \textcolor{red}{\bf Atom~2}}
		      		   		      		             							\end{array}
		      		   		      		   \right.
		         \end{array}\ee	
Here $\mathcal{K}^{\alpha \alpha}(\pm \omega_{0})$ for $\alpha \in \{1,2\}$ are Hilbert transformations of two-point Wightmann functions discussed in previous section. Also, $\omega_0$ represents the natural frequency of the two identical atoms. In this context, the atoms are characterised by the label $\alpha\in [1,2]$ and  $\sigma^{\alpha}_{i}\forall i \in [1,2,3]$ are the Pauli matrices.

 The Hilbert space of such a system is bipartite in nature, i. e. 
 \be {\cal H}_{\bf Total}:={\cal H}_{\bf System}\otimes {\cal H}_{\bf Bath},\ee where ${\cal H}_{\bf System}$ and ${\cal H}_{\bf Bath}$ are the corresponding Hilbert spaces of the system and bath. Also, ${\cal H}_{\bf Total}$ represents the Hilbert space corresponding to the combined configuration of the system and the bath.
This entangled two atomic OQS can be represented by the following total Hamiltonian:
\be
H_{\bf total}(\tau)=H_{\bf System}(\tau)\otimes {\bf I}_{\bf Bath}+{\bf I}_{\bf System}\otimes H_{\bf Bath}(\tau)+H_{\bf Int}(\tau)
\ee
where ${\bf I}_{\bf System}$ and ${\bf I}_{\bf Bath}$ are the identity operators defined for the system and bath. When we are accessing the system we don't see anything from the bath and converse is also true. Here, the identity operators as appearing in the system and bath corresponds to the no access. More precisely , if one observer is sitting on the system, made by \textcolor{red}{\bf atom 1} and \textcolor{red}{\bf atom 2}, then that observer will not feel any further effect from the thermal bath environment. On the other hand, once the observer is sitting at the reference frame of thermal bath, that observe will not see any further effect from the two entangled OQS. However, in the interaction term both the system and thermal bath explicitly contribute and due to the entanglement one cannot separate their contribution from this term in the Hamiltonian. Here $\tau$ represents the proper time in the co-moving frame of two atoms. To make things clear, we would like to highlight the fact that instead of proper time one could use the static time cooardinate($t$). In that case, one might have to deal with the inherent curvature of de-Sitter spacetime explicitly in every calculation. However, the use of proper time incorporates this curvature effect directly through it's relation with static time, as will be discussed later. More details of each of the terms of the total Hamiltonian is as follows:

\begin{enumerate}
\item \underline{\textcolor{red}{\bf System:}}\\
The system Hamiltonian of the two entangled atoms are described by the linear combination of the individual contributions coming for each atom~\footnote{ Here we define the following direction cosines for two entangled atoms:
\bea \cos(\alpha^{\alpha}_1)=\cos(\alpha^{\alpha}),~~
\cos(\alpha^{\alpha}_2)=\cos(\beta^{\alpha}),~~
\cos(\alpha^{\alpha}_3)=\cos(\gamma^{\alpha})~~~~~~~\alpha=(1\Rightarrow\textcolor{red}{\bf atom~1},~2\Rightarrow\textcolor{red}{\bf atom~2}).
\eea}:
\be\begin{array}{lll}
\displaystyle~H_{\bf System}(\tau)=\sum_{\alpha=1}^{2}\frac{\omega}{2}\hat{n^\alpha}.\vec{\sigma^\alpha}=\frac{\omega}{2}\sum_{\alpha=1}^{2} \left[ n^{\alpha}_{1}\sigma^{\alpha}_{1}\cos \alpha ^{\alpha}+n^{\alpha}_{2}\sigma^{\alpha}_{2}\cos \beta^{\alpha}+n^{\alpha}_{3}\sigma^{\alpha}_{3}\cos \gamma ^{\alpha} \right] 
\end{array}\ee
where, the sum is taken over two individual contribution appearing from the Hamiltonians of the two atoms. In the above expression the normal vectors for two atoms are represented by $n^{\alpha}_{i}~\forall \alpha =1,2, \& ~i=1,2,3$ and the corresponding projection of Pauli-matrices are characterised by the direction cosine of Euler angles $\alpha$, $\beta$ and $\gamma$ respectively. 

\item \underline{\textcolor{red}{\bf Thermal bath:}}\\
The thermal bath Hamiltonian in the present context is characterised by a free rescaled field $\Phi(x)=a(\tau)\chi(x)$, where the original massless scalar field $\chi(x)$ is conformally coupled with De Sitter background with scale factor $a(\tau)$. The Hamiltonian in the de Sitter background can be written as:

\bea
H_{\bf Bath}(\tau)&=&\int^{\infty}_{0} dr~\int ^{\pi}_{0}d\theta~\int^{2\pi}_{0}d\phi~\left[\frac{\Pi^2_{\Phi}(\tau,r,\theta,\phi)}{2}\right.\nonumber\\
&& \left.+\frac{r^2\sin^2\theta}{2}\left\{r^2~(\partial_{r}\Phi(\tau,r,\theta,\phi))^2+ \frac{\left((\partial_{\theta}\Phi(\tau,r,\theta,\phi))^2+\frac{(\partial_{\phi}\Phi(\tau,r,\theta,\phi))^2}{\sin^2\theta}\right)}{\displaystyle\left(1-\frac{r^2}{\alpha^2}\right)}\right\}\right]~~~~~~~
\eea 
where $\Pi_{\Phi}(\tau,r,\theta,\phi)$ is the canonically conjugate momentum of the scalar field $\Phi(x)$. 
\item \underline{\textcolor{red}{\bf System-thermal bath interaction:}}\\
The interaction Hamiltonian is characterised by the following expression:
\be
\small{H_{\bf Int}(\tau)= \mu \sum_{\alpha=1}^{2}\hat{n^\alpha}.\vec{\sigma^\alpha}\phi(x^{\alpha})=\mu \sum_{\alpha=1}^{2}[n^{\alpha}_{1}\sigma^{\alpha}_{1}\cos \alpha ^{\alpha}+n^{\alpha}_{2}\sigma^{\alpha}_{2}\cos \beta^{\alpha}+n^{\alpha}_{3}\sigma^{\alpha}_{3}\cos \gamma ^{\alpha}]\phi(x^{\alpha})}
\ee
where, $\mu$ represents the coupling between the pair of atoms and the external massless scalar field placed at the thermal bath. While deriving the reduced dynamics of the two atomic detectors, we consider the weak coupling limiting approximation between the pair of atoms and the external massless probe scalar field. We assume that the coupling parameter $\mu$ is sufficiently small so that perturbation theory is applicable in the open quantum mechanical system under consideration. 
\end{enumerate}
\subsection{\textcolor{blue}{\bf \large Time dynamics of two atomic OQS}}
 To construct the effective Hamiltonian from the time dynamics of the present system, we first consider that there is no correlation between the pair of atoms with the the external probe free scalar field.  Hence, the thermal density-matrix of the combined system and thermal environment can be expressed in the following form:
\be 
\rho_{\bf Total}(0)=\underbrace{\rho_{\bf System}(0)}_{\textcolor{red}{\bf System}}~~\otimes \underbrace{\rho_{\bf Bath}(0)}_{\textcolor{red}{\bf Thermal~bath} }
\ee 
where $\rho_{\bf System}(0)$ and $\rho_{\bf Bath}=| 0\rangle \langle 0|$ represent the initial density matrix for the pair of atoms (system) and environment (bath) degrees of freedom. Here, $|0\rangle$ characterise the vacuum state of the external free scalar field. 

Now it is important to note that, in the interaction picture of OQS, the time evolution of the total density matrix can be written in the following form:
\be
\frac{\partial \rho_{\bf Total}(\tau)}{\partial \tau}=-i[H_{\bf Total}(\tau),\rho_{\bf Total}(\tau)]
\ee

One can further write the most general structure of the time evolved version of the total density matrix of the combined system and thermal bath, as given by:
\be 
\rho_{\bf Total}(\tau)=\underbrace{\rho_{\bf System}(\tau)}_{\textcolor{red}{\bf System}}~~\otimes \underbrace{\rho_{\bf Bath}(\tau)}_{\textcolor{red}{\bf Thermal~bath} }+\underbrace{\rho_{\bf correlation}(\tau)}_{\textcolor{red}{\bf Interaction} }~.\ee
Now, we have already started with the assumption that the non-adiabatic interaction between the system and thermal bath is switched on at $t=0$ (from initial structure of the total density matrix) and prior to that the interaction between the system and the thermal bath is extremely weak or absent for which there is no correlation exists in the OQS under consideration i.e.
\be 
\rho_{\bf correlation}(0)=0 \ee 
The idea of having no correlation at the initial stage is not very restrictive in nature, since for the computational purpose we can always find a time prior to which the system and environment did not interact. In the weak coupling limiting situation, the time evolved density matrix can be further simplified as:
\be 
\rho_{\bf Total}(\tau)\approx\underbrace{\rho_{\bf System}(\tau)}_{\textcolor{red}{\bf System}}~~\otimes \underbrace{\rho_{\bf Bath}(\tau)}_{\textcolor{red}{\bf Thermal~bath} }~,\ee
which is valid through a time scale on which the perturbation theory in OQS is valid. Furthermore, we also assume that the time scale corresponding to the correlation, $\tau_{\bf Bath}$, which is sometimes identified to be the {\it relaxation time scale} of the thermal bath, assumed to be extremely weak.  Here one can write:
\be \textcolor{red}{\bf Time~evolution~of ~the ~bath~for~time~scale~\tau>>\tau_{\bf Bath}:}~~~\rho_{\bf Bath}(\tau)\approx \rho_{\bf Bath}(0)\ee
In this context, the time evolved  total density matrix can be represented as:
\be
\rho_{\bf Total}(\tau)=U_{\bf Total}\rho_{\bf Total}(0)U^{\dagger}_{\bf Total}=U_{\bf Total}\left({\underbrace{\rho_{\bf System}(0)}_{\textcolor{red}{\bf System}}~~\otimes \underbrace{\rho_{\bf Bath}(0)}_{\textcolor{red}{\bf Thermal~bath} }}\right)U^{\dagger}_{\bf Total}
\ee
where $U_{\bf Total}$ represents the time evolution operator of OQS under consideration.  

Since, we are only interested in the reduced dynamics of the density matrix of the entangled atoms only, therefore, we trace out the external field (thermal bath) degrees of freedom from the total combined open quantum two atomic system. Consequently, the reduced density matrix of such a two atomic system can be expressed as:
\be 
\rho_{\bf System}(\tau)={\rm Tr}_{\bf Bath}[\rho_{\bf Total}(\tau)] \ee 
Further, doing simplifications one can write the reduced density matrix of the system after performing the partial trace operation over the bath degrees of freedom in the following simplified form:
\begin{equation} 
\footnotesize{\rho_{\bf System}(\tau)=\sum_{k}\langle k|U_{\bf T}\rho_{\bf T}(0)U^{\dagger}_{\bf T} |k\rangle=\sum_{k}\langle k|U_{\bf T}|0\rangle~\underbrace{\rho_{\bf S}(0)}_{\textcolor{red}{\bf System}}~\langle 0|U^{\dagger}_{\bf T} |k\rangle =\sum_{k}{\cal M}_{k}~\underbrace{\rho_{\bf S}(0)}_{\textcolor{red}{\bf System}}~{\cal M}^{\dagger}_{k}}
\end{equation}
\textcolor{red}{The subscript T  in the above equations refers to the term Total}.
where $|k\rangle$ represents all possible orthonormal basis states of the thermal bath defined in the Hilbert space ${\cal H}_{\bf Bath}$. Here, additionally we define the operator ${\cal M}_{k}$ as:
\be {\cal M}_{k}=\langle k|U_{\bf Total}|0\rangle={\rm Tr}_{\bf Bath}\left[|0\rangle \langle k|U_{\bf Total}\right]\ee
which is defined on the Hilbert space corresponding to the system under consideration i.e. ${\cal H}_{\bf System}$.   

Since in the present context, the time evolution operator $U_{\bf T}$ is unitary i.e. $U^{\dagger}_{\bf T}U_{\bf T}={\bf I}_{\bf T}$, which further implies that:
 
 \be 
 \footnotesize{\sum_{k}{\cal M}^{\dagger}_{k}{\cal M}_{k}=\sum_{k}\langle 0|U^{\dagger}_{\bf T}|k\rangle \langle k|U_{\bf T}|0\rangle=\langle 0|U^{\dagger}_{\bf T}\underbrace{\left(\sum_{k}|k\rangle \langle k|\right)}_{\textcolor{red}{\bf \equiv {\bf I}_{\bf Bath} }} U_{\bf T}|0\rangle=\langle 0|U^{\dagger}_{\bf T}U_{\bf T}|0\rangle=\langle 0|0\rangle=\textcolor{red}{{\bf I}_{\bf System}}}
 \ee
 Now, we explicitly mention about the properties of the system density matrix:
 \begin{enumerate}
 \item \underline{\textcolor{red}{\bf Hermiticity:}}\\
 The system density matrix is hermitian in nature which can be tested as:
 \be 
 \footnotesize{\rho^{\dagger}_{\bf System}(\tau)=\left(\sum_{k}{\cal M}_{k}~\underbrace{\rho_{\bf System}(0)}_{\textcolor{red}{\bf System}}~{\cal M}^{\dagger}_{k}\right)^{\dagger}
 =\sum_{k}{\cal M}_{k}~\underbrace{\rho_{\bf System}(0)}_{\textcolor{red}{\bf System}}~{\cal M}^{\dagger}_{k}=\rho_{\bf System}(\tau)}
 \ee
The last step can be justified using the fact that the density matrix at time $t=0$ is hermitian.
 \item \underline{\textcolor{red}{\bf Positivity:}}\\
 The system density matrix also satisfies positivity property.
 \end{enumerate}
 
Now, the  non-unitary time evolution of the reduced density matrix in the weak coupling limiting situation can expressed in terms of the {\it Gorini–Kossakowski–Sudarshan–Lindblad {\rm (GKSL)} master equation} in OQS, as given by:
\be
\frac{\partial \rho_{\bf System}(\tau)}{\partial \tau}=-i[H_{\bf eff},\rho_{\bf System}(\tau)]+\mathcal{L}[\rho_{\bf System}(\tau)]
\ee
where $H_{\bf eff}$ is the effective Hamiltonian of the two atomic system under consideration, which incorporates the effect of inter atomic interaction {\it aka} {\it Resonant Casimir Polder Interaction} ({\bf RCPI}). Also, the last term in the above mentioned evolution is known as the {\it Lindbladian}~\footnote{In the {\it Gorini–Kossakowski–Sudarshan–Lindblad {\rm (GKSL)} master equation} in OQS, the {it Lindbladian} operator is sometimes known as the {\it Dissipator} which captures the effect of dissipation within the framework of quantum mechanics.}, which describes the dissipative contribution due to the influence of the thermal bath on the two entangled atomic system. In the following subsections, we discuss about the effective Hamiltonian and the Lindbadian with greater detail.


\subsubsection{\textcolor{blue}{\bf \large Effective Hamiltonian construction }}

 As we are primarily interested in entanglement properties of atoms we trace out the bath degrees of freedom. When one does that in the path integral formalism, one obtains an effective Hamiltonian.
 
 For our setup of a two qubit system interacting with the scalar field this is given by:
\be\label{Heff} 
H_{\bf eff}=H_{\bf System}+H_{\bf Lamb~Shift}= \underbrace{\sum_{\alpha=1}^{2}\frac{\omega}{2}\hat{n^\alpha}.\vec{\sigma^\alpha}}_{\textcolor{red}{\bf System}} -\underbrace{ \frac{i}{2}\sum_{\alpha,\beta=1}^{2}\sum_{i,j=1}^{3}H^{(\alpha \beta)}_{ij}( n^\alpha_i .\sigma^{\alpha}_{i}) (n^\beta_j .\sigma^{\beta}_{j})}_{\textcolor{red}{\bf Lamb~shift= Heisenberg~spin~chain}}
\ee
where the first term in the effective Hamiltonian represents the system Hamiltonian of the two atomic system which are interacting with each other and one can treat this term to be a self interaction at the level of Hamiltonian. The specific details of this issue we have already mentioned earlier.  The second part necessarily captures the interaction between the two atoms as it has contribution from both atoms.  In condensed matter theory ($e.g.$ in Ising model), one usually considers nearest neighbour interactions to study the quantum mechanical behaviour and phase transition.  Our second part of the above Hamiltonian resembles with that well known Ising model Hamiltonian and mimics the role of an spin-spin interaction Hamiltonian in Heisenberg spin chain model which arises from the interaction between the two atoms and the external field.  The terminology {\it Lamb shift} was first used in the context of a hydrogen-like atom where in the lowest order approximation in the fine structure constant energy level shift was determined from an effective Hamiltonian.  Following the same,  we and also in other refs.\cite{Tian:2016uwp,Tian:2014jha,Tian:2014hwa} people have used the terminology {\it Lamb shift} in the present context to describe the energy level shift or spectroscopic shift of the effective Hamiltonian describing spin-spin self interaction.

Hence for usual practice,  we will name this specific part of Hamiltonian,  in our case also,  as Lamb Shift Hamiltonian.  
Here, $\hat{n}$ is the the normal unit vector for individual atom under consideration. Also the angles between normal vector and the Pauli matrices are characterised by the three Euler angles, $\alpha$, $\beta$ and $\gamma$. Therefore,  the Lamb Shift Hamiltonian \cite{Tian:2016uwp,Tian:2014jha,Tian:2014hwa} can be simplified in terms of the Eulerian angles as:
\be
\footnotesize{H_{\bf Lamb~ shift} = - \frac{i}{2}\sum_{\alpha,\beta=1}^{2}\sum_{i,j=1}^{3}H^{(\alpha \beta)}_{ij}(n^\alpha_i.\sigma^{\alpha}_{i})(n^\beta_j.\sigma^{\beta}_{j}) = -\frac{i}{2}\sum_{\alpha,\beta=1}^{2}\sum_{i,j=1}^{3}H^{(\alpha\beta)}_{ij}\cos(\alpha_i^{\alpha})\cos(\alpha_j^{\beta})\sigma_{i}^{\alpha}\sigma_{j}^{\beta}} 
\ee

Now, to know the explicit contribution in the effective Hamiltonian in the present two entangled atomic set up we define the following set of Pauli operators:
\bea
\textcolor{red}{\bf Atom~1}:\Rightarrow~~~~\sigma^{1}_{i} = \sigma_{i}\otimes\sigma_{0},~~~~~~
\textcolor{red}{\bf Atom~2}:\Rightarrow~~~~\sigma^{2}_{i} = \sigma_{0}\otimes\sigma_{i}~~~~~~~
\eea
which is actually expressed in terms of the tensor product of $2\times2$ identity matrix $\sigma_{0}$ and the three $2\times 2$ Pauli matrices $\sigma_i, \ \forall i=1,2,3$, which satisfy the following conditions:
\be \textcolor{red}{\bf Pauli~matrix~algebra:}~~~~~[\sigma_i,\sigma_j]=2i\epsilon_{ijk}\sigma_k,~~~~
\left\{\sigma_i,\sigma_j\right\}=2\delta_{ij}\sigma_0~.\ee
Here also it is obvious from the mathematical structures of these Pauli operators obey the following algebra:
\be 
\left[\sigma^\alpha_i,\sigma^\beta_j\right]= 2i\delta^{\alpha \beta} \epsilon_{ijk} \sigma^\alpha_k \ \ ,~
 \left\{\sigma^1_i,\sigma^1_j\right\} = \left\{\sigma^2_i,\sigma^2_j\right\} = 2 \delta_{ij} \sigma_0 \otimes \sigma_0 \ \ ,~ \left\{\sigma^1_i,\sigma^2_j\right\} = 2 \sigma_i \otimes \sigma_j~~~ \ee
where $(i,j)=1,2,3$ and $(\alpha,\beta)=1,2$.
With these set of definition of the Pauli matrices the Pauli operators can be expressed in terms of the tensor product as:
\bea
\sigma^{1}_{1} & =& \left(\begin{array}{cc} 0 & ~\sigma_{0} \\ ~\sigma_{0} & 0 \end{array}\right),~~~~~ 
\sigma^{2}_{1}  = \left(\begin{array}{cc} ~\sigma_{1} & 0 \\  0 & ~\sigma_{1} \end{array}\right),~~~\\
\sigma^{1}_{2}  &=& \left(\begin{array}{cc} 0 &-i\sigma_{0} \\ i\sigma_{0} & 0 \end{array}\right),~~~
\sigma^{2}_{2}  = \left(\begin{array}{cc} ~\sigma_{2} & 0 \\  0 & ~\sigma_{2} \end{array}\right),~~~\\
\sigma^{1}_{3}  &=& \left(\begin{array}{cc} ~\sigma_{0} & 0 \\  0 & ~-\sigma_{0} \end{array}\right),~~~
\sigma^{2}_{3}  = \left(\begin{array}{cc} ~\sigma_{3} & 0 \\  0 & ~\sigma_{3} \end{array}\right).
\eea

In this context, one can obtain the matrix elements of $H^{\alpha \beta}$ appearing in the Lamb shift Hamiltonian from the two point correlator which is given as:
\be
G^{\alpha \beta}(\Delta \tau=\tau-\tau')=\langle \Phi(\bf{x_{\alpha}},\tau)\Phi(\bf{x_{\beta}},\tau')\rangle
\ee
The Fourier-transform of the above two point propagator in frequency ($\omega_0$) space can be written as:
\be
\mathcal{G}^{\alpha \beta}(\pm \omega_{0})=\int_{-\infty}^{\infty}d\Delta \tau~ e^{\pm i\omega_{0}  \Delta \tau}~G^{\alpha \beta}(\Delta \tau)~,
\ee
where $\omega_{0}$ represents the difference between the energy levels of the ground and the excited states of the atoms respectively. In turn, the elements of the effective Hamiltonian matrix $H^{\alpha \beta}_{ij}$ are given by the following Hilbert transform of the Wightman function (two point correlator) as given by:
\be \label{hil}
\small{\mathcal{K}^{\alpha \beta}(\pm \omega_{0})=\frac{P}{\pi i}\int_{-\infty}^{\infty}d\omega~ \frac{\mathcal{G}^{\alpha \beta}(\pm \omega)}{\omega \pm\omega_{0}}=\frac{P}{\pi i}\int_{-\infty}^{\infty}d\omega~ \frac{1}{\omega \pm\omega_{0}}~\int_{-\infty}^{\infty}d\Delta \tau~ e^{\pm i\omega \Delta \tau}~G^{\alpha \beta}(\Delta \tau)}
\ee
where $P$ is the Principal value of the integral.

Now, the elements of co-efficient matrix $H^{(\alpha \beta)}_{ij}$ of the effective Hamiltonian can be explicitly represented by the following expression:
\be \label{H}
H^{(\alpha \beta)}_{ij}=\mathcal{A}^{(\alpha \beta)}\delta_{ij}-i\mathcal{B}^{(\alpha \beta)}\epsilon_{ijk}\delta_{3k}-{\mathcal{A}}^{(\alpha \beta)}\delta_{3i}\delta_{3j}
\ee 
where, the quantities ${\mathcal{A}}^{(\alpha \beta)}$ and ${\mathcal{B}}^{(\alpha \beta)}$ for the two atomic system are defined as:
\be
\begin{aligned}
	\mathcal{A}^{(\alpha \beta)} &= \frac{\mu^{2}}{4}[\mathcal{K}^{(\alpha \beta)}(\omega_{0})+\mathcal{K}^{(\alpha \beta)}(-\omega_{0})] \\
	\mathcal{B}^{(\alpha \beta)} &= \frac{\mu^{2}}{4}[\mathcal{K}^{(\alpha \beta)}(\omega_{0})-\mathcal{K}^{(\alpha \beta)}(-\omega_{0})] 
\end{aligned}
\ee
Here, $\mu$ is the coupling parameter which represents the interaction strength between the system and the external thermal bath (conformally coupled scalar field) degrees of freedom. Determining the structure of the elements of the co-efficient matrix $H^{(\alpha \beta)}_{ij}$ in terms of the two atomic two point correlation function (Wightman function) of external free conformally coupled massless scalar field in de Sitter background actually fixes the structure of the effective Hamiltonian in the present context. 

We define the following uantities (equality sign holds because of symmetry of Hilbert transformations as seen in Appendix \ref{coeff}.
\be
\mathcal{A}^{11} = \mathcal{A}^{22} (\equiv {\cal A}_1) \ \ \ \ \ \mathcal{A}^{12} = \mathcal{A}^{21} (\equiv {\cal A}_2)\ \ \ \ \mathcal{B}^{11} = \mathcal{B}^{22} (\equiv {\cal B}_1)\ \ \ \ \ \mathcal{B}^{12} = \mathcal{B}^{21} (\equiv {\cal B}_2)
\ee

From equation \ref{H}, the non-vanishing Hamiltonian matrix elemnts are as follows :

\be \label{hamco} 
\footnotesize{\begin{aligned}
H^{(11)}_{11}&=H^{(22)}_{11}={\cal A}_1~~~~
H^{(11)}_{12}=H^{(22)}_{12}=-i{\cal B}_1~~~~
H^{(11)}_{21}=H^{(22)}_{21}=i{\cal B}_1~~~~
H^{(11)}_{22}=H^{(22)}_{22}={\cal A}_1 \\
H^{(12)}_{11}&=H^{(21)}_{11}={\cal A}_2 ~~~~
H^{(12)}_{12}=H^{(21)}_{12}=-i{\cal B}_2 ~~~~
H^{(12)}_{21}=H^{(21)}_{21}=i{\cal B}_2 ~~~~
H^{(12)}_{22}=H^{(21)}_{22}={\cal A}_2
\end{aligned}}
\ee
and these will explicitly contribute to the final expression for the expectation values of the Lamb shift Hamiltonian.
\subsubsection{\textcolor{blue}{\bf \large Lindbladian construction }}
The fluctuation-dissipation into the system is introduced by the additional contribution in the time-evolution equation of the reduce density-matrix often known as the {\it Lindbladian} \cite{Gorini:1975nb} or the {\it Lindblad operator} in OQS. The second significant term in the {\it Gorini–Kossakowski-Sudarshan–Lindblad} {\rm (GKSL)} master equation is actually characterised as the {\it Lindbladian} or {\it Quantum Dissipator} of the OQS, which is for the present two atomic model can be written as:
\be
\footnotesize{\mathcal{L}[\rho_{\bf System}(\tau)]=\frac{1}{2}\sum_{i,j=1}^{3}\sum_{\alpha,\beta=1}^{2}C^{\alpha\beta}_{ij}\left[2(n^{\beta}_j .\sigma^{\beta}_{j}) \rho_{\bf System}(\tau) (n^{\alpha}_i.\sigma^{\alpha}_{i})-\left\{(n^{\alpha}_i.\sigma^{\alpha}_{i})(n^{\beta}_j.\sigma^{\beta}_{j}),\rho_{\bf System}(\tau)\right\}\right]}
\ee
where, $\rho_{\bf System}(\tau)$ is the reduced density matrix of the two entangled atomic system where we have trace over the external bath scalar field degrees of freedom. The co-efficient matrix $C^{\alpha \beta}_{ij}$ is known as the {\it Gorini–Kossakowski–Sudarshan–Lindblad matrix}, which is constructed under the weak coupling limiting approximation on the coupling parameter $\mu$, as appearing in the interaction Hamiltonian~\footnote{In the case of closed quantum mechanical system all the entries of the {\it Gorini–Kossakowski–Sudarshan–Lindblad} matrix is zero. Consequently,  the time evolution of the reduced density matrix is described by the quantum {\it Liouville equation}, which is the quantum mechanical analog of the classical {\it Liouville equation}. }. In the context of any OQS, {\it Lindbladian} captures the effect of dissipation implicitly describing by the system operator $(n^{\alpha}_i.\sigma^{\alpha}_{i})$, which one can treat as an influence of the thermal bath on the two entangled atomic system under consideration. 

Following the same procedure performed in the previous section to compute the co-efficient matrix element , $H^{(\alpha \beta)}_{ij}$, in the present context the elements of the {\it Gorini–Kossakowski Sudarshan–Lindblad} matrix, $C^{(\alpha \beta)}_{ij}$, as appearing in the expression for the {\it Linbladian} can be expressed as:
\be \label{C}
C^{(\alpha \beta)}_{ij}=\tilde{\mathcal{A}}^{(\alpha \beta)}\delta_{ij}-i\tilde{\mathcal{B}}^{(\alpha \beta)}\epsilon_{ijk}\delta_{3k}-\tilde{\mathcal{A}}^{(\alpha \beta)}\delta_{3i}\delta_{3j}~,
\ee 
where, the quantities $\tilde{\mathcal{A}}^{(\alpha \beta)}$ and $\tilde{\mathcal{B}}^{(\alpha \beta)}$ for the two atomic system are defined as:
\be
\begin{aligned}
	\tilde{\mathcal{A}}^{(\alpha \beta)} &= \frac{\mu^{2}}{4}[\mathcal{G}^{(\alpha \beta)}(\omega_{0})+\mathcal{G}^{(\alpha \beta)}(-\omega_{0})] \\ 
	\tilde{\mathcal{B}}^{(\alpha \beta)} &= \frac{\mu^{2}}{4}[\mathcal{G}^{(\alpha \beta)}(\omega_{0})-\mathcal{G}^{(\alpha \beta)}(-\omega_{0})] 
\end{aligned} 
\ee
In the later half of the paper we will explicitly compute the entries of the  {\it Gorini–Kossakowski Sudarshan–Lindblad matrix}, $C^{(\alpha \beta)}_{ij}$ to fix the mathematical structure of the {\it Lindbladian} operator in the present two entangled atomic OQS under consideration. This can be done once we compute the all of the possible combinations of two body Wightman (two point) correlation function for the external probe scalar field placed in the thermal bath.

The non-vanishing elements of the {\it Gorini–Kossakowski–Sudarshan–Lindblad matrix}, $C^{(\alpha \beta)}_{ij}$ elements are obtained from equation \ref{C}; 

\be \label{linco}
\footnotesize{\begin{aligned}
		C^{(11)}_{11}&=C^{(22)}_{11}=\tilde{{\cal A}}_1~~~~
		C^{(11)}_{12}=C^{(22)}_{12}=-i\tilde{\cal B}_1~~~~
		C^{(11)}_{21}=C^{(22)}_{21}=i\tilde{\cal B}_1~~~~
		C^{(11)}_{22}=C^{(22)}_{22}=\tilde{{\cal A}}_1\\
		C^{(12)}_{11}&=C^{(21)}_{11}=\tilde{\cal A}_2~~~~
		C^{(12)}_{12}=C^{(21)}_{12}=-i\tilde{\cal B}_2~~~~
		C^{(12)}_{21}=C^{(21)}_{21}=i\tilde{\cal B}_2~~~~
		C^{(12)}_{22}=C^{(21)}_{22}=\tilde{\cal A}_2
\end{aligned}}
\ee

In this paper, one of our prime objectives are to find out the expectation value of the of the Lamb Shift Hamiltonian of the OQS described by two atoms, which are entangled with each other. Fixing the co-efficient matrix $H^{(\alpha \beta)}_{ij}$ in terms of the two atomic two point correlation function (Wightman function) of external free conformally coupled massless scalar field in De Sitter background helps us to compute the analytical expression for the energy shift explicitly. To compute this expression the main ingredient is the all possible quantum mechanical states which we have to construct in the present context from the ground and excited states of the two atoms. Using these atomic states we construct four possible quantum mechanical two atomic entangled states for the combined system (two atomic system+thermal bath), which are ground state, excited state, symmetric state and antisymmetric state respectively.  We explicitly do this computation in the later part of the paper.

\section{\textcolor{blue}{\bf \large Lamb Shift spectroscopy from OQS of two entangled atoms}}
\label{lss}
Lamb shifts are obtained by taking the expectation value of Lamb shift Hamiltonian in all the states of the two atomic system. Here, we first present all the states for the atomic system formed from the individual atomic states.

Set of eigenstates ($|g_1\rangle,|e_1\rangle$ and $|g_2\rangle,|e_2\rangle$) of the two atomic OQS is given as:

\be
\small{\begin{aligned}
\underline{\textcolor{red}{\bf A. For~atom~1:}} \\
&\underline{\textcolor{red}{\bf Ground~state}}\Rightarrow \\
&|g_1\rangle =\frac{1}{\sqrt{2}}\sqrt{1+\cos(\gamma^{1})}\begin{pmatrix} 
	\displaystyle -\frac{\left(\cos(\alpha^{1})-i\cos(\beta^{1})\right)}{1+\cos(\gamma^{1})} \\
	1
\end{pmatrix}\Rightarrow\textcolor{red}{\bf Eigenvalue}~~E^{(2)}_G=-\frac{\omega}{2}\\
&\underline{\textcolor{red}{\bf Excited~state}}\Rightarrow \nonumber\\
& |e_1\rangle =\frac{1}{\sqrt{2}}\sqrt{1+\cos(\gamma^{1})}\begin{pmatrix} 
	\displaystyle 1\\
	\displaystyle  \frac{\left(\cos(\alpha^{1})+i\cos(\beta^{1})\right)}{1+\cos(\gamma^{1})} ~~
\end{pmatrix}\Rightarrow\textcolor{red}{\bf Eigenvalue}~~E^{(2)}_E=\frac{\omega}{2} \\
\end{aligned}}
\ee
\be
\small{\begin{aligned}
\underline{\textcolor{red}{\bf B. For~atom~2:}} \\
&\underline{\textcolor{red}{\bf Ground~state}}\Rightarrow \\
&|g_2\rangle =\frac{1}{\sqrt{2}}\sqrt{1+\cos(\gamma^{2})}\begin{pmatrix} 
	\displaystyle -\frac{\left(\cos(\alpha^{2})-i\cos(\beta^{2})\right)}{1+\cos(\gamma^{2})} \\
	1
\end{pmatrix}\Rightarrow\textcolor{red}{\bf Eigenvalue}~~E^{(2)}_G=-\frac{\omega}{2}\\
&\underline{\textcolor{red}{\bf Excited~state}}\Rightarrow\nonumber\\
& |e_2\rangle =\frac{1}{\sqrt{2}}\sqrt{1+\cos(\gamma^{2})}\begin{pmatrix} 
	1\\
	\displaystyle \frac{\left(\cos(\alpha^{2})+i\cos(\beta^{2})\right)}{1+\cos(\gamma^{2})} ~~
\end{pmatrix}\Rightarrow\textcolor{red}{\bf Eigenvalue}~~E^{(2)}_E=\frac{\omega}{2}
\end{aligned}}
\ee

In the collective state representation of OQS, the ground state ($|{G}\rangle$), excited state ($|{E}\rangle$), symmetric state ($|{S}\rangle$) and the antisymmetric state ($|{A}\rangle$) of the two-entangled atomic OQS \cite{Dicke:1954zz} can be expressed as:

\be
\footnotesize{\begin{split}
&\underline{\textcolor{red}{\bf 1. Ground ~state:}} \\
&|G \rangle = |g_1 \rangle \otimes| g_2 \rangle =\frac{1}{2}\sqrt{(1+\cos(\gamma^{1}))(1+\cos(\gamma^{2}))}
\begin{pmatrix} 
	\displaystyle -\frac{\left(\cos(\alpha^{1})-i\cos(\beta^{1})\right)}{1+\cos(\gamma^{1})}\frac{\left(\cos(\alpha^{2})-i\cos(\beta^{2})\right)}{1+\cos(\gamma^{2})} \\
	\displaystyle -\frac{\left(\cos(\alpha^{1})-i\cos(\beta^{1})\right)}{1+\cos(\gamma^{1})}\\
	\displaystyle -\frac{\left(\cos(\alpha^{2})-i\cos(\beta^{2})\right)}{1+\cos(\gamma^{2})} \\
	1 
\end{pmatrix} 
\end{split}}
\ee
\be
\footnotesize{\begin{split}
&\underline{\textcolor{red}{\bf 2.Excited ~state:}} \\
&|E \rangle = |e_1 \rangle \otimes |e_2 \rangle  = \frac{1}{2}\sqrt{(1+\cos(\gamma^{1}))(1+\cos(\gamma^{2}))}
\begin{pmatrix} 
	\displaystyle 1 \\
	\displaystyle \frac{\left(\cos(\alpha^{2})+i\cos(\beta^{2})\right)}{1+\cos(\gamma^{2})}\\
	\displaystyle \frac{\left(\cos(\alpha^{1})+i\cos(\beta^{1})\right)}{1+\cos(\gamma^{1})} \\
	\displaystyle \frac{\left(\cos(\alpha^{1})+i\cos(\beta^{1})\right)}{1+\cos(\gamma^{1})}\frac{\left(\cos(\alpha^{2})+i\cos(\beta^{2})\right)}{1+\cos(\gamma^{2})}
\end{pmatrix}
\end{split}}
\ee

\be \label{sym}
\footnotesize{\begin{split}
&\underline{\textcolor{red}{\bf 3.Symmetric ~state:}} \\
&|S \rangle = \frac{1}{\sqrt{2}} [ |e_1 \rangle \otimes |g_2 \rangle + |g_1 \rangle \otimes |e_2 \rangle]  \\
&~~~~~~~= \frac{1}{2\sqrt{2}}\sqrt{(1+\cos(\gamma^{1}))(1+\cos(\gamma^{2}))}
\begin{pmatrix} 
	\displaystyle -\frac{\left(\cos(\alpha^{1})-i\cos(\beta^{1})\right)}{1+\cos(\gamma^{1})}-\frac{\left(\cos(\alpha^{2})-i\cos(\beta^{2})\right)}{1+\cos(\gamma^{2})} \\
	\displaystyle 1-\frac{\left(\cos(\alpha^{1})-i\cos(\beta^{1})\right)}{1+\cos(\gamma^{1})}\frac{\left(\cos(\alpha^{2})+i\cos(\beta^{2})\right)}{1+\cos\gamma^{2}} \\
	\displaystyle 1-\frac{\left(\cos(\alpha^{1})+i\cos(\beta^{1})\right)}{1+\cos(\gamma^{1})}\frac{\left(\cos(\alpha^{2})-i\cos(\beta^{2})\right)}{1+\cos(\gamma^{2})}  \\
	\displaystyle \frac{\left(\cos(\alpha^{1})+i\cos(\beta^{1})\right)}{1+\cos(\gamma^{1})}+\frac{\left(\cos(\alpha^{2})+i\cos(\beta^{2})\right)}{1+\cos(\gamma^{2})} 
\end{pmatrix} 
\end{split}}
\ee
\be \label{asym}
\footnotesize{\begin{split}
&\underline{\textcolor{red}{\bf 4.Antisymmetric ~state:}} \\
&|A \rangle = \frac{1}{\sqrt{2}} [ |e_1 \rangle \otimes |g_2 \rangle - |g_1 \rangle \otimes |e_2 \rangle]  \\
&~~~~~~~= \frac{1}{2\sqrt{2}}\sqrt{(1+\cos(\gamma^{1}))(1+\cos(\gamma^{2}))}
\begin{pmatrix} 
	\displaystyle \frac{\left(\cos(\alpha^{1})-i\cos(\beta^{1})\right)}{1+\cos(\gamma^{1})}-\frac{\left(\cos(\alpha^{2})-i\cos(\beta^{2})\right)}{1+\cos(\gamma^{2})} \\
	\displaystyle 1+\frac{\left(\cos(\alpha^{1})-i\cos(\beta^{1})\right)}{1+\cos(\gamma^{1})}\frac{\left(\cos(\alpha^{2})+i\cos(\beta^{2})\right)}{1+\cos(\gamma^{2})} \\
	\displaystyle -1-\frac{\left(\cos(\alpha^{1})+i\cos(\beta^{1})\right)}{1+\cos(\gamma^{1})}\frac{\left(\cos(\alpha^{2})-i\cos(\beta^{2})\right)}{1+\cos\gamma^{2}}  \\
	\displaystyle \frac{\left(\cos(\alpha^{1})+i\cos(\beta^{1})\right)}{1+\cos(\gamma^{1})}-\frac{\left(\cos(\alpha^{2})+i\cos(\beta^{2})\right)}{1+\cos(\gamma^{2})} 
\end{pmatrix}
\end{split}}
\ee

Now, the expectation value of the Lamb shift Hamiltonian $H_{LS}$ with respect to the ground state ($|{G}\rangle$), excited state ($|{E}\rangle$), symmetric state ($|{S}\rangle$) and the antisymmetric state ($|{A}\rangle$) are given by (see Appendix \ref{LScalc}): 

\bea
\begin{aligned} 
\delta E_{G} &= \langle G| H_{LS} | G\rangle \\
	& = -\frac{\mu^2 P}{8\pi^2} ~[-\left\{\cos(\alpha^{1})\cos(\beta^{1})\cos(\gamma^{1})+\cos(\alpha^{2})\cos(\beta^{2})\cos(\gamma^{2})\right\} \Delta_1  \\ 
	& ~~~~~~~~~~~~~~~ + \left\{\cos[2](\alpha^{1})+\cos[2](\beta^{1})+\cos[2](\alpha^{2})+\cos[2](\beta^{2})\right\} \Delta_2 \\
	& ~~~~~~~~~~~~~~~ + \left\{\cos[2](\alpha^{1})\cos[2](\alpha^{2}) + \cos[2](\beta^{1})\cos[2](\beta^{2})\right\}\Delta_3] \\
\delta E_{E} &= \langle E| H_{LS} | E\rangle \\
	&=-\frac{\mu^2 P}{8\pi^2} ~[\left\{\cos(\alpha^{1})\cos(\beta^{1})\cos(\gamma^{1})+\cos(\alpha^{2})\cos(\beta^{2})\cos(\gamma^{2})\right\} \Delta_1 \\
	& ~~~~~~~~~~~~~~~~ + \left\{\cos[2](\alpha^{1})+\cos[2](\beta^{1})+\cos[2](\alpha^{2})+\cos[2](\beta^{2})\right\} \Delta_2 \\
	& ~~~~~~~~~~~~~~~~ + \left\{\cos[2](\alpha^{1})\cos[2](\alpha^{2}) + \cos[2](\beta^{1})\cos[2](\beta^{2})\right\} \Delta_3] \\
\delta E_{S} &= \langle S| H_{LS} | S\rangle \\
	& = -\frac{\mu^2 P}{8\pi^2} ~ [ \Omega (B^{2}+C^{2}-A^{2}-D^{2}) \cos(\alpha^{1})\cos(\alpha^{2}) \Delta_3 \\
	& ~~~~~~~~~~~~~~~ + \Omega (A^{2}+B^{2}+C^{2}+D^{2}) \cos(\beta^{1})\cos(\beta^{2})\Delta_3  \\
	&  ~~~~~~~~~~~~~~~ + \left\{\cos^2(\alpha^{1})+\cos^2(\alpha^{2}) +\cos^2(\beta^{1})+\cos^2(\beta^{2})\right\} \Delta_2] \\
\delta E_{A} &= \langle A| H_{LS} | A\rangle \\
	& = -\frac{\mu^2 P}{8\pi^2} [\Omega (\tilde  D^{2}+\tilde A^{2}-\tilde B^{2}-\tilde C^{2}) \cos(\alpha^{1})\cos(\alpha^{2}) \Delta_3 \\
	& ~~~~~~~~~~~~~~- \Omega (\tilde  A^{2}+\tilde B^{2}+\tilde C^{2}+\tilde D^{2}) \cos(\beta^{1})\cos(\beta^{2})\Delta_3  \\
	& ~~~~~~~~~~~~~~~  + \left\{\cos^2(\alpha^{1})+\cos^2(\alpha^{2}) +\cos^2(\beta^{1})+\cos^2(\beta^{2})\right\}\Delta_2]
\end{aligned}
\eea

where $\Delta_1$, $\Delta_2$ and $\Delta_3$ are the spectroscopic integrals calculated in Appendix \ref{int}. Here, we write those again to maintain consistency

\begin{enumerate}
	\item \underline{\textcolor{red}{\bf Spectroscopic ~Integral ~I:}}\be \Delta_{1}:=\int^{\infty}_{-\infty}d\omega~\frac{2\omega_0~\omega}{\left(1-e^{-2\pi k\omega}\right)\left(\omega+\omega_0\right)\left(\omega-\omega_0\right)} = 0\ee
	
	\item \underline{\textcolor{red}{\bf Spectroscopic ~Integral ~II:}}\be \Delta_{2}:=\int^{\infty}_{-\infty}d\omega~\frac{\omega^2 }{\left(1-e^{-2\pi k\omega}\right)\left(\omega+\omega_0\right)\left(\omega-\omega_0\right)} = 0\ee
	
	\item \underline{\textcolor{red}{\bf Spectroscopic ~Integral ~III:}}
	\be 
	\footnotesize{\Delta_{3}:=\int^{\infty}_{-\infty}d\omega~\frac{2~\omega^2 f(\omega,L/2)}{\left(1-e^{-2\pi k\omega}\right)\left(\omega+\omega_0\right)\left(\omega-\omega_0\right)} = \frac{2\pi} {L\sqrt{1+\left(\frac{L}{2k}\right)^2}}\cos(2k\omega_0 \sinh^{-1}\left(\frac{L}{2k}\right))}
	\ee
\end{enumerate}

\subsection{\textcolor{blue}{\bf \large Bethe regularised Lamb Shift Spectra}}

Using these spectroscopic integrals we obtain the simplified result for Lamb shifts

\begin{enumerate}
\item \underline{\textcolor{red}{\bf Bethe regularised energy shift from Ground \& Excited states:}} \\ \\ 
Here, we evaluate the contributions coming from the ground state and excited state expectation values, as given by:

\be
\begin{aligned}
	\delta E_{Y} &= -\frac{\mu^2 P}{8\pi^2} \left\{\cos[2](\alpha^{1})\cos[2](\alpha^{2}) + \cos[2](\beta^{1})\cos[2](\beta^{2})\right\} \Delta_3 \\
	&= -\frac{\mu^2 P}{4\pi} \frac{1}{L\sqrt{1+\left(\frac{L}{2k}\right)^2}} \left\{\cos[2](\alpha^{1}) \cos[2](\alpha^{2}) + \cos[2](\beta^{1})\cos[2](\beta^{2})\right\} \\
	&~~~~~~~~~~~~~~~~~~~~~~~~~~~~~~~~~~~~ \cos\left(2\omega_0 k\sinh^{-1}\left(\frac{L}{2k}\right)\right)~~~~~~~~ \forall ~~Y=(G,E)
\end{aligned}
\ee
Under a certain case for which 
$$\left\{\cos[2](\alpha^{1})\cos[2](\alpha^{2}) + \cos[2](\beta^{1})\cos[2](\beta^{2})\right\} = 0$$
we have, 
\be
\textcolor{red}{ \delta E_{G}=0=\delta E_{E}}~
\ee
This directly implies that in this particular orientation of the qubits guided by the values of the $\alpha$'s and $\beta$'s satisfying the above equation the ground and excited states does not contribute in the Lamb shift spectroscopy of two entangled qubits in OQS.

\item \underline{\textcolor{red}{\bf Bethe regularised energy shift from Symmetric \& Antisymmetric states:}}\\ \\
Now, from the present analysis we observe that the {\bf RCPI} between the two entangled atoms in the De Sitter background is being contributed only by the symmetric and antisymmetric part of the Hamiltonian $H_{LS}$ as it consists of the term $f(\omega,L/2)$,  which contains a measure of the Euclidean distance $L$  between the two entangled atoms. This actually contributes in the shift in the energy levels or more precisely the inter atomic interaction energy computed from the symmetric ($|S\rangle$) and antisymmetric ($|A\rangle$) quantum state constructed out of two entangled atoms in De Sitter space. On the other hand, we have already seen that there is no such term present in the shift in the energy levels between the ground state and excited state constructed solely from the quantum states $|G\rangle$ and $|E\rangle$ for two atoms. This is appearing due to the non inter atomic interaction appearing between the uncorrelated two atomic quantum states in the second order of perturbation theory of OQS under consideration in this work. The presence of the Euclidean distance dependent term in the expression of the symmetric and the antisymmetric part of the shift in the energy level defines the gradient of a potential between the two atoms in the curved de Sitter space and is the manifestation of the RCPI between them. Therefore, only the terms which contributes towards the RCPI between the two entangled atoms are given by the following expression: 
\be
\footnotesize{\begin{aligned}
\delta E_{S}(L) &= -\frac{\mu^2 P}{8\pi^2} [\Omega\Gamma_1(\alpha^1, \alpha^2,\beta^1, \beta^2) \Delta_3 + \left\{\cos^2(\alpha^{1})+\cos^2(\alpha^{2})+\cos^2(\beta^{1})+\cos^2(\beta^{2})\right\} \Delta_2]
\\
\delta E_{A}(L) &= -\frac{\mu^2 P}{8\pi^2} [\Omega\Gamma_1(\alpha^1, \alpha^2,\beta^1, \beta^2) \Delta_3 + \left\{\cos^2(\alpha^{1})+\cos^2(\alpha^{2})+\cos^2(\beta^{1})+\cos^2(\beta^{2})\right\} \Delta_2]
\end{aligned}} 
\ee
 where we have introduced two functions $\Gamma_1(\alpha^1, \alpha^2,\beta^1, \beta^2)$ and $\Gamma_2(\alpha^1, \alpha^2,\beta^1, \beta^2)$, which are defined as:
 
\begin{equation*}
\footnotesize{ \begin{aligned}
 \Gamma_1(\alpha^1, \alpha^2,\beta^1, \beta^2) & \equiv (B^{2}+C^{2}-A^{2}-D^{2}) \cos(\alpha^{1})\cos(\alpha^{2}) +  (A^{2}+B^{2}+C^{2}+D^{2})\cos(\beta^{1})\cos(\beta^{2})\\
 \Gamma_2(\alpha^1, \alpha^2,\beta^1, \beta^2) & \equiv (\tilde  D^{2}+\tilde A^{2}-\tilde B^{2}-\tilde C^{2}) \cos(\alpha^{1})\cos(\alpha^{2}) -  (\tilde  A^{2}+\tilde B^{2}+\tilde C^{2}+\tilde D^{2})\cos(\beta^{1})\cos(\beta^{2}) 
 \end{aligned}}
 \end{equation*}

Using the result obtained in Appendix \ref{int}, we get the following simplified result for the inter atomic interaction energy shift:
\be
\begin{aligned}
\delta E_{S}(L) &= -\frac{\mu^2 \Omega}{8\pi L\sqrt{1+\left(\frac{L}{2k}\right)^2}} \cos\left(2\omega_0 k\sinh^{-1}\left(\frac{L}{2k}\right)\right)\Gamma_1(\alpha^1, \alpha^2,\beta^1, \beta^2) \\
\delta E_{A}(L) &= \frac{\mu^2\Omega }{8\pi L\sqrt{1+\left(\frac{L}{2k}\right)^2}}\cos\left(2\omega_0 k\sinh^{-1}\left(\frac{L}{2k}\right)\right)\Gamma_2(\alpha^1, \alpha^2,\beta^1, \beta^2)
\end{aligned}
\ee
\end{enumerate}

We get the following characteristic features:

\begin{enumerate}
\item If we increase the value of the Euclidean distance ($L$), then the oscillation in the Lamb Shift increase with respect to the surface gravity in the region $L\leq k$.

\item We also found that the magnitude of the saturation value of the Lamb Shift decrease with increasing value of the Euclidean distance ($L$) in the region $L>k$.
\end{enumerate}

Further, it is important to note that, the final results of the inter atomic energy shifts depend on the background De Sitter metric through the following relation:
\be 
k=\sqrt{g_{00}}\alpha=\sqrt{\alpha^2-r^2}=\sqrt{\frac{3}{\Lambda}-r^2}
\ee
This directly implies that the parameter $k$ is directly related to the positive cosmological constant of De Sitter space. Consequently, one can theoretically probe De Sitter space using a pair of entangled atoms in OQS in presence of {\bf RCPI}. Here it is important to note that the result obtained for inter atomic energy level shift for two detectors (two entangled atoms) can be interpreted as the energy level shift obtained for a single detector immersed in a thermal bath with temperature:
\be 
T=\frac{1}{2\pi k}=\frac{1}{2\pi \sqrt{\alpha^2-r^2}}=\frac{1}{2\pi \sqrt{\frac{3}{\Lambda}-r^2}} 
\ee
which is interpreted as the \textit{Unruh Temperature}. A freely falling observer under a steady acceleration observes this temperature in de-Sitter space. In this case, the inter atomic interaction exhibits non thermal behaviour and carrying non thermal fluctuation.

Now, to understand the detailed physical features of the obtained result for energy level shift from RCPI in de Sitter space we consider two limiting situations, as given by:

\begin{enumerate}
\item \underline{\textcolor{red}{\bf Case~I:}}\\
When the inter atomic distance is much larger than characteristic length scale $k$ i.e. $L>>k$. In this case the two entangled atomic system is placed near to the cosmological horizon. In this limit, the energy-level shift from RCPI can be simplified as:

\be
\begin{aligned}
	\delta E_{S}(L) &= -\frac{\mu^2 \Omega}{8\pi L^2}\cos\left(2\omega_0 k\sinh^{-1}\left(\frac{L}{2k}\right)\right)\Gamma_1(\alpha^1, \alpha^2,\beta^1, \beta^2) \\
	\delta E_{A}(L) &= \frac{\mu^2 \Omega}{8\pi L^2}\cos\left(2\omega_0 k\sinh^{-1}\left(\frac{L}{2k}\right)\right)\Gamma_2(\alpha^1, \alpha^2,\beta^1, \beta^2)
\end{aligned}
\ee

This result shows that the energy level shift is non trivially dependent on the parameter $k$ because of that fact that in this case curvature of the De Sitter space is significant. Here we observe that the RCPI in the limit $L>>k$ falls as $1/L^{2}$, which shows that the first order correction to the energy maintains the inverse square law in De Sitter space.

\item \underline{\textcolor{red}{\bf Case~II:}}\\
When the inter atomic distance is of the order of the characteristic length scale $k$ i.e. $L\sim k$. In this case the two entangled atomic system is placed exactly at the cosmological horizon. In this limit, the energy-level shift from RCPI can be simplified as:

\be
\begin{aligned}
	\delta E_{S}(L) &= -\frac{\mu^2\Omega }{4\sqrt{5}\pi L}\cos\left(2\omega_0 L\sinh^{-1}\left(\frac{1}{2}\right)\right)\Gamma_1(\alpha^1, \alpha^2,\beta^1, \beta^2) \\
	\delta E_{A}(L) &= \frac{\mu^2\Omega }{4\sqrt{5}\pi L}\cos\left(2\omega_0 L\sinh^{-1}\left(\frac{1}{2}\right)\right)\Gamma_2(\alpha^1, \alpha^2,\beta^1, \beta^2)
\end{aligned}
\ee

This result shows that the energy level shift is dependent on the parameter $k$ because of that fact that in this case curvature of the de Sitter space is comparable to the atomic distance, which will give rise to the following constraint condition:
\be 
L=2r\sin\left(\frac{\theta-\theta'}{2}\right)\sim k
\ee
 Here we observe that the \textbf{RCPI} in the limit $L\sim k$ falls as $1/L$, which shows that the first order correction to the energy maintains the inverse law in Minkowski space.
 
\item \underline{\textcolor{red}{\bf Case~III:}}\\
When the inter atomic distance is much smaller than characteristic length scale $k$ i.e. $L<<k$. In this case the two entangled atomic system is placed far from the cosmological horizon. In this case it is possible to find a local inertial frame of reference where all physical principles coincides with that in Minkowski space. In this limit, the energy-level shift from RCPI can be simplified as:

\be
\begin{aligned}
	\delta E_{S}(L) &= -\frac{\mu^2 \Omega}{8\pi L}\Gamma_1(\alpha^1, \alpha^2,\beta^1, \beta^2)\cos\left(\omega_0 L\right) \\
	\delta E_{A}(L) &= \frac{\mu^2 \Omega}{8\pi L}\Gamma_2(\alpha^1, \alpha^2,\beta^1, \beta^2)\cos\left(\omega_0 L \right) 
\end{aligned}
\ee

This result shows that the energy level shift is independent on the parameter $k$ because of that fact that in this case curvature of the De Sitter space is negligibly small. This result exactly matches with the result obtained for Minkowski space. Here we observe that the RCPI in the limit $L\sim k$ falls as $1/L$, which shows that the first order correction to the energy maintains the inverse law in De Sitter space and exact Minkowski space.
\end{enumerate}

Additionally, in all of these physical limits we found the following features:
\begin{itemize}
\item  \underline{\textcolor{red}{\bf Feature~I:}}\\
We also observe that the RCPI contains the Eulerian angles within it in both the limiting results. This suggests that the shift in energy is described by the Eulerian angles which means that the RCPI is a function of the direction along which the spin of the atoms is directed along. In de Sitter space the RCPI is dependent on how the spin of the two atoms are oriented. The orientation of the spins of the two atoms which is determined by the Euler angles of rotation $\alpha^{i}$, $\beta^{i}$ and $\gamma^{i} ~\forall i=1,2,3$, quantifies the RCPI along arbitrary direction of spin projection.

\item \underline{\textcolor{red}{\bf Feature~II:}}\\
The Euler angles of rotation determines the manifestation of the thermal environment that the atoms see in their comoving frame. Furthermore, we observe that the pre-factors in the energy shifts determines the parameter $k$ associated with the temperature \cite{Hu:2013ypa,Hu:2012ed,Yu:2011eq,Hu:2015lda,Gibbons:1977mu,Zhou:2010nb} of the thermal bath:
\be
\small{T=\frac{1}{2\pi k}=\sqrt{T^2_{\bf GH}+T^{2}_{\bf Unruh}}=\frac{1}{2\pi\alpha}\sqrt{1+\frac{r^2}{\left(\alpha^2-r^2\right)}}=\frac{1}{2\pi}\sqrt{\frac{\Lambda}{3}}\sqrt{1+\frac{r^2}{\left(\frac{3}{\Lambda}-r^2\right)}}}
\ee
where, the \textbf{Gibbons-Hawking temperature} and  \textbf{Unruh temperature} are defined through the following expressions:
\bea 
T_{\bf GH}&=&\frac{1}{2\pi \alpha}=\frac{1}{2\pi}\sqrt{\frac{\Lambda}{3}} \\
T_{\bf Unruh}&=&\frac{a}{2\pi}=\frac{1}{2\pi \alpha}\frac{r}{\sqrt{\alpha^2-r^2}}=\frac{1}{2\pi}\sqrt{\frac{\Lambda}{3}}\frac{r}{\sqrt{\frac{3}{\Lambda}-r^2}} 
\eea 
with the proper acceleration given by:
 \be a=\frac{1}{\alpha}\frac{r}{\sqrt{\alpha^2-r^2}}=\sqrt{\frac{\Lambda}{3}}\frac{r}{\sqrt{\frac{3}{\Lambda}-r^2}}
 \ee
 which is defined in the co-moving frame of the two entangled atoms for the given OQS under consideration. 
 \item \underline{\textcolor{red}{\bf Feature~III:}}\\
 Now, it is important to note that in De Sitter space the curvature can be quantified though the Ricci scalar, which can be further expressed in terms of the cosmological constant as:
 \be
 R_{\bf DS}= \frac{12}{\alpha}=12\sqrt{\frac{\Lambda}{3}}
 \ee
 As a result, the \textbf{Gibbons-Hawking temperature} and  \textbf{Unruh temperature} can be expressed in terms of the Curvature of De Sitter space as:
 \bea 
 T_{\bf GH}&=&\frac{R_{\bf DS}}{24\pi} \\
T_{\bf Unruh}&=&\frac{R_{\bf DS}}{24\pi}\frac{R_{DS}r}{\sqrt{144-(R_{\bf DS}r)^2}}
\eea 
Consequently, the temperature of thermal bath can be expressed in terms of the curvature of de Sitter space as:
\be 
T=\frac{1}{2\pi k}=\frac{R_{\bf DS}}{24\pi}\frac{1}{\sqrt{1-\left(\frac{R_{\bf DS}r}{12}\right)^2}}
\ee

In this case, the RCPI can be expressed in terms of the curvature of the de Sitter space as: 

\be
\scalebox{0.9}[1]{$\footnotesize{\begin{aligned}
	\delta E_{S}(L) &= -\frac{\mu^2 \Omega~~\Gamma_1(\alpha^1, \alpha^2,\beta^1, \beta^2)}{8\pi L\sqrt{1+\left(\frac{LR_{\bf DS}}{24\sqrt{1-\left(\frac{R_{\bf DS}r}{12}\right)^2}}\right)^2}}\cos\left(\frac{24\omega_0 }{R_{\bf DS}}\sqrt{1-\left(\frac{R_{DS}r}{12}\right)^2}\sinh^{-1}\left(\frac{LR_{\bf DS}}{24\sqrt{1-\left(\frac{Rr}{12}\right)^2}}\right)\right)\\
	\delta E_{A}(L) &= \frac{\mu^2 \Omega~~\Gamma_2(\alpha^1, \alpha^2,\beta^1, \beta^2)}{8\pi L\sqrt{1+\left(\frac{LR_{\bf DS}}{24\sqrt{1-\left(\frac{R_{\bf DS}r}{12}\right)^2}}\right)^2}}\cos\left(\frac{24\omega_0 }{R_{\bf DS}}\sqrt{1-\left(\frac{R_{DS}r}{12}\right)^2}\sinh^{-1}\left(\frac{LR_{\bf DS}}{24\sqrt{1-\left(\frac{Rr}{12}\right)^2}}\right)\right)
\end{aligned}}$}
\ee

In the limiting situation, $L>>k=12\sqrt{1-\left(\frac{R_{DS}r}{12}\right)^2}/R_{\bf DS}$, the RCPI can be expressed in terms of the curvature of the de Sitter space as:

\bea
\scalebox{0.8}[1]{$\footnotesize{\begin{aligned}
\delta E_{S}\left(L>>k\right) &= -\frac{3\mu^2 \Omega}{R_{\bf DS}L^2\pi}\cos\left(\frac{24\omega_0 }{R_{\bf DS}}\sqrt{1-\left(\frac{R_{\bf DS}r}{12}\right)^2}\sinh^{-1}\left(\frac{LR_{\bf DS}}{24\sqrt{1-\left(\frac{R_{\bf DS}r}{12}\right)^2}}\right)\right) \Gamma_1(\alpha^1, \alpha^2,\beta^1, \beta^2) \\
\delta E_{A}\left(L>>k\right) &= \frac{3\mu^2 \Omega}{R_{\bf DS}L^2\pi}\cos\left(\frac{24\omega_0 }{R_{\bf DS}}\sqrt{1-\left(\frac{R_{\bf DS}r}{12}\right)^2}\sinh^{-1}\left(\frac{LR_{\bf DS}}{24\sqrt{1-\left(\frac{R_{\bf DS}r}{12}\right)^2}}\right)\right) \Gamma_2(\alpha^1, \alpha^2,\beta^1, \beta^2)
\end{aligned}}$}
\eea

Here one can consider another limiting situation, where $L\sim k=12\sqrt{1-\left(\frac{R_{DS}r}{12}\right)^2}/R_{DS}$. For this case, RCPI can be expressed in terms of the curvature of the de Sitter space as: 

\bea
\scalebox{0.9}[1]{$\footnotesize{\begin{aligned}
			\delta E_{S}\left(L\sim k\right) &= -\frac{R_{\bf DS} \mu^2 \Omega}{48 \sqrt{5}\pi \sqrt{1-\left(\frac{R_{\bf DS}r}{12}\right)^2}}\cos\left(\frac{24\omega_0 }{R_{\bf DS}}\sqrt{1-\left(\frac{R_{\bf DS}r}{12}\right)^2}\sinh^{-1}\left(\frac{1}{2}\right)\right)\Gamma_1(\alpha^1, \alpha^2,\beta^1, \beta^2) \\
			\delta E_{A}\left(L\sim k\right) &= \frac{R_{\bf DS} \mu^2 \Omega}{48 \sqrt{5}\pi \sqrt{1-\left(\frac{R_{\bf DS}r}{12}\right)^2}}\cos\left(\frac{24\omega_0 }{R_{\bf DS}}\sqrt{1-\left(\frac{R_{\bf DS}r}{12}\right)^2}\sinh^{-1}\left(\frac{1}{2}\right)\right)\Gamma_2(\alpha^1, \alpha^2,\beta^1, \beta^2)
		\end{aligned}}$}
\eea

\item \underline{\textcolor{red}{\bf Feature~IV:}}\\
Now, if we take the limit $R_{\bf DS}\rightarrow 0$ ( i.e. $\Lambda\rightarrow 0$ or $\alpha\rightarrow \infty$), then we get:
\be 
\lim_{R_{\bf DS}\rightarrow 0} T_{\bf GH}=0,~~\lim_{R_{\bf DS}\rightarrow 0} T_{\bf Unruh}=0~~ \Longrightarrow~~ \lim_{R_{\bf DS}\rightarrow 0} T=0~~ \Longrightarrow~~ k\rightarrow \infty
\ee
In this case te {\bf RCPI} will be reduced to the result obtained in the limiting situation $L<<k$, which is exactly same result as obtained for the Minkowski space. 
\item \underline{\textcolor{red}{\bf Feature~V:}}\\
Here additionally it is important to note that, only if the {\bf Unruh temperature} vanishes, the corresponding proper acceleration of static atom vanishes i.e. $a=0$ and this can be obtained when the atoms are localised at $r=0$. As a result, kinematic contribution will not appear in the expression for RCPI. However, in this case, the RCPI is still can be expressed in terms of the curvature of the De Sitter space, as in this case temperature of the thermal bath is quantified by the non vanishing \textbf{Gibbons-Hawking temperature} i.e.
\be 
T=\frac{1}{2\pi k}=T_{\bf GH}=\frac{R_{DS}}{24\pi}
\ee
In this case, the RCPI can be expressed in terms of the curvature of the de Sitter space as: 

\be
\small{\begin{aligned}
	\delta E_{S}(L) &= -\frac{\mu^2 \Omega}{8\pi L\sqrt{1+\left(\frac{LR_{\bf DS}}{24}\right)^2}}\cos\left(\frac{24\omega_0 }{R_{\bf DS}} \sinh^{-1}\left(\frac{LR_{\bf DS}}{24}\right)\right)\Gamma_1(\alpha^1, \alpha^2,\beta^1, \beta^2) \\
	\delta E_{A}(L) &= \frac{\mu^2 \Omega}{8\pi L\sqrt{1+\left(\frac{LR_{\bf DS}}{24}\right)^2}}\cos\left(\frac{24\omega_0 }{R_{\bf DS}} \sinh^{-1}\left(\frac{LR_{\bf DS}}{24}\right) \right)\Gamma_2(\alpha^1, \alpha^2,\beta^1, \beta^2)
\end{aligned}}
\ee

In the limiting situation, $L>>k=12/R_{\bf DS}$, the RCPI can be expressed in terms of the curvature of the de Sitter space as:

\be
\footnotesize{\begin{aligned}
	\delta E_{S}(L>>k=12/R_{\bf DS}) &= -\frac{3\mu^2\Omega  }{R_{\bf DS} L^2\pi }\cos\left(\frac{24\omega_0 }{R_{\bf DS}} \sinh^{-1}\left(\frac{LR_{\bf DS}}{24}\right)\right)\Gamma_1(\alpha^1, \alpha^2,\beta^1, \beta^2)\\
	\delta E_{A}(L>>k=12/R_{\bf DS}) &= \frac{3\mu^2 \Omega}{R_{\bf DS} L^2\pi}\cos\left(\frac{24\omega_0 }{R_{\bf DS}} \sinh^{-1}\left(\frac{LR_{\bf DS}}{24}\right) \right)\Gamma_2(\alpha^1, \alpha^2,\beta^1, \beta^2)
\end{aligned}}
\ee

Here one can consider another limiting situation, where $L\sim k=12/R_{\bf DS}$. For this case, {\bf RCPI} can be expressed in terms of the curvature of the De Sitter space as:

\be
\small{\begin{aligned}
	\delta E_{S}(L\sim k=12/R_{\bf DS}) &= -\frac{R_{\bf DS} \mu^2 \Omega}{48\sqrt{5}\pi} \cos\left(\frac{24\omega_0 }{R_{\bf DS}} \sinh^{-1}\left(\frac{1}{2}\right)\right)\Gamma_1(\alpha^1, \alpha^2,\beta^1, \beta^2) \\
	\delta E_{A}(L\sim k=12/R_{\bf DS}) &= \frac{R_{\bf DS} \mu^2 \Omega}{48\sqrt{5}\pi}\cos\left(\frac{24\omega_0 }{R_{\bf DS}} \sinh^{-1}\left(\frac{1}{2}\right) \right)\Gamma_2(\alpha^1, \alpha^2,\beta^1, \beta^2)
\end{aligned}}
\ee

\item \underline{\textcolor{red}{\bf Feature~VI:}}\\
Now, we compare the obtained results for Lamb Shift in De Sitter space with the result corresponding to the Minkowski space. For this purpose we consider a specific situation where two static atoms are interacting with the environment, where it is represented by the  massless scalar field in OQS. In this system, the two point field correlation can be expressed in terms of the Wightman function given by:
\bea
\begin{split}
G^{11}(x,x')&=G^{22}(x,x') \\ &=-\frac{1}{4\pi^2}\sum^{\infty}_{q=-\infty}\frac{1}{\left(\tau-\tau'-i\left\{\frac{q}{T}+\epsilon\right\}\right)^2} \\
&= -\frac{1}{4\pi^2}\sum^{\infty}_{q=-\infty}\frac{1}{\left(\tau-\tau'-i\left\{2\pi k q+\epsilon\right\}\right)^2} \\
&= \frac{1}{16 \pi ^2 k^2}~{\rm cosec}^2\left(\frac{\epsilon+i (\tau-\tau')}{2 k}\right) \\
G^{12}(x,x')&= G^{21}(x,x') \\
&=-\frac{1}{4\pi^2}\sum^{\infty}_{q=-\infty}\frac{1}{\left(\tau-\tau'-i\left\{\frac{q}{T}+\epsilon\right\}\right)^2-L^2} \\
&=-\frac{1}{4\pi^2}\sum^{\infty}_{q=-\infty}\frac{1}{\left(\tau-\tau'-i\left\{2\pi k q+\epsilon\right\}\right)^2-L^2} \\
&=\frac{1}{16 \pi ^2 k L}~\left[2 \left\{{\rm Floor}\left( \frac{\arg \left(\frac{\epsilon+i (t-t'+L)}{k}\right)}{2 \pi }\right)-{\rm Floor}\left( \frac{\arg \left(\frac{\epsilon+i (t-t'-L)}{k}\right)}{2 \pi }\right)\right\} \right.\\
& \left.~~~~~~~~~~~ +i \left\{\cot \left(\frac{\epsilon+i (t-t'+L)}{2 k}\right)-\cot \left(\frac{\epsilon+i (t-t'-L)}{2 k}\right)\right\}\right]
\end{split} 
\eea
where $L$ is the Euclidean distance between two atoms in OQS. Using this Wightman function we can carry forward the similar calculation for Lamb Shift from the RCPI in Minkowsi space, which will finally give rise to the following expression:
\be
\begin{aligned}
	\delta E^{(M)}_{S} &= -\frac{\mu^2 \Omega}{8\pi L} \Gamma_1(\alpha^1, \alpha^2,\beta^1, \beta^2) cos\left(\omega_0 L\right) \\
	\delta E^{(M)}_{A} &= \frac{ \mu^2 \Omega}{8\pi L} \Gamma_2(\alpha^1, \alpha^2,\beta^1, \beta^2) \cos\left(\omega_0 L\right) 
\end{aligned} 
\ee
From the above mentioned result it is clearly observed that the Lamb shift obtained from RCPI in Minkowski space not containing any contribution from the temperature of the thermal bath, $T=1/2\pi k$ and only depends on Euler angles and the Euclidean distance $L$.  Also we found that this result exactly matches with the result obtained for the case $L<<k$ with two inertial atoms in the static patch of de Sitter space.  This result additionally implies that the inter-atomic interaction between two atoms behave differently in Minkowski space and in the static patch of de Sitter space.  For two or more atoms one can construct the ground,  excited,  symmetric and antisymmetric entangled states,  out of which symmetric and antisymmetric entangled quantum states will give rise to non-zero spectroscopic shifts in the two limiting situation,  $L\gg k$ and $L\ll k$.   Using these limiting results one can able to probe RCPI and distinguish between the static de Sitter and Minkowski flat space-time in a very precise manner.  But instead of using two atoms if we use only one single atom within the framework of OQS then one cannot implement the methodology quantum mechanical entanglement in the present scenario.  For the single atomic non-entangled case one can only construct the ground state and excited state,  but there will be no symmetric and antisymmetric states exist,  which are .the necessary ingredient to probe RCPI using which one can able to distinguish the effects of Minkowski flat space-time and static patch of de Sitter space-time.  For two or more atomic scenario the atoms are entangled with each other and interacting with a thermal bath which is modelled with scalar field embedded in the static patch of de Sitter space.  In this scenario,  RCPI can be used to study the spectroscopic shift from the spin-spin self interaction from the effective part of the Hamiltonian.  On the other hand,  when we describe the present scenario with a single atom which is interacting with a thermal bath modelled by a scalar field in static de Sitter background due to the absence of having any quantum entanglement RCPI will be replaced by a very simple self interacting quadratic term.  Consequently,  using the spectroscopic shift for single atom-bath system one cannot distinguish between the Minkowski space-time and static patch of the de Sitter space.  Here two or more atomic OQS is the only feasible option using which one can precisely distinguish the behaviour of RCPI in Minkowski space and in the static patch of de Sitter space due to having quantum entanglement between the atoms which are interacting with the thermal bath. 
In short,  the phenomena of quantum entanglement play very significant role in the present context of discussion as this can be only explained through two or more atomic systems to probe the curvature of the space-time through RCPI. in the spectroscopic shift computation.  But as the quantum entanglement is not there in single atomic system,  one cannot probe the curvature of the space-time in the present context. 

\item To understand the obtained results in a more better way and to compare with the previously obtained results in the same literature we discuss now two limiting situations, which are appended below:
\begin{enumerate}
\item {\bf \textcolor{blue}{Limit~I:}}\\
In this case we fix the position of the two detectors at the angular positions $(\alpha^{1}=\pi/2,\beta^{1}=\pi/2,\gamma^{1}=0)$ and $(\alpha^{2}=\pi/2,\beta^{2}=\pi/2,\gamma^{1}=0)$ respectively. In this limit, $H_A$ matches with the model studied in previous works, but $H_{LS}$ does not. 

\item {\bf \textcolor{blue}{Limit~II:}}\\
In this case we fix the position of the two detectors at the angular positions $(\alpha^{1}=0,\beta^{1}=0,\gamma^{1}=0)$ and $(\alpha^{2}=0,\beta^{2}=0,\gamma^{1}=0)$ respectively. In this limit, $H_A$ does not match the model studied in previous works, but $H_{LS}$ does.

\end{enumerate}

\end{itemize}

\begin{table}[htb]
\begin{center}
	\begin{tabular}{|m{6.9cm}|m{4.9cm}|m{4.9cm}|} 
		\hline\hline \hline
		 Our ~work& Limit I & Limit II \\ 
		\hline
	   $ \displaystyle H_A=\frac{\omega}{2}\sum^{2}_{\alpha=1}\hat{\bf n}.{\bf \sigma}^{\alpha}$& $\displaystyle H_A=\frac{\omega}{2}\sum^{2}_{\alpha=1}\sigma^{\alpha}_3$ & $\displaystyle H_A=\frac{\omega}{2}\sum^{2}_{\alpha=1}({\sigma}^{\alpha}_1 + {\sigma}^{\alpha}_2 + {\sigma}^{\alpha}_3 ) $  \\ 
	   & (\textcolor{red}{\bf Previous ~work})  &  \\
	   \hline 
 $\displaystyle H_{LS}$=-$\displaystyle\frac{i}{2}\sum_{\alpha,\beta=1}^{2}\sum_{i,j=1}^{3}H^{(\alpha \beta)}_{ij}( n^\alpha_i .\sigma^{\alpha}_{i}) (n^\beta_j .\sigma^{\beta}_{j})$ &  $\displaystyle H_{LS}$=-$\displaystyle\frac{i}{2}\sum_{\alpha,\beta=1}^{2}\sum_{i,j=1}^{3}H^{(\alpha \beta)}_{ij} \sigma^{\alpha}_{3} \sigma^{\beta}_{3}$& $\displaystyle H_{LS}$=-$\displaystyle \frac{i}{2}\sum_{\alpha,\beta=1}^{2}\sum_{i,j=1}^{3}H^{(\alpha \beta)}_{ij}\sigma^{\alpha}_{i} \sigma^{\beta}_{j}$  \\ 
 	   &  & (\textcolor{red}{\bf Previous ~work})  \\
		\hline \hline \hline
	\end{tabular}
		\caption{ Comparison between our work and previous works.}
\end{center}
\label{tab1x}
\end{table}

\begin{figure}
	\centering
	\includegraphics[width=17cm,height=12.7cm]{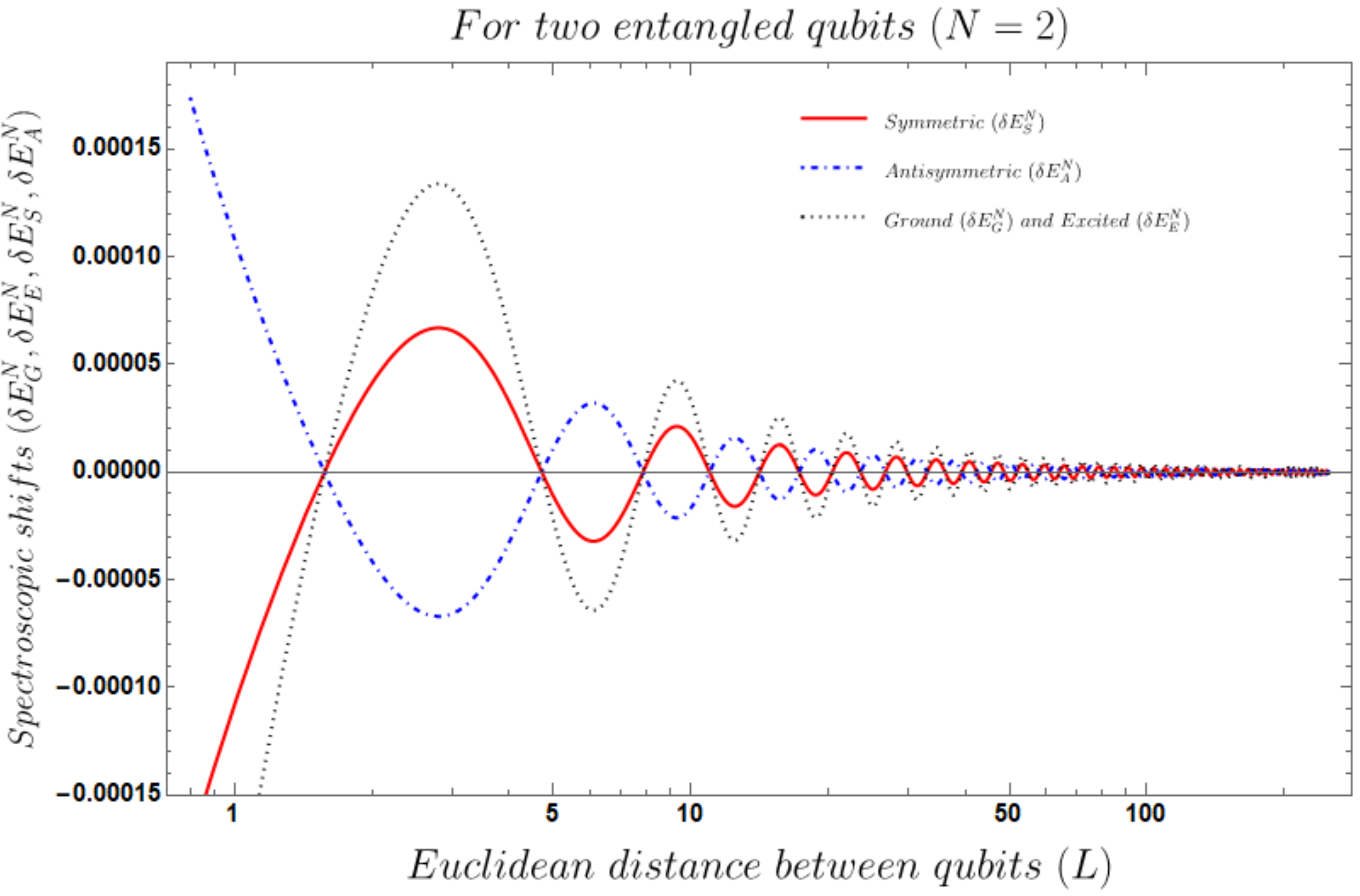}
	\caption{Behaviour of spectroscopic shifts wrt the euclidean distance between the qubits. The other parameters are kept fixed. ($\omega$=1, k$\approx$~$10^{61}$, $\mu$=0.1, $\omega_{0}$=1)}
	\label{spectralshiftwrtL}
\end{figure}

\begin{figure}
	\centering
	\includegraphics[width=17cm,height=12.7cm]{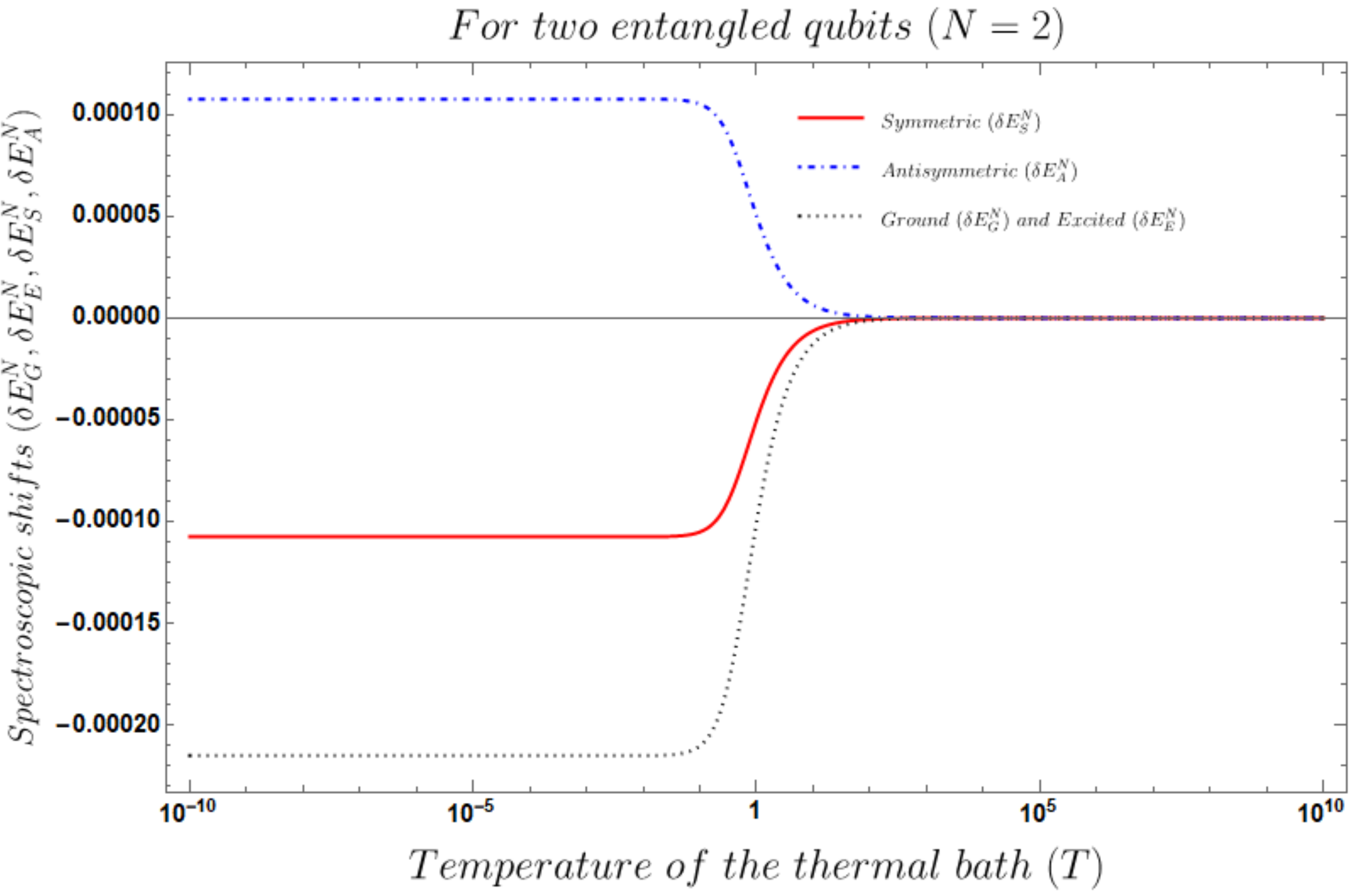}
	\caption{Behaviour of spectroscopic shifts wrt the temperature of the thermal bath. The other parameters are kept fixed. ($\omega$=0.5, L=1, $\mu$=0.1, $\omega_{0}$=1)}
	\label{spectralshiftwrtT}
\end{figure}

\begin{figure}
	\centering
	\includegraphics[width=17cm,height=12.7cm]{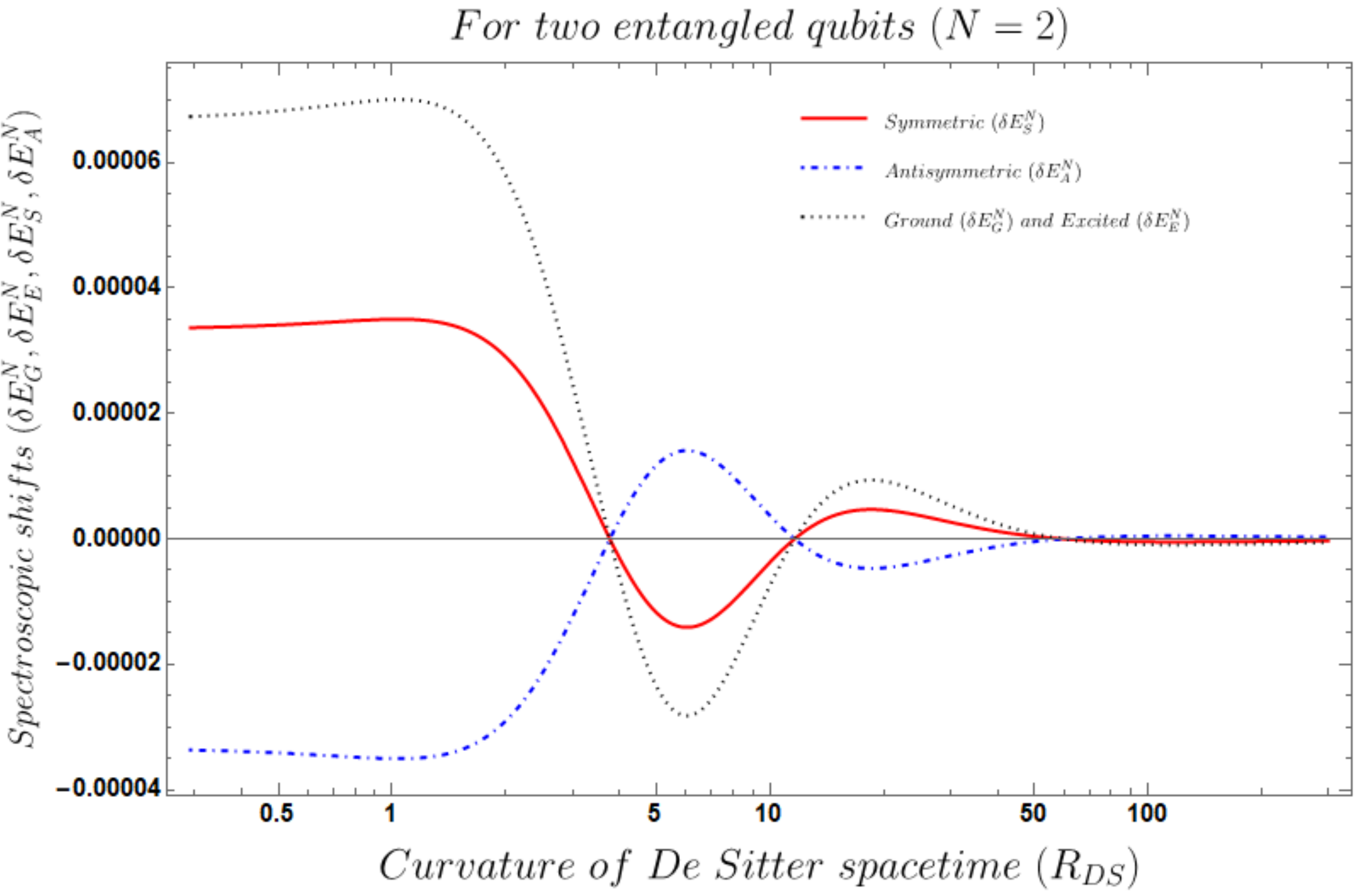}
	\caption{Behaviour of spectroscopic shifts wrt the curvature of the De-Sitter spacetime. The other parameters are kept fixed. ($\omega$=1, L=1, $\mu$=0.1, $\omega_{0}$=1)}
	\label{spectralshiftwrtR}
\end{figure}
In table \ref{tab1x},  we have explicitly mentioned the comparison between our work and previous works.

In fig: \ref{spectralshiftwrtL}, the behaviour of the spectroscopic Lamb shifts have been plotted with respect to the Euclidean distance between the two qubits. It is clearly observed that for smaller values of $L$, the shifts fluctuate with a very large amplitude. The rate of fluctuation increases with increasing $L$ with a decreasing amplitude and for asymptotically large value of $L$ the shifts diminish to zero. A naive interpretation from this plot can be that to measure a finite shift, smaller values of $L$ is preferred to larger values.

In fig: \ref{spectralshiftwrtT}, the spectroscopic Lamb shifts are plotted against the temperature of the thermal bath. It can be seen that for lower temperature of the bath a finite value of the spectral shift is obtained. However, after a certain characteristic temperature the spectral shifts start decaying and finally reach zero. This finite value of the spectral shifts for smaller values of the temperature of the thermal bath suggests that by this indirect mechanism one can expect a large value of the inverse curvature parameter k for the static patch of de Sitter space, which is related to the temperature of the thermal bath via the relation $T=1/2\pi k$.

In fig: \ref{spectralshiftwrtR}, we have plotted the characteristics of the spectroscopic Lamb shifts with respect to the curvature of the de Sitter background. It can be seen that for lower values of the curvature a finite value of the spectral shift is obtained. However, after a certain characteristic value of $R_{DS}$ the spectral shifts start diminishing in an oscillatory fashion and gradually go to zero. For larger values of $R_{DS}$, the spectral shifts oscillates rapidly to give negligible value. Although from this indirect detection mechanism, it is not possible to determine the exact value of the de Sitter curvature but one can predict that its value is smaller than a certain characteristic value and can never be larger than that.

\section{\textcolor{blue}{\bf \large Conclusion}}

In this paper, we have established a direct connection between the curvature of De Sitter space and atomic spectroscopy for an OQS described by two entangled atoms. For this purpose, without doing any terrestrial and space observations designing a laboratory atomic experiment is very useful to establish the connection between space time geometry and quantum mechanics. To summarise, in this work, we have addressed the following issues to  implement the above mentioned idea:

\begin{itemize}
\item To begin with, we have started our discussion with an OQS characterised by two entangled atoms. We have considered the two body quantum entanglement as in this situation 
it is allowed to exchange energy through Lamb Shift in terms of the geometry of De Sitter space-time. In this theoretical construction these two atomic pair represents {\it Unruh-De-Witt detectors}, which are considered to be conformally coupled to a background scalar field in thermal bath. 

\item The non-adiabatic interaction between the detectors and the thermal bath in OQS is characterised by {\bf RCPI}, which plays the key ingredient to determine the curvature of De Sitter space from the Lamb Shift spectroscopy. 

\item In order to study the full dynamics of the two entangled atoms in OQS for any arbitrary position of two detectors, we have used a generalised Hamiltonian described by Pauli operators which include contributions from the Euler rotation angles due to projection in any arbitrary direction. This direction of projection actually playing the role of direction of observation of the atoms in the atomic detectors. In this discussion, the Lamb Shift Hamiltonian includes a term that arises from the interaction between the atomic detectors with the background test scalar field. This is the most significant term which can be experimentally probed using atomic spectroscopy to detect to geometry of De Sitter space-time. In the time dynamics of the reduced density matrix a generalised expression for this contributions actually sourced by the interaction between the atomic detectors with the test scalar field, encapsulated by the Linbaldian operator in two body quantum entangled open system. This study actually helps us to know about the time evolution of the OQS from the perspective of experimentalist in the atomic detector's side. 

\item Apart from solving this prime issue, another significant motivation of our work is to quantify the two point correlation function (i.e Wightman function) between the two entangled atoms for OQS in the De Sitter background. This results in expressions for four possibilities of the Wightman function that would now directly relate the quantum fluctuations in the background geometry of De Sitter space to the Lamb Shift in atomic spectroscopy. 

\item To compare between the geometrical features of De Sitter and Minkowski flat space we have also computed the Lamb Shift from the two body entangled OQS set up in the thermal state for the flat case. In the case of De Sitter space the Lamb Shift is described by inverse square power law dependence on the Euclidean distance ($L$), which is characterised by the length scale associated with the breaking of local inertial description of the two entangled atomic OQS. On the other hand, in the Minkowski flat case we do not get any temperature dependence and in this case the the spectroscopic Lamb Shift is described by $L^{-1}$ behaviour.
From this discussion it is evident that, even both thermal Minkowski and De Sitter space satisfy similar kind of properties and cannot explicitly discriminated by a single external probe field, but using  non adiabatic {\bf RCPI} it is possible to differentiate between these two geometrical space-times.

\item From the obtained result for the Lamb Shift it is evident that, if the geometry of the space-time is curved, particularly if it is De Sitter space then in the context of two body entangled OQS the non-adiabatic inter-atomic {\bf RCPI} is purely characterised by three important contributions which are appended bellow:
\begin{enumerate}
\item The amplitude of the Lamb Shift is mainly characterised by the $L^{-2}$ factor, which indicates the inverse square power law decay in the limiting situation $L>>k$, where the inter atomic distance is larger than the characteristic length scale $k$. 

\item In the amplitude of the Lamb Shift another angular modulation factor contribute, given by the following expression lying within the following window:
$$0<(\cos^2(\alpha^{1})+\cos^2(\alpha^{2})+\cos^2(\beta^{1})+\cos^2(\beta^{2}))<4.$$ This factor is appearing in the coefficient of the cut-off dependent contribution after applying Bethe rugularisation procedure. However, in the limiting situation where the Bethe cut-off frequency is smaller than the natural frequency of the two entangled atomic system i.e. $\omega_c<<\omega_0$, such contribution will not contribute in the amplitude of Lamb Shift. 

\end{enumerate}
\item On the other hand, in case of flat space-time the amplitude of the Lamb Shift is proportional to $L^{-1}$, and the angular modulation factor ${\cal D}$ lying within the same window as mentioned above.  Most importantly, in the final expression for the Lamb Shift as appearing in the case of flat space no signature of the actual origin of quantum state i.e. whether it is non-thermal or thermal, can be observable.

\item Finally, we have added a short discussion regarding the {\it Gibbons Hawking temperature} and {\it Unruh temperature} that these detectors would measure when accelerating through the De Sitter background space-time. We have included this discussion regarding the equilibrium temperature of the thermal bath as it can be directly expressed in terms of the curvature of the De Sitter space and consequently the energy shift can be expressed in terms of these fundamental quantities.

\end{itemize}

The future prospects of our work are appended point-wise.
\begin{itemize}

\item It is completely true fact that in quantum field theory in de Sitter space (e.g. inflationary perturbations), 2-point functions encode the power spectrum while 3-point functions encode aspects of non-gaussianities.  But to explain this quantum entanglement is not the necessary ingredient.  Now one can consider the three atomic entanglement where one needs to consider pair wise entanglement and third atom is not entanglement.  Considering this possibility one can construct ground,  excited,  symmetric and antisymmetric entangled states using which one can study the perturbation in static de Sitter background and after path integrating out the bath degrees of freedom (scalar field in our consideration) one can explicitly compute two, point function and the associated power spectrum and three point functions and the associated non-gaussianities in the present context.  If this type computation is performed in the present context,  then that will be surely very helpful to probe the underlying physics of quantum mechanical entanglement at the level of correlation functions.  It is expected that the result obtained from one un-entangled atom and the entangled two/three or more atom at the level finding the quantum correlation function will be different and distinguishable.  Now if one can able to test these possibilities through various observational probes,  then one can say that whether quantum mechanical entanglement is at all be tested in near future or not.  By following the referees suggestion regarding this issue we are now thinking of doing the same computation in near future and we believe this analysis will be helpful to explore various unknowns in this context.  We have added this point in the discussion of future direction of the present work in the conclusion section~\footnote{We are thankful to the referee for pointing this issue.}.

\item Last but not the least,  in the present context one can also study the de Sitter entropy from the gravitational entanglement in the static patch of de Sitter space-time.  In this connection,  one can use generalizations of Ryu-Takayanagi holographic entanglement \cite{Ryu:2006bv, Ryu:2006ef} by following the references \cite{Lewkowycz:2019xse, Narayan:2020nsc} and related works in this area.  Also one can use,  the concept of the well known ER=EPR \cite{Maldacena:2013xja},  and related ideas on geometry-entanglement in the context of de Sitter space.  Till now we have not been used the concept and computational techniques of the gravitational entanglement to compute the de Sitter entropy.  We have used the tools and techniques of quantum field theory to study the result of de Sitter entropy in various different contexts.  It would be really good and interesting if the calculation can be performed in the gravitational side as well to give complete understanding and overview of the full theory~\footnote{We are thankful to the referee for reminding us this issue.}.  

\item In future we will extend our discussion for multi entangled (even and odd number of atoms) OQS to understand the connection between curvature of De Sitter space and Lamb Shift. 

\item Also, our plan also is to study the non-unitary dissipative time dynamics of the system from which one can compute the expression for the equilibrium temperature of the bath, which help us to check the consistency about the expression for the temperature obtained from gravity sector. 

\item One can also study various quantum information theoretic measure  from the present multi entangled open quantum theory set up to know more about many body quantum entanglement. 

\item Study of quantum fluctuations and related inflationary perturbations from the OQS is not well established yet in the context of early universe cosmology. For this reason it is good to study the cosmological consequences from the correlation functions and comparison with the various observables from OQS to know about many more unknown physical facts. 

\item Many body localisation and eigenstate thermalisation, study of tensor networks \cite{Bao:2017qmt,Biamonte:2017dgr,Orus:2013kga} and physics of quantum chaos \cite{Choudhury:2018rjl,Choudhury:2018lcb,Choudhury:2020yaa,Bhagat:2020pcd,Bhargava:2020fhl,Choudhury:2020lja,Choudhury:2020hil,Choudhury:2021qod} from the open quantum set up are also unexplored issues which one can study in detail from the present set up as well.
\end{itemize}

\section*{\textcolor{blue}{\bf \large Acknowledgements}}

SC would like to thank Quantum Gravity and Unified Theory and Theoretical Cosmology
Group, Max Planck Institute for Gravitational Physics, Albert Einstein Institute for providing the Post-Doctoral Research Fellowship.  SC also thank NISER , Bhubaneswar to provide the work friendly environment.  Resarch of SC is funded by Professor Sudhakar Panda's Jagadish Chandra Bose Fellowship. SC thank the organisers of Summer School on Cosmology 2018, ICTP, Trieste, 15 th Marcel Grossman Meeting, Rome, The European Einstein Toolkit meeting 2018, Centra, Instituto Superior Tecnico, Lisbon and The Universe as a Quantum Lab, APC, Paris, Nordic String Meeting 2019, AEI, Potsdam, Tensor networks: from simulations to holography, DESY Zeuthen and AEI, Potsdam, XXXI Workshop Beyond the Standard Model, Physikzentrum Bad Honnef, Workshop in String Theory and Cosmology, NISER, Bhubansewar for providing the local
hospitality during the work. We also thank Satyaki Chowdhury, Anjan Kumar Sarkar, Vaishak Prasad who are the members of our newly formed virtual group ``Quantum Structures of the Space-Time \& Matter" for elaborative discussions and suggestions to improve the presentation of the article. SP acknowledges the J. C. Bose National Fellowship for
support of his research. Last but not the least, we would like to acknowledge our debt to
the people belonging to the various part of the world for their generous and steady support for research in natural sciences.
\newpage

\appendix 

\section{\textcolor{blue}{\bf Coefficients of the Lamb shift Hamiltonian and Lindbladian}}
\label{coeff}

In this appendix, we will calculate the co-efficient matrix elements of the Lamb shift Hamiltonian and Lindbladian. For this purpose, we will use the Fourier transform of the two point correlation functions as given in the equation \ref{fourier}. For clarity, we re write those equations here again.


\bea
\mathcal{G}^{11}(\omega) = \mathcal{G}^{22}(\omega)&=& -\int_{-\infty}^{\infty}d\Delta \tau~\frac{e^{i\omega \Delta \tau}}{16\pi^{2}k^{2}\sinh[2](\frac{\Delta \tau}{2k}-i\epsilon)} \nonumber\\
&=&\frac{1}{2\pi}\frac{\omega}{1-e^{-2\pi k \omega}},
\\
\mathcal{G}^{12}(\omega) =\mathcal{G}^{21}(\omega) & =&  -\int_{-\infty}^{\infty}d \Delta \tau~\frac{1}{16\pi^{2}k^{2}}\frac{e^{i\omega \Delta \tau}}{\sinh^2(\frac{\Delta \tau}{2k}-i\epsilon)-\frac{r^{2}}{k^{2}}\sin^2(\frac{\Delta \theta}{2})}\nonumber\\
& =& \frac{1}{2\pi}\frac{\omega}{1-e^{-2\pi k \omega}}f(\omega,L/2),~~~~~~~~
\eea
where, we define the spectral function $f(\omega,L/2)$ as:
\be
 f(\omega,L/2)=\frac{1}{L\omega\sqrt{1+\left(\frac{L}{2k}\right)^2}}\sin(2k\omega \sinh^{-1}\left(\frac{L}{2k}\right)) 
\ee

Now, the elements of co-efficient matrix $H^{(\alpha \beta)}_{ij}$ of the effective Hamiltonian can be explicitly represented by the following expression:
\be
H^{(\alpha \beta)}_{ij}=\mathcal{A}^{(\alpha \beta)}\delta_{ij}-i\mathcal{B}^{(\alpha \beta)}\epsilon_{ijk}\delta_{3k}-{\mathcal{A}}^{(\alpha \beta)}\delta_{3i}\delta_{3j}
\ee

where

\be \label{ab}
\begin{aligned}
\mathcal{A}^{(\alpha \beta)} &= \frac{\mu^{2}}{4}[\mathcal{K}^{(\alpha \beta)}(\omega_{0})+\mathcal{K}^{(\alpha \beta)}(-\omega_{0})] \\
\mathcal{B}^{(\alpha \beta)} &= \frac{\mu^{2}}{4}[\mathcal{K}^{(\alpha \beta)}(\omega_{0})-\mathcal{K}^{(\alpha \beta)}(-\omega_{0})]
\end{aligned} 
\ee
 ${\cal K}^{\alpha\beta}(\pm \omega_0)\forall (\alpha,\beta=1,2)$ represents the Hilbert transform of the two point function in Fourier space, which we have defined in the equation \ref{hil}. These are given as follows:
\be \label{K}
\begin{aligned}
\mathcal{K}^{11}(\omega_{0})  = \mathcal{K}^{22}(\omega_{0}) &= \frac{P}{2\pi^2 i}\int_{-\infty}^{\infty}d\omega~\frac{1}{\omega - \omega_{0}}\frac{\omega}{1-e^{2\pi k\omega}} \\
\mathcal{K}^{12}(\omega_{0})  = \mathcal{K}^{21}(\omega_{0}) &= \frac{P}{2\pi^2 i}\int_{-\infty}^{\infty}d\omega~\frac{1}{\omega - \omega_{0}}\frac{\omega }{1-e^{2\pi k\omega}}f(\omega,L/2)
\end{aligned}
\ee


From the symmetry of Hilbert transformed correlations, it is easy to note that

\be
\mathcal{A}^{11} = \mathcal{A}^{22} \ \ \ \ \ \mathcal{A}^{12} = \mathcal{A}^{21} \ \ \ \ \mathcal{B}^{11} = \mathcal{B}^{22} \ \ \ \ \ \mathcal{B}^{12} = \mathcal{B}^{21} 
\ee

From equation \ref{ab}

\be \label{AB}
\begin{aligned}
\mathcal{A}_{1} \equiv \mathcal{A}^{11} = \mathcal{A}^{22} &= \frac{\mu^2}{4} \left[{\cal K}^{(11)}(\omega_0)+{\cal K}^{(11)}(-\omega_0)\right] \\
\mathcal{B}_{1} \equiv \mathcal{B}^{11} = \mathcal{B}^{22} &= \frac{\mu^2}{4}\left[{\cal K}^{(11)}(\omega_0)-{\cal K}^{(11)}(-\omega_0)\right] \\
\mathcal{A}_{2} \equiv \mathcal{A}^{12} = \mathcal{A}^{21} &= \frac{\mu^2}{4} \left[{\cal K}^{(12)}(\omega_0)+{\cal K}^{(12)}(-\omega_0)\right] \\
\mathcal{B}_{2} \equiv \mathcal{B}^{12} = \mathcal{B}^{21} &= \frac{\mu^2}{4}\left[{\cal K}^{(11)}(\omega_0)-{\cal K}^{(11)}(-\omega_0)\right] 
\end{aligned}
\ee

Further using equation (\ref{K}) in equation (\ref{AB}) we get the following simplified expression for $\mathcal{A}_{1}$, $\mathcal{B}_{1}$, $\mathcal{A}_{2}$ and $\mathcal{B}_{2}$:
\bea \label{Aint}
\mathcal{A}_{1}
&=&
=\frac{\mu^{2}P}{4\pi^2 i}\int_{-\infty}^{\infty}d\omega\frac{\omega^2}{(\omega+\omega_0)(\omega-\omega_0)(1-e^{-2\pi k\omega})},\\
\mathcal{B}_{1}
&=&
=\frac{\mu^{2}P}{4\pi^2 i}\int_{-\infty}^{\infty}d\omega\frac{\omega_0\omega}{(\omega+\omega_0)(\omega-\omega_0)(1-e^{-2\pi k\omega})},\\
\mathcal{A}_{2}
&=&
=\frac{\mu^{2}P}{4\pi^2 i}\int_{-\infty}^{\infty}d\omega\frac{\omega^2~f(\omega,L/2)}{(\omega+\omega_0)(\omega-\omega_0)(1-e^{-2\pi k\omega})},\\
\mathcal{B}_{2}
&=&
=\frac{\mu^{2}P}{4\pi^2 i}\int_{-\infty}^{\infty}d\omega\frac{\omega_0\omega~f(\omega,L/2)}{(\omega+\omega_0)(\omega-\omega_0)(1-e^{-2\pi k\omega})}.
\eea

We will calculate the above spectroscopic integrals explicitly in the next section. \\

Similarly, the elements of the  {\it Gorini–Kossakowski–Sudarshan–Lindblad matrix}, $C^{(\alpha \beta)}_{ij}$, as appearing in the expression for the {\it Linbladian} can be expressed as:

\be
C^{(\alpha \beta)}_{ij}=\tilde{\mathcal{A}}^{(\alpha \beta)}\delta_{ij}-i\tilde{\mathcal{B}}^{(\alpha \beta)}\epsilon_{ijk}\delta_{3k}-\tilde{\mathcal{A}}^{(\alpha \beta)}\delta_{3i}\delta_{3j}~,
\ee 

where, the quantities $\tilde{\mathcal{A}}^{(\alpha \beta)}$ and $\tilde{\mathcal{B}}^{(\alpha \beta)}$ for the two atomic system are defined as:
\bea
\tilde{\mathcal{A}}^{(\alpha \beta)}=\frac{\mu^{2}}{4}[\mathcal{G}^{(\alpha \beta)}(\omega_{0})+\mathcal{G}^{(\alpha \beta)}(-\omega_{0})] \\ 
\tilde{\mathcal{B}}^{(\alpha \beta)}=\frac{\mu^{2}}{4}[\mathcal{G}^{(\alpha \beta)}(\omega_{0})-\mathcal{G}^{(\alpha \beta)}(-\omega_{0})] 
\eea

Again from symmetry of Fourier transformed correlations it can be seen that

\be
\tilde{\mathcal{A}}^{11} = \tilde{\mathcal{A}}^{22} (\equiv \tilde{{\cal A}}_1) \ \ \ \ \ \tilde{\mathcal{A}}^{12} = \tilde{\mathcal{A}}^{21} (\equiv \tilde{{\cal A}}_2)\ \ \ \ \tilde{\mathcal{B}}^{11} = \tilde{\mathcal{B}}^{22} (\equiv \tilde{{\cal B}}_1)\ \ \ \ \ \tilde{\mathcal{B}}^{12} = \tilde{\mathcal{B}}^{21} (\equiv \tilde{{\cal B}}_2)
\ee

Now
\bea
\tilde{{\cal A}}_1&=&\frac{\mu^2}{4}\left[{\cal G}^{(11)}(\omega_0)+{\cal G}^{(11)}(-\omega_0)\right]\nonumber\\
&=&\frac{\mu^2}{4}\left[{\cal G}^{(22)}(\omega_0)+{\cal G}^{(22)}(-\omega_0)\right]\nonumber\\
&=&\frac{\mu^{2}}{8\pi}\omega_0\left[\frac{1}{1-e^{-2\pi k\omega_0}}-\frac{1}{1-e^{2\pi k\omega_0}}\right],\\
\tilde{{\cal B}}_1&=&\frac{\mu^2}{4}\left[{\cal G}^{(11)}(\omega_0)-{\cal G}^{(11)}(-\omega_0)\right]\nonumber\\
&=&\frac{\mu^2}{4}\left[{\cal G}^{(22)}(\omega_0)-{\cal G}^{(22)}(-\omega_0)\right]\nonumber\\
&=&\frac{\mu^{2}}{8\pi}\omega_0\left[\frac{1}{1-e^{-2\pi k\omega_0}}+\frac{1}{1-e^{2\pi k\omega_0}}\right],\eea\bea
\tilde{{\cal A}}_2&=&\frac{\mu^2}{4}\left[{\cal G}^{(12)}(\omega_0)+{\cal G}^{(12)}(-\omega_0)\right]\nonumber\\
&=&\frac{\mu^2}{4}\left[{\cal G}^{(21)}(\omega_0)+{\cal G}^{(21)}(-\omega_0)\right]\nonumber\\
&=&\frac{\mu^{2}}{8\pi}\omega_0\left[\frac{f(\omega_0,L/2)}{1-e^{-2\pi k\omega_0}}-\frac{f(-\omega_0,L/2)}{1-e^{2\pi k\omega_0}}\right],\\
\tilde{{\cal B}}_2&=&\frac{\mu^2}{4}\left[{\cal G}^{(12)}(\omega_0)-{\cal G}^{(12)}(-\omega_0)\right]\nonumber\\
&=&\frac{\mu^2}{4}\left[{\cal G}^{(21)}(\omega_0)-{\cal G}^{(21)}(-\omega_0)\right]\nonumber\\
&=&\frac{\mu^{2}}{8\pi}\omega_0\left[\frac{f(\omega_0,L/2)}{1-e^{-2\pi k\omega_0}}+\frac{f(-\omega_0,L/2)}{1-e^{2\pi k\omega_0}}\right].
\eea

\section{\textcolor{blue}{\bf \large Calculation of Bethe regularised spectroscopic integrals}}
\label{int}
In the following subsections we explicitly compute the {\it Bethe regularised} integrals, which are very useful to compute the expressions for the energy shift from ground, excited, symmetric and antisymmetric state respectively. These integrals are appended below:
\bea 
\underline{\textcolor{red}{\bf Integral ~I:}}~~~~~~~~~~~~~\Delta_{1}:&&=\int^{\infty}_{-\infty}d\omega~\frac{2\omega_0~\omega}{\left(1-e^{-2\pi k\omega}\right)\left(\omega+\omega_0\right)\left(\omega-\omega_0\right)} \\
\underline{\textcolor{red}{\bf Integral ~II:}}~~~~~~~~~~~~\Delta_{2}:&&=\int^{\infty}_{-\infty}d\omega~\frac{\omega^2 }{\left(1-e^{-2\pi k\omega}\right)\left(\omega+\omega_0\right)\left(\omega-\omega_0\right)} \\ 
\underline{\textcolor{red}{\bf Integral
~III:}}~~~~~~~~~~~\Delta_{3}:&&=\int^{\infty}_{-\infty}d\omega~\frac{2\omega^2 f(\omega,L/2)}{\left(1-e^{-2\pi k\omega}\right)\left(\omega+\omega_0\right)\left(\omega-\omega_0\right)}
\eea

\subsection{\textcolor{blue}{\bf \large Spectroscopic Integral~I}}
In this subsection we explicitly compute the finite contribution from the following integral:
\bea \textcolor{red}{\Delta_{1}:=\int^{\infty}_{-\infty}d\omega~\frac{2\omega_0~\omega}{\left(1-e^{-2\pi k\omega}\right)\left(\omega+\omega_0\right)\left(\omega-\omega_0\right)}}~.\eea
It is important to note that in the large frequency range, $-\infty<\omega<\infty$, one can further expand the integrand by taking large $\omega$ approximation as:
\be {\cal F}(\omega_0,\omega,k):=~\frac{2\omega_0~\omega}{\left(1-e^{-2\pi k\omega}\right)\left(\omega+\omega_0\right)\left(\omega-\omega_0\right)}\xrightarrow[\omega\to \infty]{}\frac{2\omega_0~\omega}{\left(\omega+\omega_0\right)\left(\omega-\omega_0\right)}:={\cal F}(\omega_0,\omega)\ee
This implies that, after taking large $\omega$ approximation the integrand of $\Delta_1$ becomes independent of the parameter $k$, which representing the surface gravity. 

Now, further using this approximation the integral $\Delta_1$ can be further simplified as:
\bea 
\Delta_{1}&&\approx \int^{\infty}_{-\infty}d\omega~{\cal F}(\omega_0,\omega)=\underbrace{\int^{0}_{-\infty}d\omega~{\cal F}(\omega_0,\omega)}_{\textcolor{red}{\equiv~ {\cal U}_1(\omega_0)}}~+~\underbrace{\int^{\infty}_{0}d\omega~{\cal F}(\omega_0,\omega)}_{\textcolor{red}{\equiv~ {\cal U}_2(\omega_0)}}
\eea
where we have decomposed the integrals into two parts, indicated by 
$\textcolor{red}{{\cal U}_1(\omega_0)}$ and $\textcolor{red}{{\cal U}_2(\omega_0)}$ in the parenthesis symbol. Now, here we see that in the large frequency range, $-\infty<\omega<\infty$, we get:
\bea 
{\cal U}_1(\omega_0)=\int^{0}_{-\infty}d\omega~{\cal F}(\omega_0,\omega)=-\int^{\infty}_{0}d\omega~{\cal F}(\omega_0,\omega)=-{\cal U}_2(\omega_0)
\eea
Now, we see here that both $\textcolor{red}{{\cal U}_1(\omega_0)}$ and $\textcolor{red}{{\cal U}_2(\omega_0)}$ gives divergent contributions in the frequency range, $-\infty<\omega<0$ and $0<\omega<\infty$. To get the finite contributions from these integrals we introduce a cut-off regulator $\omega_c$, by following {\it Bethe regularisation} technique. After introducing this cut-off we get:
\be
\begin{split}
{\cal U}_1(\omega_0,\omega_c)&=\int^{0}_{-\omega_c}d\omega~{\cal F}(\omega_0,\omega)=-\int^{\omega_c}_{0}d\omega~{\cal F}(\omega_0,\omega) \\
&=-{\cal U}_2(\omega_0,\omega_c)=-\omega_0\ln\left[1-\left(\frac{\omega_c}{\omega_0}\right)^2\right]
\end{split}
\ee
Consequently, we get the following expression for the integral $\Delta_1$, as given by:
\be
{\textcolor{red}{\Delta_{1}=\textcolor{red}{{\cal U}_1(\omega_0,\omega_c)}+\textcolor{red}{{\cal U}_2(\omega_0,\omega_c)}=\omega_0\ln\left[1-\left(\frac{\omega_c}{\omega_0}\right)^2\right]-\omega_0\ln\left[1-\left(\frac{\omega_c}{\omega_0}\right)^2\right]=0}}
\ee

\subsection{\textcolor{blue}{\bf \large Spectroscopic Integral~II}}
In this subsection we explicitly compute the finite contribution from the following integral:
\bea {\textcolor{red}{\Delta_{2}:=\int^{\infty}_{-\infty}d\omega~\frac{\omega^2}{\left(1-e^{-2\pi k\omega}\right)\left(\omega+\omega_0\right)\left(\omega-\omega_0\right)}}}\eea
It is important to note that in the large frequency range, $-\infty<\omega<\infty$, one can further expand the integrand by taking large $\omega$ approximation as:
\be {\cal E}(\omega_0,\omega,k):=~\frac{\omega^2}{\left(1-e^{-2\pi k\omega}\right)\left(\omega+\omega_0\right)\left(\omega-\omega_0\right)}\xrightarrow[\omega\to \infty]{}\frac{\omega^2}{\left(\omega+\omega_0\right)\left(\omega-\omega_0\right)}:={\cal E}(\omega_0,\omega)\ee
This implies that, after taking large $\omega$ approximation the integrand of $\Delta_2$ becomes independent of the parameter $k$, which representing the surface gravity. 

Now, further using this approximation the integral $\Delta_2$ can be further simplified as:
\bea \Delta_{2}&&\approx \int^{\infty}_{-\infty}d\omega~{\cal E}(\omega_0,\omega)=\underbrace{\int^{0}_{-\infty}d\omega~{\cal E}(\omega_0,\omega)}_{\textcolor{red}{\equiv~ {\cal W}_1(\omega_0)}}~+~\underbrace{\int^{\infty}_{0}d\omega~{\cal E}(\omega_0,\omega)}_{\textcolor{red}{\equiv~ {\cal W}_2(\omega_0)}} \eea
where we have decomposed the integrals into two parts, indicated by 
$\textcolor{red}{{\cal W}_1(\omega_0)}$ and $\textcolor{red}{{\cal W}_2(\omega_0)}$ in the parenthesis symbol. Now, here we see that in the large frequency range, $-\infty<\omega<\infty$, we get:
\bea {\textcolor{red}{{\cal W}_1(\omega_0)}=\int^{0}_{-\infty}d\omega~{\cal E}(\omega_0,\omega)=-\int^{\infty}_{0}d\omega~{\cal E}(\omega_0,\omega)=-\textcolor{red}{{\cal W}_2(\omega_0)}} \eea
Now, we see here that both $\textcolor{red}{{\cal W}_1(\omega_0)}$ and $\textcolor{red}{{\cal W}_2(\omega_0)}$ gives divergent contributions in the frequency range, $-\infty<\omega<0$ and $0<\omega<\infty$. To get the finite contributions from these integrals we introduce a cut-off regulator $\omega_c$, by following {\it Bethe regularisation} technique. After introducing this cut-off we get:
\be
\begin{split}
 {\cal W}_1(\omega_0,\omega_c)&=\int^{0}_{-\omega_c}d\omega~{\cal E}(\omega_0,\omega)=\int^{\omega_c}_{0}d\omega~{\cal E}(\omega_0,\omega) \\
 &={\cal W}_2 (\omega_0,\omega_c) =\omega_c-\omega_0\tanh^{-1}\left(\frac{\omega_c}{\omega_0}\right)
\end{split}
\ee
Consequently, we get the following expression for the integral $\Delta_2$, as given by:
\bea {\textcolor{red}{~~~\Delta_{2}=\textcolor{red}{{\cal W}_1(\omega_0,\omega_c)}+\textcolor{red}{{\cal W}_2(\omega_0,\omega_c)}=2\left\{\omega_c-\omega_0\tanh^{-1}\left(\frac{\omega_c}{\omega_0}\right)\right\}}}\eea
Now, if we further use the approximation that, $\omega_c<<\omega_0$ i.e. the {\it Bethe regularised} cut-off is smaller than the natural frequency of the two entangled atomic system under consideration, then we get~\footnote{In the limit, $\omega_c<<\omega_0$ we can approximate the Taylor series expansion of the following function as:
	\be \tanh^{-1}\left(\frac{\omega_c}{\omega_0}\right)=\left(\frac{\omega_c}{\omega_0}\right)+\frac{1}{3}\left(\frac{\omega_c}{\omega_0}\right)^3+\cdots\approx \left(\frac{\omega_c}{\omega_0}\right).\ee}:

\be {\textcolor{red}{\Delta_{2}={\cal W}_1(\omega_0,\omega_c)+{\cal W}_2(\omega_0,\omega_c) =2\left\{\omega_c-\omega_0\left(\frac{\omega_c}{\omega_0}\right)\right\}=0}}
\ee
\subsection{\textcolor{blue}{\bf \large Spectroscopic Integral~III}}
In this subsection we explicitly compute the finite contribution from the following integral:
\bea {\textcolor{red}{\Delta_{3}:=\int^{\infty}_{-\infty}d\omega~\frac{2\omega^2~f(\omega,L/2)}{\left(1-e^{-2\pi k\omega}\right)\left(\omega+\omega_0\right)\left(\omega-\omega_0\right)}}}\eea
where, we define the spectral function $f(\omega,L/2)$ as:
\bea  f(\omega,L/2)=\frac{1}{L\omega\sqrt{1+\left(\frac{L}{2k}\right)^2}}\sin(2k\omega \sinh^{-1}\left(\frac{L}{2k}\right))\eea
It is important to note that in the large frequency range, $-\infty<\omega<\infty$, one can further expand the integrand by taking large $\omega$ approximation as:
\be {\cal O}(\omega_0,\omega,k):=~\frac{2\omega^2~f(\omega,L/2)}{\left(1-e^{-2\pi k\omega}\right)\left(\omega+\omega_0\right)\left(\omega-\omega_0\right)}\xrightarrow[\omega\to \infty]{}\frac{2\omega^2~f(\omega,L/2)}{\left(\omega+\omega_0\right)\left(\omega-\omega_0\right)}:=\widetilde{{\cal O}(\omega_0,\omega,k)}\ee
This implies that, after taking large $\omega$ approximation the integrand of $\Delta_3$ becomes not independent of the parameter $k$, which representing the surface gravity. 

Now, further using this approximation the integral $\Delta_3$ can be further simplified as:
\bea \Delta_{3}&&\approx \int^{\infty}_{-\infty}d\omega~\widetilde{{\cal O}(\omega_0,\omega,k)}=\underbrace{\int^{0}_{-\infty}d\omega~\widetilde{{\cal O}(\omega_0,\omega,k)}}_{\textcolor{red}{\equiv~ {\cal Q}_1(\omega_0,k)}}~+~\underbrace{\int^{\infty}_{0}d\omega~\widetilde{{\cal O}(\omega_0,\omega,k)}}_{\textcolor{red}{\equiv~ {\cal Q}_2(\omega_0,k)}}\eea
where we have decomposed the integrals into two parts, indicated by 
$\textcolor{red}{{\cal Q}_1(\omega_0,k)}$ and $\textcolor{red}{{\cal Q}_2(\omega_0,k)}$ in the parenthesis symbol. Now, here we see that in the large frequency range, $-\infty<\omega<\infty$, we get:
\be
\begin{split}
 {\cal Q}_1(\omega_0,k)&=\int^{0}_{-\infty}d\omega~\widetilde{{\cal O}(\omega_0,\omega,k)}=\int^{\infty}_{0}d\omega~\widetilde{{\cal O}(\omega_0,\omega,k)} \\
 &={\cal Q}_2(\omega_0,k)=\frac{\pi}{L \sqrt{1+\left(\frac{L}{2k}\right)^2}} \cos(2k\omega_0 \sinh^{-1}\left(\frac{L}{2k}\right))
\end{split}
\ee
Consequently, we get the following expression for the integral $\Delta_3$, as given by:
\bea {\textcolor{red}{\Delta_{3}=\textcolor{red}{{\cal Q}_1(\omega_0,k)}+\textcolor{red}{{\cal Q}_2(\omega_0,k)}=\frac{2\pi}{L\sqrt{1+\left(\frac{L}{2k}\right)^2}}\cos(2k\omega_0 \sinh^{-1}\left(\frac{L}{2k}\right))}}~.\eea

\section{\textcolor{blue}{\bf \large Calculation of Lamb Shift spectroscopy from two entangled atomic OQS}}
\label{LScalc}
In this appendix, we explicitly compute the expectation value of Lamb Shift Hamiltonian which will contribute to the atomic spectroscopy.
To serve this purpose let us first express the Lamb Shift Hamiltonian in terms of the Pauli operators (defined earlier):
\be
H_{\bf Lamb~shift}\equiv H_{LS} 
=\sum^{2}_{\alpha,\beta=1}H^{(\alpha\beta)}_{LS}=H^{(11)}_{LS}+H^{(22)}_{LS}+H^{(12)}_{LS}+H^{(21)}_{LS}
\ee
where the Lamb Shift Hamiltonian corresponding to all possible allowed interaction between two atoms are represented by the following expressions:
\begin{eqnarray*} H^{(11)}_{LS}&=-\frac{i}{2}[H^{(11)}_{11}\sigma^{1}_{1}\sigma^{1}_{1}\cos^2(\alpha^{1})+H^{(11)}_{12}\sigma^{1}_{1}\sigma^{1}_{2}\cos(\alpha^{1})\cos(\beta^{1})+H^{(11)}_{13}\sigma^{1}_{1}\sigma^{1}_{3}\cos(\alpha^{1})\cos(\gamma^{1}) \\
&~~~~~+H^{(11)}_{21}\sigma^{1}_{2}\sigma^{1}_{1}\cos(\beta^{1})\cos(\alpha^{1})+H^{(11)}_{22}\sigma^{1}_{2}\sigma^{1}_{2}\cos^2(\beta^{1})+H^{(11)}_{23}\sigma^{1}_{2}\sigma^{1}_{3}\cos(\beta^{1})\cos(\gamma^{1}) \\
&~~~~~+H^{(11)}_{31}\sigma^{1}_{3}\sigma^{1}_{1}\cos(\gamma^{1})\cos(\alpha^{1})+H^{(11)}_{32}\sigma^{1}_{3}\sigma^{1}_{2}\cos(\gamma^{1})\cos(\beta^{1})+H^{(11)}_{33}\sigma^{1}_{3}\sigma^{1}_{3}\cos^2(\gamma^{1})] \\ H^{(22)}_{LS}&=-\frac{i}{2}[H^{(22)}_{11}\sigma^{2}_{1}\sigma^{2}_{1}\cos^2(\alpha^{2})+H^{(22)}_{12}\sigma^{2}_{1}\sigma^{2}_{2}\cos(\alpha^{2})\cos(\beta^{2})+H^{(22)}_{13}\sigma^{2}_{1}\sigma^{2}_{3}\cos(\alpha^{2})\cos(\gamma^{2})\\
&~~~~~+H^{(22)}_{21}\sigma^{2}_{2}\sigma^{2}_{1}\cos(\beta^{2})\cos(\alpha^{2})+H^{(22)}_{22}\sigma^{2}_{2}\sigma^{2}_{2}\cos^2(\beta^{2})+H^{(22)}_{23}\sigma^{2}_{2}\sigma^{2}_{3}\cos(\beta^{2})\cos(\gamma^{2}) \\
&~~~~~+H^{(22)}_{31}\sigma^{2}_{3}\sigma^{2}_{1}\cos(\gamma^{2})\cos(\alpha^{2})+H^{(22)}_{32}\sigma^{2}_{3}\sigma^{2}_{2}\cos(\gamma^{2})\cos(\beta^{2})+H^{(22)}_{33}\sigma^{2}_{3}\sigma^{2}_{3}\cos^2(\gamma^{2})] 
\end{eqnarray*}

\begin{eqnarray*}
H^{(12)}_{LS}&=-\frac{i}{2}[H^{(12)}_{11}\sigma^{1}_{1}\sigma^{2}_{1}\cos(\alpha^{1})\cos(\alpha^{2})+H^{(12)}_{12}\sigma^{1}_{1}\sigma^{2}_{2}\cos(\alpha^{1})\cos(\beta^{2})+H^{(12)}_{13}\sigma^{1}_{1}\sigma^{2}_{3}\cos(\alpha^{1})\cos(\gamma^{2}) \\
&~~~~~~+H^{(12)}_{21}\sigma^{1}_{2}\sigma^{2}_{1}\cos(\beta^{1})\cos(\alpha^{2})+H^{(12)}_{22}\sigma^{1}_{2}\sigma^{2}_{2}\cos(\beta^{1})\cos(\beta^{2})+H^{(12)}_{23}\sigma^{1}_{2}\sigma^{2}_{3}\cos(\beta^{1})\cos(\gamma^{2}) \\
&~~~~~~+H^{(12)}_{31}\sigma^{1}_{3}\sigma^{2}_{1}\cos(\gamma^{1})\cos(\alpha^{2})+H^{(12)}_{32}\sigma^{1}_{3}\sigma^{2}_{2}\cos(\gamma^{1})\cos(\beta^{2})+H^{(12)}_{33}\sigma^{1}_{3}\sigma^{2}_{3}\cos(\gamma^{1})\cos(\gamma^{2})] \\ H^{(21)}_{LS}&=-\frac{i}{2}[H^{(21)}_{11}\sigma^{2}_{1}\sigma^{1}_{1}\cos(\alpha^{2})\cos(\alpha^{1})+H^{(21)}_{12}\sigma^{2}_{1}\sigma^{1}_{2}\cos(\alpha^{2})\cos(\beta^{1})+H^{(21)}_{13}\sigma^{2}_{1}\sigma^{1}_{3}\cos(\alpha^{2})\cos(\gamma^{1}) \\
&~~~~~~+H^{(21)}_{21}\sigma^{2}_{2}\sigma^{1}_{1}\cos(\beta^{2})\cos(\alpha^{1})+H^{(21)}_{22}\sigma^{2}_{2}\sigma^{1}_{2}\cos(\beta^{2})\cos(\beta^{1})+H^{(21)}_{23}\sigma^{2}_{2}\sigma^{1}_{3}\cos(\beta^{2})\cos(\gamma^{1}) \\
&~~~~~~+H^{(21)}_{31}\sigma^{2}_{3}\sigma^{1}_{1}\cos(\gamma^{2})\cos(\alpha^{1})+H^{(21)}_{32}\sigma^{2}_{3}\sigma^{1}_{2}\cos(\gamma^{2})\cos(\beta^{1})+H^{(21)}_{33}\sigma^{2}_{3}\sigma^{1}_{3}\cos(\gamma^{2})\cos(\gamma^{1})]
\end{eqnarray*}

All the entangled states are constructed out of all possible quantum states of two atoms $|g_1\rangle$, $|g_2\rangle$, $|e_1\rangle$ and $|e_2\rangle$ in section \ref{lss}. Here we derive the explicit contributions of the ground state, excited state, symmetric and antisymmetric state to the expectation value of Lamb shift Hamiltonian.

\subsection{\textcolor{blue}{\bf \large For Ground state}}
For the ground state of two entangled atoms ($|G\rangle $) the expectation values of all the possible Pauli tensor operators, which are explicitly contributing in the part of the Lamb Shift Hamiltonian ($H^{(11)}_{LS}$) are given by:
\begin{equation*}
\begin{split}
&\langle G | \sigma_1^1 \sigma_1^1 | G \rangle = 1 ~~~~~~~~~~~~~~ \langle G | \sigma_1^1 \sigma_2^1 | G \rangle = -i \cos(\gam^{1}) ~~~~~~~~ \langle G | \sigma_1^1 \sigma_3^1 | G \rangle = i \cos (\bg^{1}) \\
&\langle G | \sigma_2^1 \sigma_1^1 | G \rangle = i\cos(\gam^{1}) ~~~~~~~ \langle G | \sigma_2^1 \sigma_2^1 | G \rangle =1 ~~~~~~~~~~~~~~~~~~~~ \langle G | \sigma_2^1 \sigma_3^1 | G \rangle = -i \cos(\ag^{1}) \\
&\langle G | \sigma_3^1 \sigma_1^1 | G \rangle = -i \cos (\bg^{1}) ~~~~~~~~ \langle G | \sigma_3^1 \sigma_2^1 | G \rangle = i\cos(\ag^{1}) ~~~~~~~~~ \langle G | \sigma_3^1 \sigma_3^1 | G \rangle = 1
\end{split}
\end{equation*}
Consequently, the expectation value of the part of the Lamb Shift Hamiltonian ($H^{(11)}_{LS}$) with respect to the ground state can be written as:
\be
\begin{aligned}
\delta E^{(11)}_{G} &=\langle G| H^{(11)}_{LS} | G\rangle \\
&= -\frac{i}{2}[H^{(11)}_{11}\langle G| \sigma^{1}_{1}\sigma^{1}_{1}| G\rangle\cos^2(\alpha^{1})+H^{(11)}_{12}\langle G| \sigma^{1}_{1}\sigma^{1}_{2}| G\rangle\cos(\alpha^{1})\cos(\beta^{1})\nonumber\\
&\nonumber ~~~~~~~~~~~~~~~+H^{(11)}_{13}\langle G| \sigma^{1}_{1}\sigma^{1}_{3}| G\rangle\cos(\alpha^{1})\cos(\gamma^{1})+H^{(11)}_{21}\langle G| \sigma^{1}_{2}\sigma^{1}_{1}| G\rangle\cos(\beta^{1})\cos(\alpha^{1})\\
& \nonumber~~~~~~~~~~~~~~~+H^{(11)}_{22}\langle G| \sigma^{1}_{2}\sigma^{1}_{2}| G\rangle\cos^2(\beta^{1})+H^{(11)}_{23}\langle G| \sigma^{1}_{2}\sigma^{1}_{3}| G\rangle\cos(\beta^{1})\cos(\gamma^{1})\\
& \nonumber~~~~~~~~~~~~~~~+H^{(11)}_{31}\langle G| \sigma^{1}_{3}\sigma^{1}_{1}| G\rangle\cos(\gamma^{1})\cos(\alpha^{1})+H^{(11)}_{32}\langle G| \sigma^{1}_{3}\sigma^{1}_{2}| G\rangle\cos(\gamma^{1})\cos(\beta^{1})\\
& ~~~~~~~~~~~~~~~+H^{(11)}_{33}\langle G| \sigma^{1}_{3}\sigma^{1}_{3}| G\rangle\cos^2(\gamma^{1})]\nonumber\\
&=-\frac{i}{2}[H_{11}^{11}\cos^{2}(\ag^{1})+H_{22}^{11}\cos^{2}(\bg^{1})+H_{33}^{11}\cos^{2}(\gamma^{1})\nonumber\\
&~~~~~~~~~ + i (- H_{12}^{11} + H_{13}^{11} + H_{21}^{11} - H_{23}^{11} - H_{31}^{11} + H_{32}^{11}) \cos(\ag^{1})\cos(\bg^{1})\cos(\gamma^{1})]
\end{aligned}
\ee
For the ground state of two entangled atoms ($|G\rangle $) the expectation values of all the possible Pauli tensor operators, which are explicitly contributing in the part of the Lamb Shift Hamiltonian ($H^{(12)}_{LS}$) are given by:

\begin{equation*}
\begin{split}
\langle G | \sigma_1^1 \sigma_1^2 | G \rangle = \cos(\ag^{1})\cos(\ag^{2}) ~~ \langle G | \sigma_1^1 \sigma_2^2 | G \rangle = -i\cos(\ag^{1})\cos(\bg^{2}) ~~ \langle G | \sigma_1^1 \sigma_3^2 | G \rangle = \cos(\ag^{1})\cos(\gam^{2}) \\
\langle G | \sigma_2^1 \sigma_1^2 | G \rangle = -i \cos(\bg^{1}) \cos(\ag^{2}) ~~ \langle G | \sigma_2^1 \sigma_2^2 | G \rangle = \cos(\bg^{1})\cos(\bg^{2}) ~~ \langle G | \sigma_2^1 \sigma_3^2 | G \rangle = \cos(\bg^{1})\cos(\gam^{2}) \\
\langle G | \sigma_3^1 \sigma_1^2 | G \rangle = \cos(\gam^{1})\cos(\ag^{2}) ~~\langle G | \sigma_3^1 \sigma_2^2 | G \rangle = \cos(\gam^{1})\cos(\bg^{2}) ~~ \langle G | \sigma_3^1 \sigma_3^2 | G \rangle = \cos(\gam^{1})\cos(\gam^{2})
\end{split}
\end{equation*}

Consequently, the expectation value of the part of the Lamb Shift Hamiltonian ($H^{(12)}_{LS}$) with respect to the ground state can be written as:
\be
\begin{aligned}
 \delta E^{(12)}_{G} &=\langle G | H^{(12)}_{LS}| G \rangle \\
 &=-\frac{i}{2}[H^{(12)}_{11}\langle G | \sigma^{1}_{1}\sigma^{2}_{1}| G \rangle\cos(\alpha^{1})\cos(\alpha^{2})+H^{(12)}_{12}\langle G | \sigma^{1}_{1}\sigma^{2}_{2}| G \rangle\cos(\alpha^{1})\cos(\beta^{2}) \\
&~~~~~~~~~~~~~+H^{(12)}_{13}\langle G | \sigma^{1}_{1}\sigma^{2}_{3}| G \rangle\cos(\alpha^{1})\cos(\gamma^{2})+H^{(12)}_{21}\langle G | \sigma^{1}_{2}\sigma^{2}_{1}| G \rangle\cos(\beta^{1})\cos(\alpha^{2})\\
& ~~~~~~~~~~~~~ +H^{(12)}_{22}\langle G | \sigma^{1}_{2}\sigma^{2}_{2}| G \rangle\cos(\beta^{1})\cos(\beta^{2})+H^{(12)}_{23}\langle G | \sigma^{1}_{2}\sigma^{2}_{3}| G \rangle\cos(\beta^{1})\cos(\gamma^{2})\\
& ~~~~~~~~~~~~~+H^{(12)}_{31}\langle G | \sigma^{1}_{3}\sigma^{2}_{1}| G \rangle\cos(\gamma^{1})\cos(\alpha^{2})+H^{(12)}_{32}\langle G | \sigma^{1}_{3}\sigma^{2}_{2}| G \rangle\cos(\gamma^{1})\cos(\beta^{2})\\
& ~~~~~~~~~~~~~+H^{(12)}_{33}\langle G | \sigma^{1}_{3}\sigma^{2}_{3}| G \rangle\cos(\gamma^{1})\cos(\gamma^{2})]\nonumber\\
&=-\frac{i}{2}[H_{11}^{12}\cos^{2}(\ag^{1})\cos^{2}(\ag^{2}) +  H_{12}^{12} (-i) \cos^{2}(\ag^{1}) \cos^{2}(\bg^{2}) + H_{13}^{12}\cos^{2}(\ag^{1})\cos^{2}(\gamma^{2}) \\
& ~~~~~~~~ +  H_{21}^{12} (-i) \cos^{2}(\bg^{1})\cos^{2}(\ag^{2}) + H_{22}^{12}\cos^{2}(\bg^{1})\cos^{2}(\bg^{2}) + H_{23}^{12}\cos^{2}(\bg^{1})\cos^{2}(\gamma^{2})\nonumber \\
& ~~~~~~~~~~ + H_{31}^{12}\cos^{2}(\gamma^{1})\cos^{2}(\ag^{2}) + H_{32}^{12}\cos^{2}(\gamma^{1})\cos^{2}(\bg^{2}) + H_{33}^{12}\cos^{2}(\gamma^{1})\cos^{2}(\gamma^{2})]
\end{aligned}
\ee
For the ground state of two entangled atoms ($|G\rangle $) the expectation values of all the possible Pauli tensor operators, which are explicitly contributing in the part of the Lamb Shift Hamiltonian ($H^{(21)}_{LS}$) are given by:
\begin{equation*}
\begin{split}
&\langle G | \sigma_1^2 \sigma_1^1 | G \rangle = \cos\ag^{1}\cos\ag^{2} ~~~~\langle G | \sigma_1^2 \sigma_2^1 | G \rangle = -i \cos\bg^{1}\cos\ag^{2}~~~~  \langle G | \sigma_1^2 \sigma_3^1 | G \rangle =  \cos\gam^{1}\cos\ag^{2} \\
&\langle G | \sigma_2^2 \sigma_1^1 | G \rangle =  -i\cos\ag^{1}\cos\bg^{2}~~~~  \langle G | \sigma_2^2 \sigma_2^1 | G \rangle =  \cos\bg^{1}\cos\bg^{2}~~~~  \langle G | \sigma_2^2 \sigma_3^1 | G \rangle =  \cos\gam^{1}\cos\bg^{2} \\
&\langle G | \sigma_3^2 \sigma_1^1 | G \rangle =  \cos\ag^{1}\cos\gam^{2}~~~~ \langle G | \sigma_3^2 \sigma_2^1 | G \rangle =  \cos\gam^{2}\cos\bg^{1}~~~~ \langle G | \sigma_3^2 \sigma_3^1 | G \rangle =  \cos\gam^{1}\cos\gam^{2} 
\end{split}
\end{equation*}
Consequently, the expectation value of the part of the Lamb Shift Hamiltonian ($H^{(21)}_{LS}$) with respect to the ground state can be written as:
\be
\begin{aligned} 
\delta E^{(21)}_{G} &=\langle G |H^{(21)}_{LS}| G \rangle \\ 
&=-\frac{i}{2}[H^{(21)}_{11}\langle G |\sigma^{2}_{1}\sigma^{1}_{1}| G \rangle\cos(\alpha^{2})\cos(\alpha^{1})+H^{(21)}_{12}\langle G |\sigma^{2}_{1}\sigma^{1}_{2}| G \rangle\cos(\alpha^{2})\cos(\beta^{1}) \\
& ~~~~~~~~~~~+H^{(21)}_{13}\langle G |\sigma^{2}_{1}\sigma^{1}_{3}| G \rangle\cos(\alpha^{2})\cos(\gamma^{1})+H^{(21)}_{21}\langle G |\sigma^{2}_{2}\sigma^{1}_{1}| G \rangle\cos(\beta^{2})\cos(\alpha^{1})\\
& ~~~~~~~~~~~+H^{(21)}_{22}\langle G |\sigma^{2}_{2}\sigma^{1}_{2}| G \rangle\cos(\beta^{2})\cos(\beta^{1})+H^{(21)}_{23}\langle G |\sigma^{2}_{2}\sigma^{1}_{3}| G \rangle\cos(\beta^{2})\cos(\gamma^{1})\\
& ~~~~~~~~~~~+H^{(21)}_{31}\langle G |\sigma^{2}_{3}\sigma^{1}_{1}| G \rangle\cos(\gamma^{2})\cos(\alpha^{1})+H^{(21)}_{32}\langle G |\sigma^{2}_{3}\sigma^{1}_{2}| G \rangle\cos(\gamma^{2})\cos(\beta^{1})\\
& ~~~~~~~~~~~+H^{(21)}_{33}\langle G |\sigma^{2}_{3}\sigma^{1}_{3}| G \rangle\cos(\gamma^{2})\cos(\gamma^{1})]\nonumber \\
&=-\frac{i}{2}[H_{11}^{21}\cos^{2}(\ag^{2})\cos^{2}(\ag^{1}) +  H_{12}^{21} (-i) \cos^{2}(\ag^{2}) \cos^{2}(\bg^{1}) + H_{13}^{21}\cos^{2}(\ag^{2})\cos^{2}(\gamma^{1}) \\
& ~~~~~~~~~~~~~ +  H_{21}^{21} (-i) \cos^{2}(\bg^{2})\cos^{2}(\ag^{1}) + H_{22}^{21}\cos^{2}(\bg^{2})\cos^{2}(\bg^{1} )+ H_{23}^{21}\cos^{2}(\bg^{2})\cos^{2}(\gamma^{1})\nonumber \\
&~~~~~~~~~~~~~~ + H_{31}^{21}\cos^{2}(\gamma^{2})\cos^{2}(\ag^{1}) + H_{32}^{21}\cos^{2}(\gamma^{2})\cos^{2}(\bg^{1}) + H_{33}^{21}\cos^{2}(\gamma^{2})\cos^{2}(\gamma^{1})]
\end{aligned}
\ee
For the ground state of two entangled atoms ($|G\rangle $) the expectation values of all the possible Pauli tensor operators, which are explicitly contributing in the part of the Lamb Shift Hamiltonian ($H^{(22)}_{LS}$) are given by:
\begin{equation*}
\begin{split}
&\langle G | \sigma_1^2 \sigma_1^2 | G \rangle = 1 ~~~~~~~~ \langle G | \sigma_1^2 \sigma_2^2 | G \rangle = -i\cos\gam^{2}~~~ \langle G | \sigma_1^2 \sigma_3^2 | G \rangle = i\cos\bg^{2} \\
&\langle G | \sigma_2^2 \sigma_1^2 | G \rangle = i\cos\gam^{2}~~~ \langle G | \sigma_2^2 \sigma_2^2 | G \rangle = 1 ~~~~~~~~~~~~~~ \langle G | \sigma_2^2 \sigma_3^2 | G \rangle = -i\cos\ag^{2} \\
&\langle G | \sigma_3^2 \sigma_1^2 | G \rangle = -i\cos\bg^{2} ~~~ \langle G | \sigma_3^2 \sigma_2^2 | G \rangle = i\cos\ag^{2} ~~~~~~~ \langle G | \sigma_3^2 \sigma_3^2 | G \rangle = 1 
\end{split}
\end{equation*}
Consequently, the expectation value of the part of the Lamb Shift Hamiltonian ($H^{(22}_{LS}$) with respect to the ground state can be written as:
\be
\begin{aligned}
\delta E^{(22)}_{G} &=\langle G| H^{(22)}_{LS}| G\rangle \\
&=-\frac{i}{2}[H^{(22)}_{11}\langle G|\sigma^{2}_{1}\sigma^{2}_{1}| G\rangle\cos^2(\alpha^{2})+H^{(22)}_{12}\langle G|\sigma^{2}_{1}\sigma^{2}_{2}| G\rangle\cos(\alpha^{2})\cos(\beta^{2}) \\
& ~~~~~~~~~~~~+H^{(22)}_{13}\langle G|\sigma^{2}_{1}\sigma^{2}_{3}| G\rangle\cos(\alpha^{2})\cos(\gamma^{2})+H^{(22)}_{21}\langle G|\sigma^{2}_{2}\sigma^{2}_{1}| G\rangle\cos(\beta^{2})\cos(\alpha^{2}) \\
& ~~~~~~~~~~~~~+H^{(22)}_{22}\langle G|\sigma^{2}_{2}\sigma^{2}_{2}| G\rangle\cos^2(\beta^{2})+H^{(22)}_{23}\langle G|\sigma^{2}_{2}\sigma^{2}_{3}| G\rangle\cos(\beta^{2})\cos(\gamma^{2}) \\
& ~~~~~~~~~~~~~~~+H^{(22)}_{31}\langle G|\sigma^{2}_{3}\sigma^{2}_{1}| G\rangle\cos(\gamma^{2})\cos(\alpha^{2})+H^{(22)}_{32}\langle G|\sigma^{2}_{3}\sigma^{2}_{2}| G\rangle\cos(\gamma^{2})\cos(\beta^{2}) \\
& ~~~~~~~~~~~~~~~~~~+H^{(22)}_{33}\langle G|\sigma^{2}_{3}\sigma^{2}_{3}| G\rangle\cos^2(\gamma^{2})] \\
&=-\frac{i}{2}[H_{11}^{22}cos^{2}\ag^{2} + H_{22}^{22}cos^{2}\bg^{2} + H_{33}^{22}cos^{2}\gamma^{2}\nonumber\\
&~~~~~~~~~ + i (- H_{12}^{22} + H_{13}^{22} + H_{21}^{22} - H_{23}^{22} - H_{31}^{22} + H_{32}^{22}) \cos\ag^{2}\cos\bg^{2}\cos\gamma^{2}]
\end{aligned}
\ee
After that, summing over all the possible contributions obtained for the ground state of two entangled atoms ($|G\rangle $) the expectation value of the Lamd Shift Hamiltonian can be expressed as:
\be
\begin{aligned}
\delta E_G &= \sum^{2}_{i=1}\sum^{2}_{j=1}\delta E^{(ij)}_G=\sum^{2}_{i=1}\sum^{2}_{j=1} \langle G| H^{(ij)}_{LS}|G \rangle\nonumber\\
&=-\frac{i}{2}[H_{11}^{11}cos^{2}\ag^{1}+H_{22}^{11}cos^{2}\bg^{1}+H_{33}^{11}cos^{2}\gamma^{1} \\
&~~~~~~~~~ + i (- H_{12}^{11} + H_{13}^{11} + H_{21}^{11} - H_{23}^{11} - H_{31}^{11} + H_{32}^{11}) \cos\ag^{1}\cos\bg^{1}\cos\gamma^{1}] \nonumber \\
&~~ -\frac{i}{2}[H_{11}^{12}\cos^{2}\ag^{1}\cos^{2}\ag^{2} +  H_{12}^{12} (-i) \cos^{2}\ag^{1} \cos^{2}\bg^{2} + H_{13}^{12}\cos^{2}\ag^{1}\cos^{2}\gamma^{2} \\
& ~~~~~~~~~~~~~ +  H_{21}^{12} (-i) \cos^{2}\bg^{1}\cos^{2}\ag^{2} + H_{22}^{12}\cos^{2}\bg^{1}\cos^{2}\bg^{2} + H_{23}^{12}\cos^{2}\bg^{1}\cos^{2}\gamma^{2}  \\
& ~~~~~~~~~~~~~~ + H_{31}^{12}\cos^{2}\gamma^{1}\cos^{2}\ag^{2} + H_{32}^{12}\cos^{2}\gamma^{1}\cos^{2}\bg^{2} + H_{33}^{12}\cos^{2}\gamma^{1}\cos^{2}\gamma^{2}]  \\
&~~ -\frac{i}{2}[H_{11}^{21}\cos^{2}\ag^{2}\cos^{2}\ag^{1} +  H_{12}^{21} (-i) \cos^{2}\ag^{2} \cos^{2}\bg^{1} + H_{13}^{21}\cos^{2}\ag^{2}\cos^{2}\gamma^{1} \\
& ~~~~~~~~~~~~ +  H_{21}^{21} (-i) \cos^{2}\bg^{2}\cos^{2}\ag^{1} + H_{22}^{21}\cos^{2}\bg^{2}\cos^{2}\bg^{1} + H_{23}^{21}\cos^{2}\bg^{2}\cos^{2}\gamma^{1} \\
& ~~~~~~~~~~~~ + H_{31}^{21}\cos^{2}\gamma^{2}\cos^{2}\ag^{1} + H_{32}^{21}\cos^{2}\gamma^{2}\cos^{2}\bg^{1} + H_{33}^{21}\cos^{2}\gamma^{2}\cos^{2}\gamma^{1}]  \\
&~~ -\frac{i}{2}[H_{11}^{22}\cos^{2}\ag^{2} + H_{22}^{22}\cos^{2}\bg^{2} + H_{33}^{22}\cos^{2}\gamma^{2} \\
&~~~~~~~~~~~ + i (- H_{12}^{22} + H_{13}^{22} + H_{21}^{22} - H_{23}^{22} - H_{31}^{22} + H_{32}^{22}) \cos\ag^{2}\cos\bg^{2}\cos\gamma^{2}]
\end{aligned}
\ee
Substituting the particular values of these Hamiltonian coefficients from equation \ref{hamco}
\begin{equation*}
\begin{aligned}
	\delta E_{G} = & -\frac{i}{2}[{\cal A}_1 cos^{2}\ag^{1} + {\cal A}_1 cos^{2}\bg^{1}  - 2  {\cal B}_1  \cos\ag^{1}\cos\bg^{1}\cos\gamma^{1}]  \\
	&-\frac{i}{2}[{\cal A}_2 \cos^{2}\ag^{1}\cos^{2}\ag^{2} - {\cal B}_2 \cos^{2}\ag^{1} \cos^{2}\bg^{2} + {\cal B}_2  \cos^{2}\bg^{1}\cos^{2}\ag^{2} + {\cal A}_2 \cos^{2}\bg^{1}\cos^{2}\bg^{2}]  \\
	&-\frac{i}{2}[{\cal A}_2 \cos^{2}\ag^{2}\cos^{2}\ag^{1} - {\cal B}_2 \cos^{2}\ag^{2} \cos^{2}\bg^{1}] + {\cal B}_2 \cos^{2}\bg^{2}\cos^{2}\ag^{1} + {\cal A}_2 \cos^{2}\bg^{2}\cos^{2}\bg^{1}]  \\
	&-\frac{i}{2}[{\cal A}_1 \cos^{2}\ag^{2} + {\cal A}_1 \cos^{2}\bg^{2} - 2 {\cal B}_1  \cos\ag^{2}\cos\bg^{2}\cos\gamma^{2}]
\end{aligned}
\end{equation*}

Using the equation \ref{Aint}

\begin{equation*}
\begin{aligned}
\delta E_{G}=-\frac{\mu^2 P}{8\pi^2} \int^{\infty}_{-\infty} d\omega~&\frac{\omega}{\left(1-e^{-2\pi k\omega}\right)\left(\omega+\omega_0\right)\left(\omega-\omega_0\right)} \\ &[-2\omega_0\left\{\cos(\alpha^{1})\cos(\beta^{1})\cos(\gamma^{1})+\cos(\alpha^{2})\cos(\beta^{2})\cos(\gamma^{2})\right\} \\
& +\omega\left\{\cos[2](\alpha^{1})+\cos[2](\beta^{1})+\cos[2](\alpha^{2})+\cos[2](\beta^{2})\right\} \\
& +2\omega f(\omega,L/2) \left\{\cos[2](\alpha^{1})\cos[2](\alpha^{2})+\cos[2](\beta^{1})\cos[2](\beta^{2})\right\}]
\end{aligned}
\end{equation*}

In abbreviated notation

\be
\begin{aligned}
\delta E_{G}= -\frac{\mu^2 P}{8\pi^2} ~[&-\left\{\cos(\alpha^{1})\cos(\beta^{1})\cos(\gamma^{1})+\cos(\alpha^{2})\cos(\beta^{2})\cos(\gamma^{2})\right\} \Delta_1  \\
& + \left\{\cos[2](\alpha^{1})+\cos[2](\beta^{1})+\cos[2](\alpha^{2})+\cos[2](\beta^{2})\right\}\Delta_2 \\
&+ \left\{\cos[2](\alpha^{1})\cos[2](\alpha^{2}) + \cos[2](\beta^{1})\cos[2](\beta^{2})\right\}\Delta_3]
\end{aligned}
\ee

\subsection{\textcolor{blue}{\bf \large For Excited state}}
For the excited state of two entangled atoms ($|E\rangle $) the expectation values of all the possible Pauli tensor operators, which are explicitly contributing in the part of the Lamb Shift Hamiltonian ($H^{(11)}_{LS}$) are given by:
\begin{equation*}
\begin{split}
&\langle E | \sigma_1^1 \sigma_1^1 | E \rangle = 1 ~~~~~~~~~~~ \langle G | \sigma_1^1 \sigma_2^1 | E \rangle = i\cos\gam^{1} ~~~~ \langle E | \sigma_1^1 \sigma_3^1 | E \rangle = -i\cos\bg^{1} \\
&\langle E | \sigma_2^1 \sigma_1^1 | E \rangle = -i\cos\gam^{1} ~~~ \langle G | \sigma_2^1 \sigma_2^1 | E \rangle = 1 ~~~~~~~~~~~ \langle E | \sigma_2^1 \sigma_3^1 | E \rangle = i\cos\ag^{1} \\
&\langle E | \sigma_3^1 \sigma_1^1 | E \rangle = i\cos\bg^{1} ~~~~~~~ \langle G | \sigma_3^1 \sigma_2^1 | E \rangle = -i\cos\ag^{1} ~~~ \langle E | \sigma_3^1 \sigma_3^1 | E \rangle = 1
\end{split}
\end{equation*}
Consequently, the expectation value of the part of the Lamb Shift Hamiltonian ($H^{(11)}_{LS}$) with respect to the excited state can be written as:
\be
\begin{aligned}
 \delta E^{(11)}_{E}&=\langle E| H^{(11)}_{LS} | E\rangle \\
 &=-\frac{i}{2}[H^{(11)}_{11}\langle E| \sigma^{1}_{1}\sigma^{1}_{1}| E\rangle\cos^2(\alpha^{1})+H^{(11)}_{12}\langle E| \sigma^{1}_{1}\sigma^{1}_{2}| E\rangle\cos(\alpha^{1})\cos(\beta^{1}) \\
& ~~~~~~~~~~~~~+H^{(11)}_{13}\langle E| \sigma^{1}_{1}\sigma^{1}_{3}| E\rangle\cos(\alpha^{1})\cos(\gamma^{1})+H^{(11)}_{21}\langle E| \sigma^{1}_{2}\sigma^{1}_{1}| E\rangle\cos(\beta^{1})\cos(\alpha^{1})\\
& ~~~~~~~~~~~~~+H^{(11)}_{22}\langle E| \sigma^{1}_{2}\sigma^{1}_{2}| E\rangle\cos^2(\beta^{1})+H^{(11)}_{23}\langle E| \sigma^{1}_{2}\sigma^{1}_{3}| E\rangle\cos(\beta^{1})\cos(\gamma^{1}) \\
& ~~~~~~~~~~~~~+H^{(11)}_{31}\langle E| \sigma^{1}_{3}\sigma^{1}_{1}| E\rangle\cos(\gamma^{1})\cos(\alpha^{1})+H^{(11)}_{32}\langle E| \sigma^{1}_{3}\sigma^{1}_{2}| E\rangle\cos(\gamma^{1})\cos(\beta^{1}) \\
& ~~~~~~~~~~~~~+H^{(11)}_{33}\langle E| \sigma^{1}_{3}\sigma^{1}_{3}| E\rangle\cos^2(\gamma^{1})] \\
&=-\frac{i}{2}[H^{(11)}_{11}\cos^2(\alpha^{1}) +  H^{(11)}_{22}\cos^2(\beta^{1}) + H^{(11)}_{33}\cos^2(\gamma^{1}) \\
& ~~~~~~~~~~~~ + i (H^{(11)}_{12} - H^{(11)}_{13} - H^{(11)}_{21} + H^{(11)}_{23} + H^{(11)}_{31} - H^{(11)}_{32} )\cos(\alpha^{1})\cos(\beta^{1}) \cos(\gamma^{1})]\nonumber\\
\end{aligned}
\ee
For the excited state of two entangled atoms ($|E\rangle $) the expectation values of all the possible Pauli tensor operators, which are explicitly contributing in the part of the Lamb Shift Hamiltonian ($H^{(12)}_{LS}$) are given by:
\begin{equation*}
\begin{split}
&\langle E | \sigma_1^1 \sigma_1^2 | E \rangle = \cos\ag^{1}\cos\ag^{2} \ \ \ \ \langle E | \sigma_1^1 \sigma_2^2 | E \rangle =  \cos\ag^{1}\cos\bg^{2} \ \ \ \ \langle E | \sigma_1^1 \sigma_3^2 | E \rangle =  \cos\ag^{1}\cos\gam^{2} \\
&\langle E | \sigma_2^1 \sigma_1^2 | E \rangle =  \cos\bg^{1}\cos\ag^{2} \ \ \ \ \langle E | \sigma_2^1 \sigma_2^2 | E \rangle =  \cos\bg^{1}\cos\bg^{2} \ \ \ \ \langle E | \sigma_2^1 \sigma_3^2 | E \rangle =  \cos\bg^{1}\cos\gam^{2} \\
&\langle E | \sigma_3^1 \sigma_1^2 | E \rangle =  \cos\gam^{1}\cos\ag^{2} \ \ \ \ \langle E | \sigma_3^1 \sigma_2^2 | E \rangle =  \cos\gam^{1}\cos\bg^{2} \ \ \ \ \langle E | \sigma_3^1 \sigma_3^2 | E \rangle =  \cos\gam^{1}\cos\gam^{2} 
\end{split}
\end{equation*}
Consequently, the expectation value of the part of the Lamb Shift Hamiltonian ($H^{(12)}_{LS}$) with respect to the excited state can be written as:
\be
\begin{aligned}
 \delta E^{(12)}_{E} &=\langle E | H^{(12)}_{LS}| E \rangle \\
 &=-\frac{i}{2}[H^{(12)}_{11}\langle E | \sigma^{1}_{1}\sigma^{2}_{1}| E \rangle\cos(\alpha^{1})\cos(\alpha^{2})+H^{(12)}_{12}\langle E | \sigma^{1}_{1}\sigma^{2}_{2}| E \rangle\cos(\alpha^{1})\cos(\beta^{2}) \\
& ~~~~~~~~~~+H^{(12)}_{13}\langle E | \sigma^{1}_{1}\sigma^{2}_{3}| E \rangle\cos(\alpha^{1})\cos(\gamma^{2})+H^{(12)}_{21}\langle E | \sigma^{1}_{2}\sigma^{2}_{1}| E \rangle\cos(\beta^{1})\cos(\alpha^{2})\\
& ~~~~~~~~~~ +H^{(12)}_{22}\langle E | \sigma^{1}_{2}\sigma^{2}_{2}| E \rangle\cos(\beta^{1})\cos(\beta^{2})+H^{(12)}_{23}\langle E | \sigma^{1}_{2}\sigma^{2}_{3}| E \rangle\cos(\beta^{1})\cos(\gamma^{2})\\
& ~~~~~~~~~~+H^{(12)}_{31}\langle E | \sigma^{1}_{3}\sigma^{2}_{1}| E \rangle\cos(\gamma^{1})\cos(\alpha^{2})+H^{(12)}_{32}\langle E | \sigma^{1}_{3}\sigma^{2}_{2}| E \rangle\cos(\gamma^{1})\cos(\beta^{2})\\
& ~~~~~~~~~~+H^{(12)}_{33}\langle E | \sigma^{1}_{3}\sigma^{2}_{3}| E \rangle\cos(\gamma^{1})\cos(\gamma^{2})]\nonumber\\
&=-\frac{i}{2}[H^{(12)}_{11} \cos^2(\alpha^{1})\cos^2(\alpha^{2}) + H^{(12)}_{12}\cos^2(\alpha^{1})\cos^2(\beta^{2}) + + H^{(12)}_{13} \cos^2(\alpha^{1})\cos^2(\gamma^{2}) \nonumber\\
& ~~~~~~~~~~~  + H^{(12)}_{21} \cos^2(\beta^{1})\cos^2(\alpha^{2}) + H^{(12)}_{22} \cos^2(\beta^{1})\cos^2(\beta^{2}) + H^{(12)}_{23} \cos^2(\beta^{1})\cos^2(\gamma^{2})\\
& ~~~~~~~~~~~ + H^{(12)}_{31} \cos^2(\gamma^{1})\cos^2(\alpha^{2}) + H^{(12)}_{32} \cos^2(\gamma^{1})\cos^2(\beta^{2}) + H^{(12)}_{33} \cos^2(\gamma^{1})\cos^2(\gamma^{2})]
\end{aligned}
\ee
For the excited state of two entangled atoms ($|E\rangle $) the expectation values of all the possible Pauli tensor operators, which are explicitly contributing in the part of the Lamb Shift Hamiltonian ($H^{(21)}_{LS}$) are given by:
\begin{equation*}
\begin{split}
&\langle E | \sigma_1^2 \sigma_1^1 | E \rangle = \cos\ag^{1}\cos\ag^{2} \ \ \ \ \langle E | \sigma_1^2 \sigma_2^1 | E \rangle =  \cos\bg^{1}\cos\ag^{2} \ \ \ \ \langle E | \sigma_1^2 \sigma_3^1 | E \rangle =  \cos\gam^{1}\cos\ag^{2} \\
&\langle E | \sigma_2^2 \sigma_1^1 | E \rangle =  \cos\ag^{1}\cos\bg^{2} \ \ \ \ \langle E | \sigma_2^2 \sigma_2^1 | E \rangle =  \cos\bg^{1}\cos\bg^{2} \ \ \ \ \langle E | \sigma_2^2 \sigma_3^1 | E \rangle = \cos\gam^{1}\cos\bg^{2} \\
&\langle E | \sigma_3^2 \sigma_1^1 | E \rangle =  \cos\ag^{1}\cos\gam^{2} \ \ \ \ \langle E | \sigma_3^2 \sigma_2^1 | E \rangle =  \cos\bg^{1}\cos\gam^{2} \ \ \ \ \langle E | \sigma_3^2 \sigma_3^1 | E \rangle =  \cos\gam^{1}\cos\gam^{2} 
\end{split}
\end{equation*}
Consequently, the expectation value of the part of the Lamb Shift Hamiltonian ($H^{(21)}_{LS}$) with respect to the excited state can be written as:
\be 
\begin{aligned}
\delta E^{(21)}_{E} &=\langle E |H^{(21)}_{LS}| E \rangle \\
&=-\frac{i}{2}[H^{(21)}_{11}\langle E |\sigma^{2}_{1}\sigma^{1}_{1}| E \rangle\cos(\alpha^{2})\cos(\alpha^{1})+H^{(21)}_{12}\langle E |\sigma^{2}_{1}\sigma^{1}_{2}| E \rangle\cos(\alpha^{2})\cos(\beta^{1}) \\
& ~~~~~~~~~~+H^{(21)}_{13}\langle E |\sigma^{2}_{1}\sigma^{1}_{3}| E \rangle\cos(\alpha^{2})\cos(\gamma^{1})+H^{(21)}_{21}\langle E |\sigma^{2}_{2}\sigma^{1}_{1}| E \rangle\cos(\beta^{2})\cos(\alpha^{1})\\
& ~~~~~~~~~~+H^{(21)}_{22}\langle E |\sigma^{2}_{2}\sigma^{1}_{2}| E \rangle\cos(\beta^{2})\cos(\beta^{1})+H^{(21)}_{23}\langle E |\sigma^{2}_{2}\sigma^{1}_{3}| E \rangle\cos(\beta^{2})\cos(\gamma^{1})\\
& ~~~~~~~~~~+H^{(21)}_{31}\langle E |\sigma^{2}_{3}\sigma^{1}_{1}| E \rangle\cos(\gamma^{2})\cos(\alpha^{1})+H^{(21)}_{32}\langle E |\sigma^{2}_{3}\sigma^{1}_{2}| E \rangle\cos(\gamma^{2})\cos(\beta^{1})\\
& ~~~~~~~~~~+H^{(21)}_{33}\langle E |\sigma^{2}_{3}\sigma^{1}_{3}| E \rangle\cos(\gamma^{2})\cos(\gamma^{1})]\nonumber \\
&=-\frac{i}{2}[H^{(21)}_{11} \cos^2(\alpha^{2})\cos^2(\alpha^{1}) + H^{(21)}_{12}\cos^2(\alpha^{2})\cos^2(\beta^{1}) + + H^{(21)}_{13} \cos^2(\alpha^{2})\cos^2(\gamma^{1}) \\
& ~~~~~~~~~~~  + H^{(21)}_{21} \cos^2(\beta^{2})\cos^2(\alpha^{1}) + H^{(21)}_{22} \cos^2(\beta^{2})\cos^2(\beta^{1}) + H^{(21)}_{23} \cos^2(\beta^{2})\cos^2(\gamma^{1})\\
& ~~~~~~~~~~~ + H^{(21)}_{31} \cos^2(\gamma^{2})\cos^2(\alpha^{1}) + H^{(21)}_{32} \cos^2(\gamma^{2})\cos^2(\beta^{1}) + H^{(21)}_{33} \cos^2(\gamma^{2})\cos^2(\gamma^{1})]
\end{aligned}
\ee
For the excited state of two entangled atoms ($|E\rangle $) the expectation values of all the possible Pauli tensor operators, which are explicitly contributing in the part of the Lamb Shift Hamiltonian ($H^{(22)}_{LS}$) are given by:
\begin{equation*}
\begin{split}
&\langle E | \sigma_1^2 \sigma_1^2 | E \rangle = 1 ~~~~~~~~~ \langle E | \sigma_1^2 \sigma_2^2 | E \rangle = i\cos\gam^{2}  ~~~ \langle E | \sigma_1^2 \sigma_3^2 | E \rangle = -i \cos\bg^{2} \\
&\langle E | \sigma_2^2 \sigma_1^2 | E \rangle = -i\cos\gam^{2}  ~~~ \langle E | \sigma_2^2 \sigma_2^2 | E \rangle = 1 ~~~~~~~~~~ \langle E | \sigma_2^2 \sigma_3^2 | E \rangle = i\cos\ag^{2} \\
&\langle E | \sigma_3^2 \sigma_1^2 | E \rangle = i\cos\beta^{2}  ~~~~~~~ \langle E | \sigma_3^2 \sigma_2^2 | E \rangle = -i \cos\ag^{2} ~~~ \langle E | \sigma_3^2 \sigma_3^2 | E \rangle = 1 
\end{split}
\end{equation*}
Consequently, the expectation value of the part of the Lamb Shift Hamiltonian ($H^{(22}_{LS}$) with respect to the excited state can be written as:
\begin{equation*}
\begin{aligned}
\delta E^{(22)}_{E} &=\langle E| H^{(22)}_{LS}| E\rangle \\
&=-\frac{i}{2}[H^{(22)}_{11}\langle E|\sigma^{2}_{1}\sigma^{2}_{1}| E\rangle\cos^2(\alpha^{2})+H^{(22)}_{12}\langle E|\sigma^{2}_{1}\sigma^{2}_{2}| E\rangle\cos(\alpha^{2})\cos(\beta^{2}) \\
& ~~~~~~~~~~~~+H^{(22)}_{13}\langle E|\sigma^{2}_{1}\sigma^{2}_{3}| E\rangle\cos(\alpha^{2})\cos(\gamma^{2})+H^{(22)}_{21}\langle E|\sigma^{2}_{2}\sigma^{2}_{1}| G\rangle\cos(\beta^{2})\cos(\alpha^{2})\\
& ~~~~~~~~~~~~+H^{(22)}_{22}\langle E|\sigma^{2}_{2}\sigma^{2}_{2}| E\rangle\cos^2(\beta^{2})+H^{(22)}_{23}\langle E|\sigma^{2}_{2}\sigma^{2}_{3}| E\rangle\cos(\beta^{2})\cos(\gamma^{2})\\
& ~~~~~~~~~~~~+H^{(22)}_{31}\langle E|\sigma^{2}_{3}\sigma^{2}_{1}| E\rangle\cos(\gamma^{2})\cos(\alpha^{2})+H^{(22)}_{32}\langle E|\sigma^{2}_{3}\sigma^{2}_{2}| E\rangle\cos(\gamma^{2})\cos(\beta^{2})\\
& ~~~~~~~~~~~~+H^{(22)}_{33}\langle E|\sigma^{2}_{3}\sigma^{2}_{3}| E\rangle\cos^2(\gamma^{2})] \\
&=-\frac{i}{2}[H^{(22)}_{11}\cos^2(\alpha^{2}) +  H^{(22)}_{22}\cos^2(\beta^{2}) + H^{(22)}_{33}\cos^2(\gamma^{2}) \\
& ~~~~~~~~~~~ + i (H^{(22)}_{12} - H^{(22)}_{13} - H^{(22)}_{21} + H^{(22)}_{23} + H^{(22)}_{31} - H^{(22)}_{32} )\cos(\alpha^{2})\cos(\beta^{2}) \cos(\gamma^{2})]
\end{aligned}
\end{equation*}
After that, summing over all the possible contributions obtained for the excited state of two entangled atoms ($|E\rangle $) the expectation value of the Lamd Shift Hamiltonian can be expressed as:

\begin{equation*}
\begin{aligned}
\delta E_E &= \sum^{2}_{i=1}\sum^{2}_{j=1}\delta E^{(ij)}_E=\sum^{2}_{i=1}\sum^{2}_{j=1} \langle E| H^{(ij)}_{LS}|E \rangle \\
&= -\frac{i}{2}[H^{(11)}_{11}\cos^2(\alpha^{1}) +  H^{(11)}_{22}\cos^2(\beta^{1}) + H^{(11)}_{33}\cos^2(\gamma^{1})  \\
& ~~~~~~~~~~ + i (H^{(11)}_{12} - H^{(11)}_{13} - H^{(11)}_{21} + H^{(11)}_{23} + H^{(11)}_{31} - H^{(11)}_{32} )\cos(\alpha^{1})\cos(\beta^{1}) \cos(\gamma^{1})] \\
&~~ -\frac{i}{2}[H^{(12)}_{11} \cos^2(\alpha^{1})\cos^2(\alpha^{2}) + H^{(12)}_{12}\cos^2(\alpha^{1})\cos^2(\beta^{2}) + + H^{(12)}_{13} \cos^2(\alpha^{1})\cos^2(\gamma^{2}) \\
& ~~~~~~~~~~  + H^{(12)}_{21} \cos^2(\beta^{1})\cos^2(\alpha^{2}) + H^{(12)}_{22} \cos^2(\beta^{1})\cos^2(\beta^{2}) + H^{(12)}_{23} \cos^2(\beta^{1})\cos^2(\gamma^{2}) \\
& ~~~~~~~~~~ + H^{(12)}_{31} \cos^2(\gamma^{1})\cos^2(\alpha^{2}) + H^{(12)}_{32} \cos^2(\gamma^{1})\cos^2(\beta^{2}) + H^{(12)}_{33} \cos^2(\gamma^{1})\cos^2(\gamma^{2})] \\
&~~ -\frac{i}{2}[H^{(21)}_{11} \cos^2(\alpha^{2})\cos^2(\alpha^{1}) + H^{(21)}_{12}\cos^2(\alpha^{2})\cos^2(\beta^{1}) + + H^{(21)}_{13} \cos^2(\alpha^{2})\cos^2(\gamma^{1}) \\
& ~~~~~~~~~~  + H^{(21)}_{21} \cos^2(\beta^{2})\cos^2(\alpha^{1}) + H^{(21)}_{22} \cos^2(\beta^{2})\cos^2(\beta^{1}) + H^{(21)}_{23} \cos^2(\beta^{2})\cos^2(\gamma^{1})\\
& ~~~~~~~~~~ + H^{(21)}_{31} \cos^2(\gamma^{2})\cos^2(\alpha^{1}) + H^{(21)}_{32} \cos^2(\gamma^{2})\cos^2(\beta^{1}) + H^{(21)}_{33} \cos^2(\gamma^{2})\cos^2(\gamma^{1})] \\
&~~ -\frac{i}{2}[H^{(22)}_{11}\cos^2(\alpha^{2}) +  H^{(22)}_{22}\cos^2(\beta^{2}) + H^{(22)}_{33}\cos^2(\gamma^{2}) \\
& ~~~~~~~~~~ + i (H^{(22)}_{12} - H^{(22)}_{13} - H^{(22)}_{21} + H^{(22)}_{23} + H^{(22)}_{31} - H^{(22)}_{32} )\cos(\alpha^{2})\cos(\beta^{2}) \cos(\gamma^{2})]
\end{aligned}
\end{equation*}

Substituting the particular values of these Hamiltonian coefficients from equation \ref{hamco}

\begin{equation*}
\begin{aligned}                   
\delta E_{E} = &-\frac{i}{2} [ {\cal A}_1 \cos^2(\alpha^{1}) +  {\cal A}_1  \cos^2(\beta^{1})  +  2  {\cal B}_1 \cos(\alpha^{1})\cos(\beta^{1}) \cos(\gamma^{1})] \\
&-\frac{i}{2}[{\cal A}_2 \cos^2(\alpha^{1})\cos^2(\alpha^{2}) - i {\cal B}_2 \cos^2(\alpha^{1})\cos^2(\beta^{2}) \\
&+ i{\cal B}_2 \cos^2(\beta^{1})\cos^2(\alpha^{2}) + {\cal A}_2 \cos^2(\beta^{1})\cos^2(\beta^{2})] \\
&-\frac{i}{2}[{\cal A}_2 \cos^2(\alpha^{2})\cos^2(\alpha^{1}) - i {\cal B}_2 \cos^2(\alpha^{2})\cos^2(\beta^{1}) \\
&+ i {\cal B}_2 \cos^2(\beta^{2})\cos^2(\alpha^{1}) + {\cal A}_2 \cos^2(\beta^{2})\cos^2(\beta^{1})] \\
&-\frac{i}{2}[{\cal A}_1 \cos^2(\alpha^{2}) + {\cal A}_1 \cos^2(\beta^{2}) + 2 {\cal B}_1  \cos(\alpha^{2})\cos(\beta^{2}) \cos(\gamma^{2})]
\end{aligned}  
\end{equation*}

Using the equation \ref{Aint}

\begin{equation*}
\begin{aligned}                    
\delta E_{E} = -\frac{\mu^2 P}{8\pi^2} \int^{\infty}_{-\infty} d\omega~&\frac{\omega}{\left(1-e^{-2\pi k\omega}\right)\left(\omega+\omega_0\right)\left(\omega-\omega_0\right)} \\ &[2\omega_0\left\{\cos(\alpha^{1})\cos(\beta^{1})\cos(\gamma^{1})+\cos(\alpha^{2})\cos(\beta^{2})\cos(\gamma^{2})\right\} \\
& +\omega\left\{\cos[2](\alpha^{1})+\cos[2](\beta^{1})+\cos[2](\alpha^{2})+\cos[2](\beta^{2})\right\} \\
& +2\omega f(\omega,L/2) \left\{\cos[2](\alpha^{1})\cos[2](\alpha^{2})+\cos[2](\beta^{1})\cos[2](\beta^{2})\right\}]
\end{aligned}
\end{equation*}

In abbreviated notation

\be
\begin{aligned}                    
\delta E_{E}=-\frac{\mu^2 P}{8\pi^2} ~[&\left\{\cos(\alpha^{1})\cos(\beta^{1})\cos(\gamma^{1})+\cos(\alpha^{2})\cos(\beta^{2})\cos(\gamma^{2})\right\} \Delta_1 \\
& + \left\{\cos[2](\alpha^{1})+\cos[2](\beta^{1})+\cos[2](\alpha^{2})+\cos[2](\beta^{2})\right\} \Delta_2 \\
& + \left\{\cos[2](\alpha^{1})\cos[2](\alpha^{2}) + \cos[2](\beta^{1})\cos[2](\beta^{2})\right\} \Delta_3]
\end{aligned}
\ee

\subsection{\textcolor{blue}{\bf \large For Symmetric state}}
For the symmetric state of two entangled atoms ($|S\rangle $) the expectation values of all the possible Pauli tensor operators, which are explicitly contributing in the part of the Lamb Shift Hamiltonian ($H^{(11)}_{LS}$) are given by:
\begin{equation*}
\begin{split}
&\langle S | \sigma_1^1 \sigma_1^1 | S \rangle = 1 \ \ \ \ \langle S | \sigma_1^1 \sigma_2^1 | S \rangle = 0 \ \ \ \ \langle S | \sigma_1^1 \sigma_3^1 | S \rangle = 0 \\
&\langle S | \sigma_2^1 \sigma_1^1 | S \rangle = 0 \ \ \ \ \langle S | \sigma_2^1 \sigma_2^1 | S \rangle = 1 \ \ \ \ \langle S | \sigma_2^1 \sigma_3^1 | S \rangle = 0 \\
&\langle S | \sigma_3^1 \sigma_1^1 | S \rangle = 0 \ \ \ \ \langle S | \sigma_3^1 \sigma_2^1 | S \rangle = 0 \ \ \ \ \langle S | \sigma_3^1 \sigma_3^1 | S \rangle = 1 
\end{split}
\end{equation*}
Consequently, the expectation value of the part of the Lamb Shift Hamiltonian ($H^{(11)}_{LS}$) with respect to the symmetric state can be written as:
\be
\begin{aligned}
\delta E^{(11)}_{S} &=\langle S| H^{(11)}_{LS} | S\rangle \\
&=-\frac{i}{2}[H^{(11)}_{11}\langle S| \sigma^{1}_{1}\sigma^{1}_{1}| S\rangle\cos^2(\alpha^{1})+H^{(11)}_{12}\langle S| \sigma^{1}_{1}\sigma^{1}_{2}| S\rangle\cos(\alpha^{1})\cos(\beta^{1})\nonumber\\
& ~~~~~~~~~~~~+H^{(11)}_{13}\langle S| \sigma^{1}_{1}\sigma^{1}_{3}| S\rangle\cos(\alpha^{1})\cos(\gamma^{1})+H^{(11)}_{21}\langle S| \sigma^{1}_{2}\sigma^{1}_{1}| S\rangle\cos(\beta^{1})\cos(\alpha^{1})\\
& ~~~~~~~~~~~~+H^{(11)}_{22}\langle S| \sigma^{1}_{2}\sigma^{1}_{2}| S\rangle\cos^2(\beta^{1})+H^{(11)}_{23}\langle S| \sigma^{1}_{2}\sigma^{1}_{3}| S\rangle\cos(\beta^{1})\cos(\gamma^{1})\\
& ~~~~~~~~~~~~+H^{(11)}_{31}\langle S| \sigma^{1}_{3}\sigma^{1}_{1}| S\rangle\cos(\gamma^{1})\cos(\alpha^{1})+H^{(11)}_{32}\langle S| \sigma^{1}_{3}\sigma^{1}_{2}| S\rangle\cos(\gamma^{1})\cos(\beta^{1})\\
& ~~~~~~~~~~~~+H^{(11)}_{33}\langle E| \sigma^{1}_{3}\sigma^{1}_{3}| E\rangle\cos^2(\gamma^{1})]\nonumber\\
&=-\frac{i}{2}[H^{(11)}_{11}\cos^2(\alpha^{1})+H^{(11)}_{22}\cos^2(\beta^{1})+H^{(11)}_{33}\cos^2(\gamma^{1})]
\end{aligned}
\ee

Now keeping in mind the equation \ref{sym}, we define few quantities important for rest of the calculation:
\begin{equation*}
\begin{split}
A&=\left[\frac{\cos \ag_{1}-i \cos \bg_{1}}{1+\cos \gamma_{1}}+\frac{\cos \ag_{2}-i \cos\bg_{2}}{1+\cos \gamma_{2}}\right] \\
B&=\left[1-\frac{\cos \ag_{1}-i \cos \bg_{1}}{1+\cos \gamma_{1}}.\frac{\cos\ag_{2}+i \cos\bg_{2}}{1+\cos \gamma_{2}}\right] \\
C&=\left[1-\frac{\cos \ag_{1}+i \cos\bg_{1}}{1+\cos \gamma_{1}}.\frac{\cos \ag_{2}-i\cos \bg_{2}}{1+\cos \gamma_{2}}\right] \\
D&=\left[\frac{\cos \ag_{1}+i \cos \bg_{1}}{1+\cos \gamma_{1}}+\frac{\cos \ag_{2}+i \cos \bg_{2}}{1+\cos \gamma_{2}}\right] \\
\Omega&=\frac{1}{2\sqrt{2}}\sqrt{(1+\cos \gamma_{1})(1+\cos \gamma_{2})}
\end{split}
\end{equation*}
For the symmetric state of two entangled atoms ($|S\rangle $) the expectation values of all the possible Pauli tensor operators, which are explicitly contributing in the part of the Lamb Shift Hamiltonian ($H^{(12)}_{LS}$) are given by:
\begin{equation*}
\begin{split}
&\langle S | \sigma_1^1 \sigma_1^2 | S \rangle = \Omega[B^{2}+C^{2}-A^{2}-D^{2}]
~~~~
\langle S | \sigma_1^1 \sigma_2^2 | S \rangle = i\Omega[D^{2}+C^{2}-B^{2}-A^{2}]
\\
&\langle S | \sigma_1^1 \sigma_3^2 | S \rangle = 2\Omega[AB-CD] 
~~~~~~~~~~~~~~~
\langle S | \sigma_2^1 \sigma_1^2 | S \rangle = i\Omega[D^{2}-A^{2}-(C^{2}-B^{2})]
\\
&\langle S | \sigma_2^1 \sigma_2^2 | S \rangle = \Omega[D^{2}+C^{2}+B^{2}+A^{2}]
~~~~
\langle S | \sigma_2^1 \sigma_3^2 | S \rangle = 2i \Omega[AB+CD] \\
&\langle S | \sigma_3^1 \sigma_1^2 | S \rangle = 2\Omega[AC-BD]
~~~~~~~~~~~~~~~
\langle S | \sigma_3^1 \sigma_2^2 | S \rangle = 2i\Omega(AC+BD) \\
&\langle S | \sigma_3^1 \sigma_3^2 | S \rangle = -2\Omega(BC+AD)
\end{split}
\end{equation*}
Consequently, the expectation value of the part of the Lamb Shift Hamiltonian ($H^{(12)}_{LS}$) with respect to the symmetric state can be written as:
\begin{equation*}
\begin{split}
 \delta E^{(12)}_{S}&=\langle S | H^{(12)}_{LS}| S \rangle \\
 &=-\frac{i}{2}\{H^{(12)}_{11}\langle S | \sigma^{1}_{1}\sigma^{2}_{1}| S \rangle\cos(\alpha^{1})\cos(\alpha^{2}) + H^{(12)}_{12}\langle S | \sigma^{1}_{1}\sigma^{2}_{2}| S \rangle\cos(\alpha^{1})\cos(\beta^{2}) \\
& ~~~~~~~~~~~+H^{(12)}_{13}\langle S | \sigma^{1}_{1}\sigma^{2}_{3}| S \rangle\cos(\alpha^{1})\cos(\gamma^{2})+H^{(12)}_{21}\langle S | \sigma^{1}_{2}\sigma^{2}_{1}| S \rangle\cos(\beta^{1})\cos(\alpha^{2})\\
& ~~~~~~~~~~~~ +H^{(12)}_{22}\langle S | \sigma^{1}_{2}\sigma^{2}_{2}| S \rangle\cos(\beta^{1})\cos(\beta^{2})+H^{(12)}_{23}\langle S | \sigma^{1}_{2}\sigma^{2}_{3}| S \rangle\cos(\beta^{1})\cos(\gamma^{2})\\
& ~~~~~~~~~~~~+H^{(12)}_{31}\langle S | \sigma^{1}_{3}\sigma^{2}_{1}| S \rangle\cos(\gamma^{1})\cos(\alpha^{2})+H^{(12)}_{32}\langle S | \sigma^{1}_{3}\sigma^{2}_{2}| S \rangle\cos(\gamma^{1})\cos(\beta^{2})\\
& ~~~~~~~~~~~~+H^{(12)}_{33}\langle S | \sigma^{1}_{3}\sigma^{2}_{3}| S \rangle\cos(\gamma^{1})\cos(\gamma^{2})\}\\
&=-\frac{i}{2} \Omega \{H^{(12)}_{11} [B^{2}+C^{2}-A^{2}-D^{2}] \cos(\alpha^{1})\cos(\alpha^{2}) \\
& ~~~~~~~~~~~~+ H^{(12)}_{12} i[D^{2}+C^{2}-B^{2}-A^{2}] \cos(\alpha^{1})\cos(\beta^{2}) 
 + H^{(12)}_{13} 2 [AB-CD] \cos(\alpha^{1})\cos(\gamma^{2})     \\
 & ~~~~~~~~~~~~ +      H^{(12)}_{21} i[D^{2}-A^{2}-(C^{2}-B^{2})] \cos(\beta^{1})\cos(\alpha^{2})  \\
  & ~~~~~~~~~~~~ + H^{(12)}_{22} [D^{2}+C^{2}+B^{2}+A^{2}] \cos(\beta^{1})\cos(\beta^{2})  \\
  & ~~~~~~~~~~~~ +  H^{(12)}_{23} 2i [AB+CD] \cos(\beta^{1})\cos(\gamma^{2})  +        H^{(12)}_{31} 2 [AC-BD]\cos(\gamma^{1})\cos(\alpha^{2})  \\
  & ~~~~~~~~~~~~ + H^{(12)}_{32} 2i [AC+BD] \cos(\gamma^{1})\cos(\beta^{2}) 
  +  H^{(12)}_{33} (-2) [BC+AD]\cos(\gamma^{1})\cos(\gamma^{2}) \}
\end{split}
\end{equation*}
For the symmetric state of two entangled atoms ($|E\rangle $) the expectation values of all the possible Pauli tensor operators, which are explicitly contributing in the part of the Lamb Shift Hamiltonian ($H^{(21)}_{LS}$) are given by:
\begin{equation*}
\begin{split}
&\langle S | \sigma_1^2 \sigma_1^1 | S \rangle = \Omega[B^{2}+C^{2}-A^{2}-D^{2}] ~~~~~ \langle S | \sigma_1^2 \sigma_2^1 | S \rangle = i\Omega[D^{2}-A^{2}-(C^{2}-B^{2})] \\ &\langle S | \sigma_1^2 \sigma_3^1 | S \rangle = 2\Omega[AC-BD] ~~~~~~~~~~~~~~~~
\langle S | \sigma_2^2 \sigma_1^1 | S \rangle = i\Omega[D^{2}+C^{2}-B^{2}-A^{2}] \\ &\langle S | \sigma_2^2 \sigma_2^1 | S \rangle = \Omega[D^{2}+C^{2}+B^{2}+A^{2}] ~~~~~ \langle S | \sigma_2^2 \sigma_3^1 | S \rangle = 2i\Omega(AC+BD)  \\
&\langle S | \sigma_3^2 \sigma_1^1 | S \rangle = 2\Omega[AB-CD]~~~~~~~~~~~~~~~~ 
\langle S | \sigma_3^2 \sigma_2^1 | S \rangle =2i \Omega[AB+CD] \\
& \langle S | \sigma_3^2 \sigma_3^1 | S \rangle = -2 \Omega[BC+AD]
\end{split}
\end{equation*}
Consequently, the expectation value of the part of the Lamb Shift Hamiltonian ($H^{(21)}_{LS}$) with respect to the symmetric state can be written as:
\begin{equation*}
\begin{split}
\delta E^{(21)}_{S} &=\langle S |H^{(21)}_{LS}| S \rangle \\
&=-\frac{i}{2} \{ H^{(21)}_{11}\langle S |\sigma^{2}_{1}\sigma^{1}_{1}| S \rangle\cos(\alpha^{2})\cos(\alpha^{1})+H^{(21)}_{12}\langle S |\sigma^{2}_{1}\sigma^{1}_{2}| S \rangle\cos(\alpha^{2})\cos(\beta^{1}) \\
& ~~~~~~~~~~~+H^{(21)}_{13}\langle S |\sigma^{2}_{1}\sigma^{1}_{3}| S \rangle\cos(\alpha^{2})\cos(\gamma^{1})+H^{(21)}_{21}\langle S |\sigma^{2}_{2}\sigma^{1}_{1}| S \rangle\cos(\beta^{2})\cos(\alpha^{1})\\
& ~~~~~~~~~~~+H^{(21)}_{22}\langle S |\sigma^{2}_{2}\sigma^{1}_{2}| S \rangle\cos(\beta^{2})\cos(\beta^{1})+H^{(21)}_{23}\langle S |\sigma^{2}_{2}\sigma^{1}_{3}| S \rangle\cos(\beta^{2})\cos(\gamma^{1})\\
& ~~~~~~~~~~~+H^{(21)}_{31}\langle S |\sigma^{2}_{3}\sigma^{1}_{1}| S \rangle\cos(\gamma^{2})\cos(\alpha^{1})+H^{(21)}_{32}\langle S |\sigma^{2}_{3}\sigma^{1}_{2}| S \rangle\cos(\gamma^{2})\cos(\beta^{1})\\
& ~~~~~~~~~~~+H^{(21)}_{33}\langle S |\sigma^{2}_{3}\sigma^{1}_{3}| S \rangle\cos(\gamma^{2})\cos(\gamma^{1}) \} \\
&=-\frac{i}{2} \Omega \{ H^{(21)}_{11} [B^{2}+C^{2}-A^{2}-D^{2}] \cos(\alpha^{2})\cos(\alpha^{1}) \\
& ~~~~~~~~~~~~+ H^{(21)}_{12} i[D^{2}-A^{2}-(C^{2}-B^{2})]  \cos(\alpha^{2})\cos(\beta^{1}) \nonumber\\
& ~~~~~~~~~~~ + H^{(21)}_{13} 2 [AC-BD]  \cos(\alpha^{2})\cos(\gamma^{1})     +      H^{(21)}_{21} i[D^{2}+C^{2}-B^{2}-A^{2}] \cos(\beta^{2})\cos(\alpha^{1}) \\
& ~~~~~~~~~~~  + H^{(21)}_{22} [D^{2}+C^{2}+B^{2}+A^{2}] \cos(\beta^{2})\cos(\beta^{1})  +  H^{(21)}_{23}  2i [AC+BD]  \cos(\beta^{2})\cos(\gamma^{1})  \\
& ~~~~~~~~~~~~ + H^{(21)}_{31} 2 [AB-CD] \cos(\gamma^{2})\cos(\alpha^{1}) + H^{(21)}_{32} 2i [AB+CD] \cos(\gamma^{2})\cos(\beta^{1}) \\
& ~~~~~~~~~~~  +  H^{(21)}_{33} (-2) [BC+AD]\cos(\gamma^{2})\cos(\gamma^{1}) \}
\end{split} 
\end{equation*}
For the symmetric state of two entangled atoms ($|E\rangle $) the expectation values of all the possible Pauli tensor operators, which are explicitly contributing in the part of the Lamb Shift Hamiltonian ($H^{(22)}_{LS}$) are given by: 
\begin{equation*}
\begin{split}
\langle S | \sigma_1^2 \sigma_1^2 | S \rangle &= 1~~~ \langle S | \sigma_1^2 \sigma_2^2 | S \rangle = 0 ~~~ \langle S | \sigma_1^2 \sigma_3^2 | S \rangle = 0 \\
\langle S | \sigma_2^2 \sigma_1^2 | S \rangle &= 0 ~~~ \langle S | \sigma_2^2 \sigma_2^2 | S \rangle = 1 ~~~ \langle S | \sigma_2^2 \sigma_3^2 | S \rangle = 0 \\
\langle S | \sigma_3^2 \sigma_1^2 | S \rangle &= 0 ~~~ \langle S | \sigma_3^2 \sigma_2^2 | S \rangle = 0 ~~~ \langle S | \sigma_3^2 \sigma_3^2 | S \rangle = 1
\end{split}
\end{equation*}
Consequently, the expectation value of the part of the Lamb Shift Hamiltonian ($H^{(22}_{LS}$) with respect to the symmetric state can be written as:
\beno
\begin{split}
\delta E^{(22)}_{S} &=\langle S| H^{(22)}_{LS}| S\rangle \\
&=-\frac{i}{2} \{ H^{(22)}_{11}\langle S|\sigma^{2}_{1}\sigma^{2}_{1}| S\rangle\cos^2(\alpha^{2})+H^{(22)}_{12}\langle S|\sigma^{2}_{1}\sigma^{2}_{2}| S\rangle\cos(\alpha^{2})\cos(\beta^{2}) \\
& ~~~~~~~~~~~~~+H^{(22)}_{13}\langle S|\sigma^{2}_{1}\sigma^{2}_{3}| S\rangle\cos(\alpha^{2})\cos(\gamma^{2})+H^{(22)}_{21}\langle S|\sigma^{2}_{2}\sigma^{2}_{1}| S\rangle\cos(\beta^{2})\cos(\alpha^{2})\\
& ~~~~~~~~~~~~~+H^{(22)}_{22}\langle S|\sigma^{2}_{2}\sigma^{2}_{2}| S\rangle\cos^2(\beta^{2})+H^{(22)}_{23}\langle S|\sigma^{2}_{2}\sigma^{2}_{3}| S\rangle\cos(\beta^{2})\cos(\gamma^{2})\\
& ~~~~~~~~~~~~~+H^{(22)}_{31}\langle S|\sigma^{2}_{3}\sigma^{2}_{1}| S\rangle\cos(\gamma^{2})\cos(\alpha^{2})+H^{(22)}_{32}\langle S|\sigma^{2}_{3}\sigma^{2}_{2}| S\rangle\cos(\gamma^{2})\cos(\beta^{2})\\
& ~~~~~~~~~~~~~+H^{(22)}_{33}\langle S|\sigma^{2}_{3}\sigma^{2}_{3}| S\rangle\cos^2(\gamma^{2}) \} \\
&=-\frac{i}{2}[H^{(22)}_{11}\cos^2(\alpha^{2})+H^{(22)}_{22}\cos^2(\beta^{2})+H^{(22)}_{33}\cos^2(\gamma^{2})]
\end{split} 
\eeno
After that, summing over all the possible contributions obtained for the symmetric state of two entangled atoms ($|S\rangle $) the expectation value of the Lamd Shift Hamiltonian can be expressed as:
\beno
\begin{split} 
\delta E_S &= \sum^{2}_{i=1}\sum^{2}_{j=1}\delta E^{(ij)}_S=\sum^{2}_{i=1}\sum^{2}_{j=1} \langle S| H^{(ij)}_{LS}|S \rangle \\
&=-\frac{i}{2}[H^{(11)}_{11}\cos^2(\alpha^{1})+H^{(22)}_{11}\cos^2(\alpha^{2})+H^{(11)}_{22}\cos^2(\beta^{1}) \\
&~~~~~~~~~~~~ + H^{(22)}_{22}\cos^2(\beta^{2}) +H^{(11)}_{33}\cos^2(\gamma^{1}) +H^{(22)}_{33}\cos^2(\gamma^{2})] \\
&~~ -\frac{i}{2} \Omega [H^{(12)}_{11} [B^{2}+C^{2}-A^{2}-D^{2}] \cos(\alpha^{1})\cos(\alpha^{2}) \\
&~~~~~~~~~~~~+ H^{(12)}_{12} i[D^{2}+C^{2}-B^{2}-A^{2}] \cos(\alpha^{1})\cos(\beta^{2}) \\
& ~~~~~~~~~ + H^{(12)}_{13} 2 [AB-CD] \cos(\alpha^{1})\cos(\gamma^{2})     +      H^{(12)}_{21} i[D^{2}-A^{2}-(C^{2}-B^{2})] \cos(\beta^{1})\cos(\alpha^{2}) \\
&  ~~~~~~~~~~ + H^{(12)}_{22} [D^{2}+C^{2}+B^{2}+A^{2}] \cos(\beta^{1})\cos(\beta^{2})  +  H^{(12)}_{23} 2i [AB+CD] \cos(\beta^{1})\cos(\gamma^{2})   \\
& ~~~~~~~~~~~~ +        H^{(12)}_{31} 2 [AC-BD]\cos(\gamma^{1})\cos(\alpha^{2}) + H^{(12)}_{32} 2i [AC+BD] \cos(\gamma^{1})\cos(\beta^{2}) \\
& ~~~~~~~~~~~~  +  H^{(12)}_{33} (-2) [BC+AD]\cos(\gamma^{1})\cos(\gamma^{2})] \\
&~~ -\frac{i}{2} \Omega [H^{(21)}_{11} [B^{2}+C^{2}-A^{2}-D^{2}] \cos(\alpha^{2})\cos(\alpha^{1}) \\
&~~~~~~~~~~~~+ H^{(21)}_{12} i[D^{2}-A^{2}-(C^{2}-B^{2})]  \cos(\alpha^{2})\cos(\beta^{1}) \\
& ~~~~~~~~~~~~ + H^{(21)}_{13} 2 [AC-BD]  \cos(\alpha^{2})\cos(\gamma^{1})     +      H^{(21)}_{21} i[D^{2}+C^{2}-B^{2}-A^{2}] \cos(\beta^{2})\cos(\alpha^{1}) \\
& ~~~~~~~~~~~~  + H^{(21)}_{22} [D^{2}+C^{2}+B^{2}+A^{2}] \cos(\beta^{2})\cos(\beta^{1})  +  H^{(21)}_{23}  2i [AC+BD]  \cos(\beta^{2})\cos(\gamma^{1}) \\
& ~~~~~~~~~~~~ +        H^{(21)}_{31} 2 [AB-CD] \cos(\gamma^{2})\cos(\alpha^{1}) + H^{(21)}_{32} 2i [AB+CD] \cos(\gamma^{2})\cos(\beta^{1}) \\
& ~~~~~~~~~~~~  +  H^{(21)}_{33} (-2) [BC+AD]\cos(\gamma^{2})\cos(\gamma^{1})]
\end{split}
\eeno

Substituting the particular values of these Hamiltonian coefficients from equation \ref{hamco}

\begin{equation*}
\begin{aligned}
\delta E_{S} = &-\frac{i}{2} {\cal A}_1 [\cos^2(\alpha^{1}) + \cos^2(\alpha^{2}) +  \cos^2(\beta^{1}) + \cos^2(\beta^{2})] \\
&-\frac{i}{2} \Omega [ {\cal A}_2  [B^{2}+C^{2}-A^{2}-D^{2}] \cos(\alpha^{1})\cos(\alpha^{2}) + {\cal B}_2 [D^{2}+C^{2}-B^{2}-A^{2}] \cos(\alpha^{1})\cos(\beta^{2}) \\
& ~~~~~~~   -    {\cal B}_2 [D^{2}-A^{2}-(C^{2}-B^{2})] \cos(\beta^{1})\cos(\alpha^{2}) + {\cal A}_2 [D^{2}+C^{2}+B^{2}+A^{2}] \cos(\beta^{1})\cos(\beta^{2})  \\
&-\frac{i}{2} \Omega [{\cal A}_2 [B^{2}+C^{2}-A^{2}-D^{2}] \cos(\alpha^{2})\cos(\alpha^{1}) + {\cal B}_2 [D^{2}-A^{2}-(C^{2}-B^{2})]  \cos(\alpha^{2})\cos(\beta^{1}) \\
& ~~~~~~~    -  {\cal B}_2 [D^{2}+C^{2}-B^{2}-A^{2}] \cos(\beta^{2})\cos(\alpha^{1}) + {\cal A}_2 [D^{2}+C^{2}+B^{2}+A^{2}] \cos(\beta^{2})\cos(\beta^{1}) ]
\end{aligned}
\end{equation*}

Using the equation \ref{Aint}

\begin{equation*}
\begin{split}
\delta E_{S} = -\frac{\mu^2 P}{8\pi^2}\int^{\infty}_{-\infty}d\omega~&\frac{\omega^2 }{\left(1-e^{-2\pi k\omega}\right)\left(\omega+\omega_0\right)\left(\omega-\omega_0\right)} \\
&[2\Omega (B^{2}+C^{2}-A^{2}-D^{2}) f(\omega,L/2) \cos(\alpha^{1})\cos(\alpha^{2}) \\
& +2\Omega (A^{2}+B^{2}+C^{2}+D^{2}) f(\omega,L/2) \cos(\beta^{1})\cos(\beta^{2}) \\
&+\{\cos^2(\alpha^{1})+\cos^2(\alpha^{2})+\cos^2(\beta^{1})+\cos^2(\beta^{2}) \}]
\end{split}
\end{equation*}

In abbreviated notation

\be
\begin{aligned}
\delta E_{S} = -\frac{\mu^2 P}{8\pi^2} ~ [&\Omega (B^{2}+C^{2}-A^{2}-D^{2}) \cos(\alpha^{1})\cos(\alpha^{2}) \Delta_3 \\
&+ \Omega (A^{2}+B^{2}+C^{2}+D^{2}) \cos(\beta^{1})\cos(\beta^{2})\Delta_3  \\
& + \{ \cos^2(\alpha^{1})+\cos^2(\alpha^{2})+\cos^2(\beta^{1})+\cos^2(\beta^{2}) \} \Delta_2]
\end{aligned}
\ee

\subsection{\textcolor{blue}{\bf \large For Antisymmetric state}}
Here, keeping in mind equation \ref{asym}, we again define few quantities to ease our presentation:
\beno
\begin{split}
\Omega &= \frac{1}{2\sqrt{2}}\sqrt{(1+\cos \gamma_{1})(1+\cos \gamma_{2})} \\
\tilde A &= \left[\frac{\cos\ag_{1}-i \cos\bg_{1}}{1+\cos \gamma_{1}}-\frac{\cos\ag_{2}-i \cos\bg_{2}}{1+\cos\gamma_{2}}\right] \\
\tilde B &= \left[1+\frac{\cos\ag_{1}-i \cos\bg_{1}}{1+\cos \gamma_{1}}\frac{\cos\ag_{2}-i \cos\bg_{2}}{1+\cos\gamma_{2}}\right] \\
\tilde C &= \left[-1-\frac{\cos\ag_{1}+ i \cos\bg_{1}}{1+\cos \gamma_{1}}-\frac{\cos\ag_{2}-i \cos\bg_{2}}{1+\cos\gamma_{2}}\right] \\
\tilde D &= \left[\frac{\cos\ag_{1}-i \cos\bg_{1}}{1+\cos \gamma_{1}}-\frac{\cos\ag_{2}+ i \cos\bg_{2}}{1+\cos\gamma_{2}}\right]
\end{split}
\eeno
For the antisymmetric state of two entangled atoms ($|A\rangle $) the expectation values of all the possible Pauli tensor operators, which are explicitly contributing in the part of the Lamb Shift Hamiltonian ($H^{(11)}_{LS}$) are given by:
\beno
\begin{split}
\langle A | \sigma_1^1 \sigma_1^1 | A \rangle &= 1 \ \ \ \ \langle A | \sigma_1^1 \sigma_2^1 | A \rangle = 0 \ \ \ \ \langle A | \sigma_1^1 \sigma_3^1 | A \rangle = 0 \\
\langle A | \sigma_2^1 \sigma_1^1 | A \rangle &= 0 \ \ \ \ \langle A | \sigma_2^1 \sigma_2^1 | A \rangle = 1 \ \ \ \ \langle A | \sigma_2^1 \sigma_3^1 | A \rangle = 0 \\
\langle A | \sigma_3^1 \sigma_1^1 | A \rangle &= 0 \ \ \ \ \langle A | \sigma_3^1 \sigma_2^1 | A \rangle = 0 \ \ \ \ \langle A | \sigma_3^1 \sigma_3^1 | A \rangle = 1 \end{split}
\eeno
Consequently, the expectation value of the part of the Lamb Shift Hamiltonian ($H^{(11)}_{LS}$) with respect to the antisymmetric state can be written as:
\beno
\begin{split}
 \delta E^{(11)}_{A} &=\langle A| H^{(11)}_{LS} | A\rangle \\
 &=-\frac{i}{2}[H^{(11)}_{11}\langle A| \sigma^{1}_{1}\sigma^{1}_{1}| A\rangle\cos^2(\alpha^{1})+H^{(11)}_{12}\langle A| \sigma^{1}_{1}\sigma^{1}_{2}| A\rangle\cos(\alpha^{1})\cos(\beta^{1})\nonumber\\
& ~~~~~~~~~~~~+H^{(11)}_{13}\langle A| \sigma^{1}_{1}\sigma^{1}_{3}| A\rangle\cos(\alpha^{1})\cos(\gamma^{1})+H^{(11)}_{21}\langle A| \sigma^{1}_{2}\sigma^{1}_{1}| A\rangle\cos(\beta^{1})\cos(\alpha^{1}) \\
& ~~~~~~~~~~~~~+H^{(11)}_{22}\langle A| \sigma^{1}_{2}\sigma^{1}_{2}| A\rangle\cos^2(\beta^{1})+H^{(11)}_{23}\langle A| \sigma^{1}_{2}\sigma^{1}_{3}| A\rangle\cos(\beta^{1})\cos(\gamma^{1}) \\
& ~~~~~~~~~~~~+H^{(11)}_{31}\langle A| \sigma^{1}_{3}\sigma^{1}_{1}| A\rangle\cos(\gamma^{1})\cos(\alpha^{1})+H^{(11)}_{32}\langle A| \sigma^{1}_{3}\sigma^{1}_{2}| A\rangle\cos(\gamma^{1})\cos(\beta^{1}) \\
& ~~~~~~~~~~~~+H^{(11)}_{33}\langle A| \sigma^{1}_{3}\sigma^{1}_{3}| A\rangle\cos^2(\gamma^{1})] \\
&=-\frac{i}{2}[H^{(11)}_{11}\cos^2(\alpha^{1})+H^{(11)}_{22}\cos^2(\beta^{1})+H^{(11)}_{33}\cos^2(\gamma^{1})]
\end{split}
\eeno
For the antisymmetric state of two entangled atoms ($|A\rangle $) the expectation values of all the possible Pauli tensor operators, which are explicitly contributing in the part of the Lamb Shift Hamiltonian ($H^{(12)}_{LS}$) are given by:

\beno
\begin{split}
&\langle A | \sigma_1^1 \sigma_1^2 | A \rangle = \Omega[\tilde  D^{2}+\tilde A^{2}-(\tilde B^{2}+\tilde C^{2})] ~~~~~ \langle A | \sigma_1^1 \sigma_2^2 | A \rangle = i\Omega [\tilde A^{2}-\tilde D^{2}+\tilde B^{2}-\tilde C^{2}] \\
&\langle A | \sigma_1^1 \sigma_3^2 | A \rangle = 2\Omega (\tilde C\tilde D-\tilde A\tilde B) ~~~~~~~~~~~~~~~~~~
\langle A | \sigma_2^1 \sigma_1^2 | A \rangle = -i \Omega[\tilde D^{2}-\tilde A^{2}+\tilde B^{2}-\tilde C^{2}] \\
&\langle A | \sigma_2^1 \sigma_2^2 | A \rangle = -\Omega[\tilde A^{2}+\tilde D^{2}+\tilde B^{2}+\tilde C^{2}] ~~~~~ \langle A | \sigma_2^1 \sigma_3^2 | A \rangle = -2i\Omega[\tilde A\tilde B+\tilde C\tilde D] \\ 
&\langle A | \sigma_3^1 \sigma_1^2 | A \rangle = 2\Omega(\tilde B\tilde D-\tilde A\tilde C) ~~~~~~~~~~~~~~~~~~ \langle A | \sigma_3^1 \sigma_2^2 | A \rangle = -2i\Omega(\tilde A\tilde C+\tilde B\tilde D) \\
&\langle A | \sigma_3^1 \sigma_3^2 | A \rangle = 2\Omega[\tilde A\tilde D+\tilde B\tilde C]
\end{split} 
\eeno
Consequently, the expectation value of the part of the Lamb Shift Hamiltonian ($H^{(12)}_{LS}$) with respect to the symmetric state can be written as:
\beno
\begin{split}
 \delta E^{(12)}_{A} &=\langle A | H^{(12)}_{LS}| A \rangle \\
 &=-\frac{i}{2} \{ H^{(12)}_{11}\langle A | \sigma^{1}_{1}\sigma^{2}_{1}| A \rangle\cos(\alpha^{1})\cos(\alpha^{2})+H^{(12)}_{12}\langle A | \sigma^{1}_{1}\sigma^{2}_{2}| A \rangle\cos(\alpha^{1})\cos(\beta^{2}) \\
& ~~~~~~~~~~~~+H^{(12)}_{13}\langle A | \sigma^{1}_{1}\sigma^{2}_{3}| A \rangle\cos(\alpha^{1})\cos(\gamma^{2})+H^{(12)}_{21}\langle A | \sigma^{1}_{2}\sigma^{2}_{1}| A \rangle\cos(\beta^{1})\cos(\alpha^{2})\\
& ~~~~~~~~~~~~ +H^{(12)}_{22}\langle A | \sigma^{1}_{2}\sigma^{2}_{2}| A \rangle\cos(\beta^{1})\cos(\beta^{2})+H^{(12)}_{23}\langle A | \sigma^{1}_{2}\sigma^{2}_{3}| A \rangle\cos(\beta^{1})\cos(\gamma^{2})\\
& ~~~~~~~~~~~~+H^{(12)}_{31}\langle A | \sigma^{1}_{3}\sigma^{2}_{1}| A \rangle\cos(\gamma^{1})\cos(\alpha^{2})+H^{(12)}_{32}\langle A | \sigma^{1}_{3}\sigma^{2}_{2}| A \rangle\cos(\gamma^{1})\cos(\beta^{2})\\
& ~~~~~~~~~~~~+H^{(12)}_{33}\langle A | \sigma^{1}_{3}\sigma^{2}_{3}| A \rangle\cos(\gamma^{1})\cos(\gamma^{2}) \} \\
&=-\frac{i}{2} \Omega  \{ H^{(12)}_{11}[\tilde  D^{2}+\tilde A^{2}-(\tilde B^{2}+\tilde C^{2})]\cos(\alpha^{1})\cos(\alpha^{2})  \\
& ~~~~~~~~~~~~ +   H^{(12)}_{12}i[\tilde A^{2}-\tilde D^{2}+\tilde B^{2}-\tilde C^{2}] \cos(\alpha^{1})\cos(\beta^{2}) \\
& ~~~~~~~~~~~~+H^{(12)}_{13}2 (\tilde C\tilde D-\tilde A\tilde B) \cos(\alpha^{1})\cos(\gamma^{2})   \\
&~~~~~~~~~~~~+   H^{(12)}_{21}(-i) [\tilde D^{2}-\tilde A^{2}+\tilde B^{2}-\tilde C^{2}]  \cos(\beta^{1})\cos(\alpha^{2})\\
& ~~~~~~~~~~~~ +  H^{(12)}_{22} (-1)[\tilde A^{2}+\tilde D^{2}+\tilde B^{2}+\tilde C^{2}] \cos(\beta^{1})\cos(\beta^{2})  \\
& ~~~~~~~~~~~~ +    H^{(12)}_{23} (-2i) [\tilde A\tilde B+\tilde C\tilde D]\cos(\beta^{1})\cos(\gamma^{2})\\
& ~~~~~~~~~~~~   +   H^{(12)}_{31} 2 (\tilde B\tilde D-\tilde A\tilde C) \cos(\gamma^{1})\cos(\alpha^{2})   +    H^{(12)}_{32} (-2i) (\tilde A\tilde C+\tilde B\tilde D) \cos(\gamma^{1})\cos(\beta^{2})\\
& ~~~~~~~~~~~~+H^{(12)}_{33} 2[\tilde A\tilde D+\tilde B\tilde C] \cos(\gamma^{1})\cos(\gamma^{2}) \}
\end{split}
\eeno
For the antisymmetric state of two entangled atoms ($|A\rangle $) the expectation values of all the possible Pauli tensor operators, which are explicitly contributing in the part of the Lamb Shift Hamiltonian ($H^{(21)}_{LS}$) are given by:

\beno
\begin{split}
&\langle A | \sigma_1^2 \sigma_1^1 | A \rangle = \Omega[\tilde D^{2}+\tilde A^{2}-(\tilde B^{2}+\tilde C^{2})] ~~~~~ \langle A | \sigma_1^2 \sigma_2^1 | A \rangle = -i \Omega[\tilde D^{2}-\tilde A^{2}+\tilde B^{2}-\tilde C^{2}] \\
&\langle A | \sigma_1^2 \sigma_3^1 | A \rangle =  2\Omega(\tilde B\tilde D-\tilde A\tilde C) ~~~~~~~~~~~~~~~~~~
\langle A | \sigma_2^2 \sigma_1^1 | A \rangle = i\Omega [\tilde A^{2}-\tilde D^{2}+\tilde B^{2}-\tilde C^{2}] \\
&\langle A | \sigma_2^2 \sigma_2^1 | A \rangle = -\Omega[\tilde A^{2}+\tilde D^{2}+\tilde B^{2}+\tilde C^{2}] ~~~~~ \langle A | \sigma_2^2 \sigma_3^1 | A \rangle = -2i\Omega(\tilde A\tilde C+\tilde B\tilde D) \\ 
&\langle A | \sigma_3^2 \sigma_1^1 | A \rangle =  2\Omega (\tilde C\tilde D-\tilde A\tilde B) ~~~~~~~~~~~~~~~~~~ \langle A | \sigma_3^2 \sigma_2^1 | A \rangle =-2i\Omega[\tilde A\tilde B+\tilde C\tilde D] \\
&\langle A | \sigma_3^2 \sigma_3^1 | A \rangle =2\Omega[\tilde A\tilde D+\tilde B\tilde C]
\end{split}
\eeno

Consequently, the expectation value of the part of the Lamb Shift Hamiltonian ($H^{(21)}_{LS}$) with respect to the antisymmetric state can be written as:
\beno
\begin{split} 
\delta E^{(21)}_{A} &=\langle A |H^{(21)}_{LS}| A \rangle \\
&=-\frac{i}{2}[H^{(21)}_{11}\langle A |\sigma^{2}_{1}\sigma^{1}_{1}| A \rangle\cos(\alpha^{2})\cos(\alpha^{1})+H^{(21)}_{12}\langle A |\sigma^{2}_{1}\sigma^{1}_{2}| A \rangle\cos(\alpha^{2})\cos(\beta^{1}) \\
& ~~~~~~~~~~+H^{(21)}_{13}\langle A |\sigma^{2}_{1}\sigma^{1}_{3}| A \rangle\cos(\alpha^{2})\cos(\gamma^{1})+H^{(21)}_{21}\langle A |\sigma^{2}_{2}\sigma^{1}_{1}| A \rangle\cos(\beta^{2})\cos(\alpha^{1})\\
& ~~~~~~~~~~+H^{(21)}_{22}\langle A |\sigma^{2}_{2}\sigma^{1}_{2}| A \rangle\cos(\beta^{2})\cos(\beta^{1})+H^{(21)}_{23}\langle A |\sigma^{2}_{2}\sigma^{1}_{3}| A \rangle\cos(\beta^{2})\cos(\gamma^{1})\\
& ~~~~~~~~~~+H^{(21)}_{31}\langle A |\sigma^{2}_{3}\sigma^{1}_{1}| A \rangle\cos(\gamma^{2})\cos(\alpha^{1})+H^{(21)}_{32}\langle A |\sigma^{2}_{3}\sigma^{1}_{2}| A \rangle\cos(\gamma^{2})\cos(\beta^{1})\\
& ~~~~~~~~~~+H^{(21)}_{33}\langle A |\sigma^{2}_{3}\sigma^{1}_{3}| A \rangle\cos(\gamma^{2})\cos(\gamma^{1})] \\
&=-\frac{i}{2} \Omega [H^{(21)}_{11}[\tilde  D^{2}+\tilde A^{2}-(\tilde B^{2}+\tilde C^{2})]\cos(\alpha^{2})\cos(\alpha^{1})  \\
& ~~~~~~~~~~~~+   H^{(21)}_{12} (-i) [\tilde D^{2}-\tilde A^{2}+\tilde B^{2}-\tilde C^{2}] \cos(\alpha^{2})\cos(\beta^{1}) \\
& ~~~~~~~~~~~~+H^{(21)}_{13} 2 (\tilde B\tilde D-\tilde A\tilde C) \cos(\alpha^{2})\cos(\gamma^{1})   +   H^{(21)}_{21} i[\tilde A^{2}-\tilde D^{2}+\tilde B^{2}-\tilde C^{2}]   \cos(\beta^{2})\cos(\alpha^{1})\\
& ~~~~~~~~~~~~ +  H^{(21)}_{22} (-1)[\tilde A^{2}+\tilde D^{2}+\tilde B^{2}+\tilde C^{2}] \cos(\beta^{2})\cos(\beta^{1})  \\
& ~~~~~~~~~~~~ +    H^{(21)}_{23} (-2i) (\tilde A\tilde C+\tilde B\tilde D) \cos(\beta^{2})\cos(\gamma^{1})\\
& ~~~~~~~~~~~~   +   H^{(21)}_{31} 2 (\tilde C\tilde D-\tilde A\tilde B)  \cos(\gamma^{2})\cos(\alpha^{1})   +    H^{(21)}_{32} (-2i) [\tilde A\tilde B+\tilde C\tilde D]  \cos(\gamma^{2})\cos(\beta^{1})\\
& ~~~~~~~~~~~~+H^{(21)}_{33} 2[\tilde A\tilde D+\tilde B\tilde C] \cos(\gamma^{2})\cos(\gamma^{1})]
\end{split}
\eeno
For the antisymmetric state of two entangled atoms ($|A\rangle $) the expectation values of all the possible Pauli tensor operators, which are explicitly contributing in the part of the Lamb Shift Hamiltonian ($H^{(22)}_{LS}$) are given by:
\beno
\begin{split}
\langle A | \sigma_1^2 \sigma_1^2 | A \rangle &= 1 \ \ \ \ \langle A | \sigma_1^2 \sigma_2^2 | A \rangle = 0 \ \ \ \ \langle A | \sigma_1^2 \sigma_3^2 | A \rangle = 0 \\
\langle A | \sigma_2^2 \sigma_1^2 | A \rangle &= 0 \ \ \ \ \langle A | \sigma_2^2 \sigma_2^2 | A \rangle = 1 \ \ \ \ \langle A | \sigma_2^2 \sigma_3^2 | A \rangle = 0 \\
\langle A | \sigma_3^2 \sigma_1^2 | A \rangle &= 0 \ \ \ \ \langle A | \sigma_3^2 \sigma_2^2 | A \rangle = 0 \ \ \ \ \langle A | \sigma_3^2 \sigma_3^2 | A \rangle = 1 
\end{split}
\eeno
Consequently, the expectation value of the part of the Lamb Shift Hamiltonian ($H^{(22}_{LS}$) with respect to the antisymmetric state can be written as:
\beno
\begin{split}
\delta E^{(22)}_{A} &=\langle A| H^{(22)}_{LS}| A\rangle \\ 
&=-\frac{i}{2}[H^{(22)}_{11}\langle A|\sigma^{2}_{1}\sigma^{2}_{1}| A\rangle\cos^2(\alpha^{2})+H^{(22)}_{12}\langle A|\sigma^{2}_{1}\sigma^{2}_{2}| A\rangle\cos(\alpha^{2})\cos(\beta^{2}) \\
& ~~~~~~~~~~~~~+H^{(22)}_{13}\langle A|\sigma^{2}_{1}\sigma^{2}_{3}| A\rangle\cos(\alpha^{2})\cos(\gamma^{2})+H^{(22)}_{21}\langle A|\sigma^{2}_{2}\sigma^{2}_{1}| A\rangle\cos(\beta^{2})\cos(\alpha^{2})\\
& ~~~~~~~~~~~~~+H^{(22)}_{22}\langle A|\sigma^{2}_{2}\sigma^{2}_{2}| A\rangle\cos^2(\beta^{2})+H^{(22)}_{23}\langle A|\sigma^{2}_{2}\sigma^{2}_{3}| A\rangle\cos(\beta^{2})\cos(\gamma^{2}) \\
& ~~~~~~~~~~~~~+H^{(22)}_{31}\langle A|\sigma^{2}_{3}\sigma^{2}_{1}| A\rangle\cos(\gamma^{2})\cos(\alpha^{2})+H^{(22)}_{32}\langle A|\sigma^{2}_{3}\sigma^{2}_{2}| A\rangle\cos(\gamma^{2})\cos(\beta^{2})\\
& ~~~~~~~~~~~~~+H^{(22)}_{33}\langle S|\sigma^{2}_{3}\sigma^{2}_{3}| S\rangle\cos^2(\gamma^{2})] \\
&=-\frac{i}{2}[H^{(22)}_{11}\cos^2(\alpha^{2})+H^{(22)}_{22}\cos^2(\beta^{2})+H^{(22)}_{33}\cos^2(\gamma^{2})]
\end{split}
\eeno
After that, summing over all the possible contributions obtained for the antisymmetric state of two entangled atoms ($|A\rangle $) the expectation value of the Lamd Shift Hamiltonian can be expressed as:
\beno
\begin{split}
\delta E_A &= \sum^{2}_{i=1}\sum^{2}_{j=1}\delta E^{(ij)}_A=\sum^{2}_{i=1}\sum^{2}_{j=1} \langle A| H^{(ij)}_{LS}|A \rangle \\
&=-\frac{i}{2}[H^{(11)}_{11}\cos^2(\alpha^{1})+H^{(22)}_{11}\cos^2(\alpha^{2})+H^{(11)}_{22}\cos^2(\beta^{1}) \\
& ~~~~~~~~~~~~ +H^{(22)}_{22}\cos^2(\beta^{2}) +H^{(11)}_{33}\cos^2(\gamma^{1}) +H^{(22)}_{33}\cos^2(\gamma^{2})] \\
&-\frac{i}{2} \Omega [H^{(12)}_{11}[\tilde  D^{2}+\tilde A^{2}-(\tilde B^{2}+\tilde C^{2})]\cos(\alpha^{1})\cos(\alpha^{2})  \\
&~~~~~~~~~~~~+   H^{(12)}_{12}i[\tilde A^{2}-\tilde D^{2}+\tilde B^{2}-\tilde C^{2}] \cos(\alpha^{1})\cos(\beta^{2}) \\
& ~~~~~~~+H^{(12)}_{13}2 (\tilde C\tilde D-\tilde A\tilde B) \cos(\alpha^{1})\cos(\gamma^{2})   +   H^{(12)}_{21}(-i) [\tilde D^{2}-\tilde A^{2}+\tilde B^{2}-\tilde C^{2}]  \cos(\beta^{1})\cos(\alpha^{2})\\
& ~~~~~~~~ +  H^{(12)}_{22} (-1)[\tilde A^{2}+\tilde D^{2}+\tilde B^{2}+\tilde C^{2}] \cos(\beta^{1})\cos(\beta^{2})  \\
&~~~~~~~~~~~~ +    H^{(12)}_{23} (-2i) [\tilde A\tilde B+\tilde C\tilde D]\cos(\beta^{1})\cos(\gamma^{2})\\
& ~~~~~~~~   +   H^{(12)}_{31} 2 (\tilde B\tilde D-\tilde A\tilde C) \cos(\gamma^{1})\cos(\alpha^{2})   +    H^{(12)}_{32} (-2i) (\tilde A\tilde C+\tilde B\tilde D) \cos(\gamma^{1})\cos(\beta^{2})\\
& ~~~~~~~~+H^{(12)}_{33} 2[\tilde A\tilde D+\tilde B\tilde C] \cos(\gamma^{1})\cos(\gamma^{2})] \\
&-\frac{i}{2} \Omega [H^{(21)}_{11}[\tilde  D^{2}+\tilde A^{2}-(\tilde B^{2}+\tilde C^{2})]\cos(\alpha^{2})\cos(\alpha^{1})  \\
&~~~~~~~~~~~~ +   H^{(21)}_{12} (-i) [\tilde D^{2}-\tilde A^{2}+\tilde B^{2}-\tilde C^{2}] \cos(\alpha^{2})\cos(\beta^{1}) \\
& ~~~~~~~~+H^{(21)}_{13} 2 (\tilde B\tilde D-\tilde A\tilde C) \cos(\alpha^{2})\cos(\gamma^{1})   +   H^{(21)}_{21} i[\tilde A^{2}-\tilde D^{2}+\tilde B^{2}-\tilde C^{2}]   \cos(\beta^{2})\cos(\alpha^{1})\\
& ~~~~~~~~~~ +  H^{(21)}_{22} (-1)[\tilde A^{2}+\tilde D^{2}+\tilde B^{2}+\tilde C^{2}] \cos(\beta^{2})\cos(\beta^{1})   \\
&~~~~~~~~~~~~+    H^{(21)}_{23} (-2i) (\tilde A\tilde C+\tilde B\tilde D) \cos(\beta^{2})\cos(\gamma^{1})\\
& ~~~~~~~~~~   +   H^{(21)}_{31} 2 (\tilde C\tilde D-\tilde A\tilde B)  \cos(\gamma^{2})\cos(\alpha^{1})   +    H^{(21)}_{32} (-2i) [\tilde A\tilde B+\tilde C\tilde D]  \cos(\gamma^{2})\cos(\beta^{1})\\
& ~~~~~~~~~~+H^{(21)}_{33} 2[\tilde A\tilde D+\tilde B\tilde C] \cos(\gamma^{2})\cos(\gamma^{1})]
\end{split}
\eeno

Substituting the particular values of these Hamiltonian coefficients from equation \ref{hamco}

\begin{equation*}
\begin{aligned}
\delta E_{A} = &-\frac{i}{2}{\cal A}_1 [\cos^2(\alpha^{1}) + \cos^2(\alpha^{2}) + \cos^2(\beta^{1}) + \cos^2(\beta^{2})] \\
&-\frac{i}{2} \Omega [{\cal A}_2[\tilde  D^{2}+\tilde A^{2}-(\tilde B^{2}+\tilde C^{2})]\cos(\alpha^{1})\cos(\alpha^{2})  +   {\cal B}_2 [\tilde A^{2}-\tilde D^{2}+\tilde B^{2}-\tilde C^{2}] \cos(\alpha^{1})\cos(\beta^{2}) \\
& ~~~~~~   +   {\cal B}_2 [\tilde D^{2}-\tilde A^{2}+\tilde B^{2}-\tilde C^{2}]  \cos(\beta^{1})\cos(\alpha^{2}) - {\cal A}_2 [\tilde A^{2}+\tilde D^{2}+\tilde B^{2}+\tilde C^{2}] \cos(\beta^{1})\cos(\beta^{2}) ]  \\ 
&-\frac{i}{2} \Omega [{\cal A}_2[\tilde  D^{2}+\tilde A^{2}-(\tilde B^{2}+\tilde C^{2})]\cos(\alpha^{2})\cos(\alpha^{1})  - {\cal B}_2 [\tilde D^{2}-\tilde A^{2}+\tilde B^{2}-\tilde C^{2}] \cos(\alpha^{2})\cos(\beta^{1}) \\
& ~~~~~~ - {\cal B}_2 [\tilde A^{2}-\tilde D^{2}+\tilde B^{2}-\tilde C^{2}]   \cos(\beta^{2})\cos(\alpha^{1}) -  {\cal A}_2 [\tilde A^{2}+\tilde D^{2}+\tilde B^{2}+\tilde C^{2}] \cos(\beta^{2})\cos(\beta^{1})]
\end{aligned}
\end{equation*}

Substituting the integral value of ${\cal A}_1$, ${\cal A}_2$, ${\cal B}_1$, ${\cal B}_2$ we get:
Using the equation \ref{Aint}

\begin{equation*}
\begin{aligned}
\delta E_{A} = -\frac{\mu^2 P}{8\pi^2}\int^{\infty}_{-\infty}d\omega~&\frac{\omega^2 }{\left(1-e^{-2\pi k\omega}\right)\left(\omega+\omega_0\right)\left(\omega-\omega_0\right)} \\
&[ 2\Omega (\tilde  D^{2}+\tilde A^{2}-\tilde B^{2}-\tilde C^{2}) f(\omega,L/2) \cos(\alpha^{1})\cos(\alpha^{2})\\
&-2\Omega (\tilde  A^{2}+\tilde B^{2}+\tilde C^{2}+\tilde D^{2}) f(\omega,L/2) \cos(\beta^{1})\cos(\beta^{2}) \\
&+ \{\cos^2(\alpha^{1})+\cos^2(\alpha^{2})+\cos^2(\beta^{1})+\cos^2(\beta^{2}) \}] 
\end{aligned}
\end{equation*}

In abbreviated notation

\be
\begin{aligned}
\delta E_A = -\frac{\mu^2 P}{8\pi^2} [&\Omega (\tilde  D^{2}+\tilde A^{2}-\tilde B^{2}-\tilde C^{2}) \cos(\alpha^{1})\cos(\alpha^{2}) \Delta_3 \\
&- \Omega (\tilde  A^{2}+\tilde B^{2}+\tilde C^{2}+\tilde D^{2}) \cos(\beta^{1})\cos(\beta^{2})\Delta_3  \\
& + \{ \cos^2(\alpha^{1})+\cos^2(\alpha^{2})+\cos^2(\beta^{1})+\cos^2(\beta^{2}) \}\Delta_2]
\end{aligned}
\ee

{}

\end{document}